\documentclass[preprint,10pt,1p]{elsarticle} 
\usepackage{graphicx}  
\usepackage{dcolumn}   
\usepackage{bm,amssymb,amsmath}        
\usepackage{sidecap}   
\usepackage{framed,float}
\usepackage{nomencl}
\usepackage[symbol]{footmisc}
\usepackage{xfrac}
\usepackage{color} 
\makenomenclature

\usepackage{fancyhdr}
\newcommand{\bom}{{\mbox{\boldmath $\omega$}}} 

\newcommand\shorttitle{\textsl{QT in trapped atomic BECs}}
\newcommand\authors{\textsl{M.C. Tsatsos et. al.}}
\numberwithin{equation}{section}
\usepackage[nottoc,notlof,notlot]{tocbibind} 

\begin{document}
\begin{frontmatter}

\title{Quantum turbulence in trapped atomic Bose-Einstein condensates}

\author{Marios C. Tsatsos\footnote{Corresponding author.}}
\ead{marios@ifsc.usp.br}
\author{Pedro E. S. Tavares,  Andr{\'e} Cidrim, Amilson R. Fritsch, M{\^o}nica A. Caracanhas}
\address{S{\~a}o Carlos Institute of Physics, University of S\~ao Paulo, PO Box 369 13560-970, S\~ao Carlos - SP, Brazil}

\author{F. Ednilson A. dos Santos}
\address{Department of Physics, Federal University of S{\~a}o Carlos, 13565-905, S{\~a}o Carlos - SP, Brazil}

\author{Carlo F. Barenghi}
\address{Joint Quantum Centre Durham-Newcastle, and School of Mathematics and Statistics, Newcastle University, Newcastle upon Tyne NE1 7RU, United Kingdom}

\author{Vanderlei S. Bagnato}
\ead{vander@ifsc.usp.br}
\address{S{\~a}o Carlos Institute of Physics, University of S\~ao Paulo, PO Box 369 13560-970, S\~ao Carlos - SP, Brazil}

\date{\today}

\begin{abstract}
Turbulence, the complicated fluid behavior of nonlinear and statistical nature, arises in many physical systems across various disciplines, from tiny laboratory scales to geophysical and astrophysical ones. The notion of turbulence in the quantum world was conceived long ago by Onsager and Feynman, but the occurrence of turbulence in ultracold gases has been studied in the laboratory only very recently. Albeit new as a field, it already offers new paths and perspectives on the problem of turbulence. Herein we review the general properties of quantum gases at ultralow temperatures paying particular attention to vortices, their dynamics and turbulent behavior. We review the recent advances both from theory and experiment. We highlight, moreover, the difficulties of identifying and characterizing turbulence in gaseous Bose-Einstein condensates compared to ordinary turbulence and turbulence in superfluid liquid helium and spotlight future possible directions.
\end{abstract}


\end{frontmatter}

\setcounter{footnote}{0}
\tableofcontents

\begin{framed}
 \printnomenclature
\end{framed}
\nomenclature[3]{\textcolor{white}{$\alpha$} \textsf{List~of~frequently~used~symbols~and~abbreviations}}{}

\section{Introduction \label{Sec:Introduction}}

\subsection{Why quantum turbulence?}
Turbulence occupies a unique position among the disciplines of physics, at the intersection of traditional fundamental problems on one hand and modern fashionable trends on the other. Influential scientists such as S. Weinberg and R. Feynman\footnote{Steven Weinberg  said that ``If we had the fundamental laws of nature tomorrow, we still wouldn't understand consciousness. We wouldn't even understand turbulence (\dots) That's an outstanding problem that has been with us for almost two centuries and we 're not very close to a solution.''\cite{FinalFrontierBook}. Richard Feynman, in a letter to a former student of his, explicitly mentioned that, among other problems, he has failed to come up with a theory of turbulence despite having spent several years on it \cite{LetterFeynman}.} considered turbulence one of the biggest unsolved problems in modern science. 
Nevertheless, there is not a single problem of turbulence. Turbulence is a context, that for many years has appeared as a mystery to mathematicians, physicists and engineers, challenging them with problems involving the nonlinear interaction of a huge number of degrees of freedom which are excited over a vast number of length scales. As the much newer field of quantum turbulence (QT) is slowly establishing itself in the turbulence community, new light might be shed on both the quantum and classical aspects of turbulence. 
In the quantum world, the recent advances in quantum trapped superfluids, possible due to techniques of controlling and trapping neutral bosonic atoms, open up a new window of possibilities for this topic. 

The ``super'' properties of ultracold quantum matter (superconductivity and superfluidity), discovered more than a century ago, furthered our understanding of the nature of atoms and molecules. The many-body theoretical approaches that emerged from the study of superfluidity and superconductivity became the most important tools in modern quantum physics. Gradually, new phenomena emerged from exploring new settings and configurations of the superfluid state. For example, with the introduction of rotation, a very peculiar response of the system was observed; if the liquid was thermal (i.e. in the normal state) then it would follow the rotation of its container. But if it was cooled down to temperatures smaller than a critical temperature $T_c$, then it would 
stand still, as if the container were not rotating. Because of the existence and the uniqueness of a macroscopic wave function, the liquid in the superfluid state can only rotate by threading itself with a number of quantized vortex lines, thin mini-tornadoes aligned along the axis of rotation. Vortex lines can be created also by stirring the liquid helium with grids or propellers, or by applying heat currents. Configurations of vortex lines are either laminar (when they are regularly distributed in space, for example rotating vortex lattices) or turbulent (when the vortex lines are tangled in space). Ordinary liquid helium ($^4$He) cooled below the critical temperature $T_c=2.17~\rm K$  (called the $\lambda$-point) is not the only system exhibiting quantum turbulence.  
The rare isotope $^3$He becomes also superfluid, but at much lower temperatures (in the $\rm{mK}$ region), and quantum turbulence is currently studied in its B-phase. This distinction is important, as topological defects in the A-phase are quite different.

In 2009, the first evidence \cite{Henn2009} of quantum turbulence in trapped, dilute atomic Bose-Einstein condensates (BECs) has opened new and exciting perspectives. Atomic condensates may constitute the ideal laboratory for testing fundamental aspects of quantum turbulence, and explore similarities and differences between quantum turbulence and turbulence in ordinary (classical) fluids. The study of the life-cycle of quantum turbulence -- from its nucleation and temporal evolution until its decay in a variety of excitations and phases -- might suggest analogies between quantum turbulence and classical turbulence.

The aim of this review is to highlight the recent and most important advances in the field of quantum turbulence in trapped gases, focusing on both theory and experiments. We try to give a self-consistent and as complete as possible presentation of our knowledge on turbulent behavior in quantum gases, as this has been boosted by the recent experimental progress. Although not intended to substitute standard textbooks in the field, much of the underlining theory is included in the paper and most concepts are explained for the layman. Our objective is to put the new findings in context and clarify core ideas of turbulence. We shall be particularly concerned with topics such as the generation of quantized vortices and turbulence, its evolution and decay, and experimental techniques to measure turbulent properties. We include in our discussion the limitations of both theory and experiment and point out new challenges.

Bose-Einstein condensation is a general phenomenon which appears not only in liquid helium and ultracold gases, but also in a variety of other non-gaseous systems, for instance magnons \cite{Demokritov2006} and exciton-polaritons \cite{Kasprzak2006}. In the present work, when referring to a BEC, we shall exclusively mean a trapped ultracold gaseous atomic BEC. 

\begin{figure}[!ht]
\begin{center}
\includegraphics[width=0.8\textwidth,natwidth=1070,natheight=220]{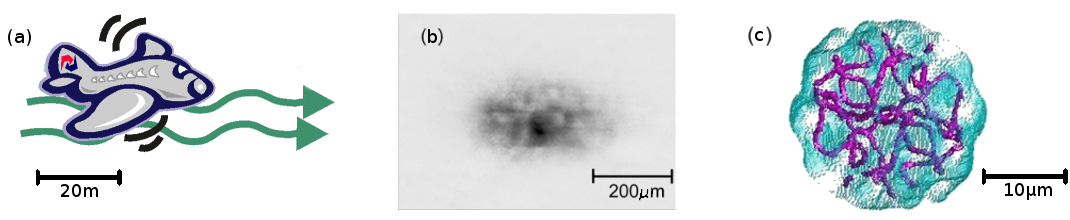}
\caption{We experience turbulence at every-day earth-scales when we travel by airplane through the turbulent air inside a cloud [panel (a)]. In this case the airplane is the probe of the turbulent cloud and the accidental bumps and jumps are caused by air currents and eddies. In the case of a microscopic quantum gas [panel (b)], turbulence must be watched from outside: one needs to release the trapped gas from its container, and from the unusual expansion of the gas deduce that the gas contains a complicated turbulent structure in the form of a tangle of vortices; panel (b) showed an experimentally expanded gas \cite{Henn2009}, panel (c) shows a computed tangle of vortices \cite{White2010}; the blue density isosurface is the condensate, the purple isosurfaces are the vortices.}
\label{Fig:TurbulenceGeneral}
\end{center}
\end{figure}

\subsection{Structure of this Review}
The plan of this Review is the following. In Sec.~\ref{Sec:Background} we present the background to the problem: we describe the basic physics of Bose-Einstein condensation, superfluidity, quantum vortices and quantum turbulence following a historic line of presentation. We then comment on the apparent similarities and differences between turbulence in ordinary fluids, turbulence in superfluid helium and turbulence in Bose-Einstein condensates. This section ends with our answer to the frequently-posed comment as to whether proper definitions of the words `turbulence' and `quantum turbulence' are necessary to make progress into our investigations.

We begin Sec.~\ref{Sec:Essentials} by exposing the reader to the basic theoretical background needed. In Secs.~\ref{SubSec:BriefCT} and \ref{SubSec:BriefQT} we briefly review the main results from classical and quantum turbulence respectively. In Sec.~\ref{SubSec:ThModels} we present the basic models known for theoretically treating trapped BECs and systems of vortices, followed by a simple exposition of the appearance of power spectra in BECs (Secs.~\ref{Sec:Hydro} and \ref{SubSec:WT}).

Section \ref{Sec:EnergyTransport} treats theoretically and experimentally the problem of energy flow and transport between different scales, a problem directly related to the problem of creation, propagation and decay of turbulence. To this end various different mechanisms are examined: vortex reconnections and Kelvin waves in three dimensions (Sec.~\ref{SubSec:Vortexreconnections}), vortex annihilation and decay in two-dimensional gases (Sec.~\ref{SubSec:2DQT}) and also phonon-mediated transport (Sec.~\ref{SubSec:Phonons}). Experimental results are intertwined with results from the theory and numerical analysis.

In Sec.~\ref{Sec:Experimental} we present the principal experimental results obtained by our group; the controlled creation of quantum turbulent and granular state in an ultracold trapped gas of rubidium. After a short presentation of the experimental setup in Sec.~\ref{SubSec:ExpSetup} and a presentation of all different types of excitations encountered (Secs.~\ref{SubSec:vortex}--\ref{SubSec:Granular}) emphasis is put on the relation of the different phases and excited structures (Secs.~\ref{SubSec:Diagram}) and also the novel phenomenon of self-similar expansion (Sec.~\ref{SubSec:self-similar}). An observed power-law in the momenta of the gas is discussed in Sec.~\ref{SubSec:MomDist}.

Section \ref{Sec:Discussion} concludes the present work. We highlight the key findings of the research to-date in quantum turbulence (Sec.~\ref{SubSec:WhatHaveWeLearnt}) and suggest some challenging steps in the near (or more distant) future needed to be taken in order to further our current knowledge in QT (Sec.~\ref{SubSec:OpenQuestions}). Related, but slightly more distant, to QT topics are discussed Sec.~\ref{SubSec:Related}. Last, Sec.~\ref{SubSec:Discuss} gives concluding remarks of the present exposition.

\nomenclature{$\rho$}{Density}
\nomenclature{$\xi$}{Healing length}
\nomenclature{$g$}{Interaction strength}
\nomenclature{$N$}{Particle number}
\nomenclature{$t_{\text{hold}}$}{Hold time: time that the BEC is held inside the trap before an absorption image is taken}

\nomenclature{QT}{Quantum turbulence}
\nomenclature{BEC}{Bose-Einstein condensate}
\nomenclature{CT}{Classical turbulence}
\nomenclature{NS}{Navier-Stokes}
\nomenclature{TF}{Thomas-Fermi}
\nomenclature{BS}{Biot-Savart}
\nomenclature{QVR}{Quantum vortex reconnection}
\nomenclature{GP}{Gross-Pitaevskii}
\nomenclature{WT}{Wave turbulence}

\nomenclature{QUIC}{Quadrupole-Ioffe configuration}
\nomenclature{TOF}{Time of flight: process that lets the gas expand and free-fall before measuring}
\nomenclature{\emph{in situ}}{A measurement has been performed while the gas is inside the trap}
\nomenclature{IEC}{Inverse energy cascade}
\nomenclature{2DQT}{Two-dimensional quantum turbulence}
\section{Background} \label{Sec:Background}
\subsection{Bose-Einstein condensation} 
Bose-Einstein condensation is the macroscopic occupation of the same quantum level by the majority of the particles of a system. Because of Pauli's principle, these particles must be bosons. The phenomenon, known since 1924, takes place if the temperature $T$ of the system is less than a certain critical value $T_c$. Initially it was believed that Bose-Einstein condensation should be limited to ideal (noninteracting) particles \cite{Bose1924}, but later it was realised that the presence of interparticle interaction 
can actually assist in the formation of the condensate \cite{Penrose1956}. In 1995 the first experimental realizations of Bose-Einstein condensation in dilute alkali gases were separately achieved by three experimental groups that cooled ensembles of sodium \cite{Anderson1995}, lithium \cite{Bradley1995} and rubidium \cite{Davis1995} to temperatures of few nK. Nowadays, BECs are routinely produced and studied in dozens of laboratories across the world. 

Upon cooling the system, Bose-Einstein condensation takes place when the mean interparticle distance $\langle l \rangle=\rho_0^{-1/3}$ becomes comparable to the de Broglie wavelength $\lambda_{dB}=h/(mv)$ of the particles. Here $\rho_0=N/V$ is the number density of $N$ particles occupying a volume $V$, $m$ and $v=\sqrt{k_B T/m}$ are the mass and the thermal velocity of the particles respectively and $k_B$ is Boltzmann constant. Assuming a homogeneous gas at constant number density $\rho_0$, the condition $\lambda_{dB}\sim \langle l \rangle$ yields 
\begin{eqnarray}
T_c &\approx& \frac{h^2 \rho_0^{2/3}}{m k_B},
\end{eqnarray}
for the critical transition temperature $T_c$. This simple qualitative argument (see \cite{KetterleWeb} for an intuitive presentation) predicts that an ensemble of bosons, if cooled below the critical temperature  $T_c$, undergoes Bose-Einstein condensation. An accurate calculation of the above $T_c$ differs only by a factor of $3.31$ \cite{Pethick2008}. 
Importantly, most experiments in BECs of trapped atoms are not done in homogeneous systems but rather in heterogeneous trapping potentials. In this case, the critical temperature depends on the external potential as  
the condensate is not uniformly distributed but centered in the potential minimum. For a harmonically trapped gas the critical temperature for condensation is \cite{Bagnato1987}
\begin{equation}
T_c = 0.15 \frac{h \overline{\omega}N^{1/3}}{k_B},
\end{equation}
where $\overline\omega=(\omega_x\omega_y\omega_z)^{1/3}$ is the geometric mean of the trap frequencies.
Note the scaling, distinct in the two cases. 

\begin{figure}
\begin{center}
\includegraphics[width=0.65\textwidth]{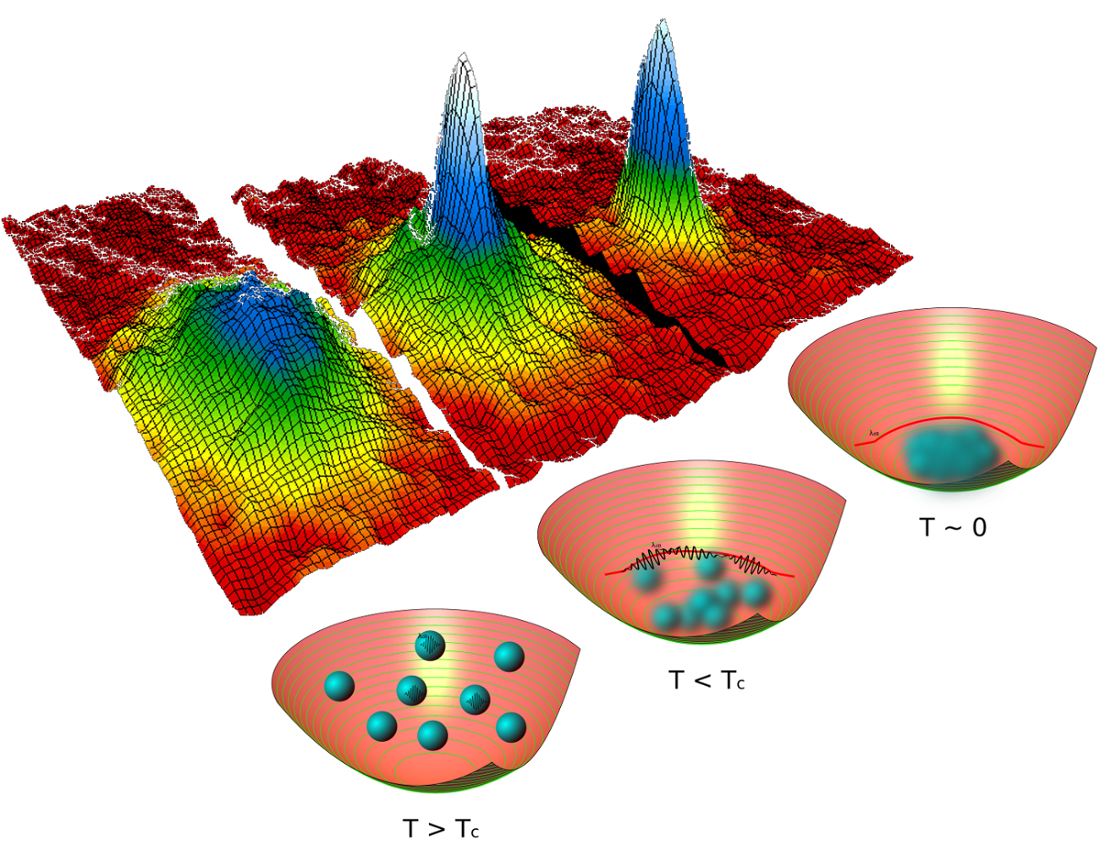}
\caption{Transition to a BEC state as the temperature is lowered. The upper row of images show momenta distributions taken in the JILA experiments (source: NIST/JILA/CU-Boulder). The lower row displays schematically
the distribution of particles in the trap and the condensation to the lowest particle state. 
\label{Fig:BEC}}
\end{center}
\end{figure}

The many-body coherent state that appears is shown in the sequence of pictures displayed in Fig. \ref{Fig:BEC}. For a gas to condense, diluteness and trapping (i.e. spatial confinement of the particles) are necessary. Typically, the number density is $10^{12}$ to $10^{15}$ particles per $\rm cm^3$. Trapping and cooling can be achieved in many different ways. While the trapping potential can be created efficiently by either magnetic or optical means, cooling is normally obtained by a combination of techniques like laser cooling and evaporation.

Research on Bose-Einstein condensation -- boosted by experimental accomplishments -- has rapidly gained a special position in modern physics \cite{Bagnato2015}. This activity has allowed us to create and study states macroscopic objects (consisting of few hundreds to few millions of particles) with properties usually associated with microscopic quantum objects. Atomic BECs serve as laboratories for testing many-body theories of physics and assist research in fields as diverse as quantum computation, metrology, foundations of quantum physics and even industrial applications. What distinguishes gaseous BECs from other condensed matter or atomic systems is the amount of control one can exert on the system: the type and sign of the interatomic interactions as well as the confining potentials can be changed at will, simply by modifying externally the applied magnetic fields \cite{Inouye1998,Kevrekidis2003}. 

More interestingly and due to the unique controllability of atomic ultracold gases, BECs can serve as platforms where other systems from condensed matter -- and beyond -- are being simulated. Examples are bosonic Josephson junctions \cite{Albiez2005,Ryu2013} the Bose-Hubbard model \cite{Jaksch1998}, vortex lattices,  synthetic gauge fields \cite{Lin2009}, hexagonal lattices and  graphene \cite{Soltan-Panahi2011}, collapsing attractive systems (nicknamed `bosenova') \cite{Donley2001} and more.

Interactions complicate the dynamics of Bose-Einstein condensates, and introduce effective nonlinearities in the mean-field equations of motion.  In ultralow temperature environments, interparticle interactions are sufficiently well described by considering $s$-wave scattering only. In the mean-field approach -- the most commonly used --  each particle of the system interacts with the mean-field generated by the other particles. As we shall see later, this approximation assumes total condensation at all times, mapping the many-body problem to a one-body problem. Nonetheless, being a gas of interacting atoms, a BEC also retains many properties typical of classical fluids. Hence, features as solitons (bright, dark or other more complicated topological excitations like skyrmions), vortices, vortex rings and other fluid properties have all been observed in atomic physics laboratories and can be nowadays routinely created and imaged. Interestingly, the above appear as stable \emph{topological excitations} of the quantum system. In contrast, ordinary classical fluids do not contain well-formed `thin' and long-lived vortices; the latter are rather fuzzy and its characterization as line filaments is hence not clear. In such a way, excitations that in viscous (normal) fluids are mathematical ideals only, in atomic BECs become real. The underlying property of such a condensate is a special kind of fluid called a `superfluid', which is the topic of the next section.

\subsection{Superfluidity and vortices}
\label{SubSec:Superfluidity}
In 1908, that is long before the first experimental realizations of Bose-Einstein condensation in atomic vapours, liquid helium was produced at Leiden by the Dutch physicist Heike Kamerlingh Onnes. Between 1927 and 1932 a group of physicists also at Leiden (W. H. Keesom, M. Wolfke, A. Keesom J.N. van der Ende and K. Clusius) \cite{Balibar2007} noticed that liquid helium has two distinct phases, named He I and He II, separated by the apparent singularity that the specific heat shows as a function of the temperature  -  the famous ``lambda point''. This singularity appears at a temperature around 2.2K at low pressure. The two liquid phases seemed very different. Liquid He I behaved like an ordinary fluid, whereas liquid He II (the low temperature phase) had unusual mechanical and thermal properties. The key step forward was the discovery of superfluidity by Kapitza \cite{Kapitza1938}, Allen and Misener \cite{Allen1938}, which stimulated much work on macroscopic quantum systems. In their experiments, they showed that liquid He II can flow through very thin tubes or slits without viscous dissipation. In other words, helium, the first quantum fluid explored, has no viscosity in the superfluid phase, in direct analogy to the lack of resistance displayed by a superconductor. Owing to its superfluid nature, He II apparently defies gravity by siphoning itself out of a container. 

In 1938 Fritz London proposed that the unusual properties of He II are actually a manifestation of Bose-Einstein condensation \cite{London1938}. He, thus, realized that a collective wave function could describe the condensed atoms if there was a macroscopic occupation of the zero momentum state. London's idea was set aside for a while, as the interest was taken by a new phenomenological theory (the two-fluid model) proposed by Tisza  \cite{Tisza1938}. According to the two-fluid model, He II consists of two fluid components, the normal fluid and the superfluid, each with its own density and velocity field. The normal fluid moves like an ordinary fluid with all of the conventional thermodynamic properties such as entropy, temperature and viscosity, while the superfluid has no entropy, is inviscid and flows without friction. Landau formalized the model, and produced a complete set of thermodynamic relations for the two-fluid system. The two-fluid model of Landau and Tisza turned out to be enormously successful in describing the properties of He II and its strange observed effects (the siphon effect, the fountain effect, the mechano-caloric and thermo-mechanical effects). The two-fluid model also led to the prediction of the existence of second sound, a mode of oscillations in which superfluid and normal fluid move in antiphase creating a temperature wave rather than the usual pressure wave of ordinary (first) sound.

One major problem with the two-fluid model, however, is that it assumes zero superfluid vorticity ($\pmb\omega=\nabla\times \mathbf v$). In fact, a zero-viscosity fluid should be dissipation-free and therefore have a conservative velocity field, $\nabla\phi=\mathbf v$, implying absence of vorticity, $\pmb\omega=\nabla\times\nabla\phi=0$. Experiments with helium in a rotating vessel, however, showed the typical parabolic surface of a rotating liquid, which would imply vorticity proportional to the angular velocity of rotation, like ordinary fluids. This observed parabolic profile was independent of temperature, in disagreement with the expectation that it should be proportional to the relative density of normal fluid which is strongly temperature dependent (as only the normal fluid can have a curl, hence rotate).

At this point it is worth examining London's association of the superfluid with a Bose-Einstein condensate where a single quantum state is collectively occupied by all particles. London realised \cite{London1938} that a collective wavefunction $\psi$ could be used for the condensed atoms if there was a macroscopic occupation of the zero momentum state,
\begin{equation}
\psi(\mathbf{r},t)=\sqrt{\rho(\mathbf{r},t)} e^{iS(\mathbf{r},t)},
\label{Eq:Wavefunction}
\end{equation}
where $\rho=|\psi|^2\equiv |\psi(\mathbf r,t)|^2$ is the condensate density, appropriately normalized
\begin{equation}
\int |\psi|^2 d\mathbf{r} = 1,
\end{equation}
and $S\equiv S(\mathbf r,t)$ is its phase. 
Following the usual quantum mechanical prescriptions, we define the probability current as
\begin{equation}
\mathbf{j}=\frac{\hbar}{2mi}(\psi^{\ast}\nabla\psi - \psi\nabla \psi^{\ast})=\frac{\hbar}{m} \rho\nabla S.
\end{equation}
This is in fact a \emph{flux} of the density $\rho$ that flows with velocity
\begin{equation}
\mathbf v=\frac{\hbar}{m} \nabla S,
\label{eq:gradS}
\end{equation}
hence the flux 
\begin{equation}
\mathbf{j}=\rho \mathbf v,
\end{equation}
takes the familiar from classical physics form. Assuming conservation of mass, we can write down the \emph{continuity equation} for the current $\mathbf j$:
\begin{equation}
\nabla\cdot \mathbf{j} + \frac{\partial \rho}{\partial t} = 0.
\label{Eq:continuityQuant}
\end{equation}
Interestingly, as we shall show in the following Section, such a continuity equation is easily derived from the (linear or nonlinear) Schr\"odinger equation by the Madelung transformation.

The identification of the velocity with the gradient of a scalar has fundamental implications on the types of motion that can occur in a condensate or superfluid. Indeed, a direct consequence of Eq.~(\ref{eq:gradS}) is that
\begin{equation}
\nabla\times\mathbf{v}=0,
\label{Eq:ZeroCurl}
\end{equation}
as envisaged by Landau. Note that the above result holds true when the field $\psi$ has continuous first and second derivatives. This is not the case if the system contains at some point a \emph{vortex line}, i.e., a line singularity along which the velocity $\mathbf v$ diverges. Hence the velocity field of a vortex-free field is irrotational. Recall now that the condensate wavefunction $\psi$ is required to be single-valued: a revolution along a closed loop $C$ should leave the value of $\psi$ unchanged, hence the phase $S$ can change at most by $2n\pi$, where $n$ is an integer. In other words, the circulation $\Gamma$ around any path $C$ yields only integer multiples of the quantum of circulation $\kappa=h/m$:
\begin{equation}
\Gamma=\oint_C  {\mathbf{ dr}}\cdot\mathbf{v}=\frac{\hbar}{m}\Delta S=n\frac{2\pi\hbar}{m}=n\frac{h}{m}.
\label{Eq:Circulation}
\end{equation}
This was the argument that led Onsager\footnote{Indeed, in 1949 L. Onsager already mentioned that ``Thus, the well-known invariant called hydrodynamic circulation is quantised; the quantum of circulation is $h/m$''\cite{Onsager1949a}. And also, ``vortices in a suprafluid are presumably quantised; the quantum of circulation is $h/m$, where $m$ is  the  mass  of  a  single  molecule'' \cite{Onsager1949d}.  Interestingly, the idea of the quantization of circulation in a superfluid, together with the quantization of flux proposed a year earlier by London, were not widely accepted until Feynman's independent work \cite{Feynman1955}. See also \cite{Eyink2006} for a discussion.} \cite{Onsager1949a} and, later on, Feynman \cite{Feynman1955}, to introduce the quantization of the vorticity in superfluid helium and to identify the quantity $h/m$ as the quantum of circulation.

As we shall see, there exist vortex solutions of the quantum mechanical equations of motion (see Sec.~\ref{SubSec:GrossPitaevskiiModel} and Sec.~\ref{SubSec:Vortexreconnections}), and these constitute simple paradigms of configurations with quantized circulation. To demonstrate that, we construct the simple wavefunction which, as we shall show, contains a straight vortex at the origin, aligned with the $z$-direction. Using cylindrical coordinates, this solution has the form 
\begin{equation}
\psi(\mathbf r)=\psi(r,z,\phi)=f(r,z)e^{i n \phi},
\label{Eq:WavefunctionII}
\end{equation}
where $f(r,z)$ is some prescribed real function and $n$ an integer. This ansatz assumes that the magnitude of the wavefunction, $\vert \psi \vert=f$, is cylindrically symmetric; the azimuthal dependence is contained only in the phase of $\psi$ and has the form $S=n\phi$. A rotation $\phi\to\phi+2n^\prime\pi$ about the $z$-axis leaves the wavefunction invariant, as it owes to. The velocity field corresponding to this wavefunction is in the azimuthal direction (parallel to the unit vector $\hat{\bm{\phi}}$): 
\begin{equation}
\mathbf v=\frac{\hbar}{m}\nabla S = \frac{\hbar}{m}\frac{1}{r}\frac{\partial S}{\partial\phi} \hat{\bm{\phi}} =  \frac{\hbar}{m}\frac{n}{r} \hat{\bm{\phi}}.
\label{vfield}
\end{equation}
Hence, following the quantization of the circulation, the velocity field is also quantized, since its magnitude $|\mathbf v|=v$ admits values that are integer multiples of $\hbar/m$. Note, moreover, that $\mathbf v$ is irrotational as long as $r \neq 0$. Formally, we write the vorticity as a (two-dimensional) delta function to show that is localized on the cylindrical axis \cite{Pethick2008}, i.e. 
\begin{equation}
\nabla\times\mathbf{v}=n\frac{h}{m}\delta^{(2)}(r,\phi) \hat{\bm{z}},
\label{Eq:Vorticity}
\end{equation}
where $\hat{\bm{z}}$ is the unit vector in the z-direction.

We conclude that in a quantum system the flow can be both inviscid and irrotational as long as it is not continuous. The velocity field  has a singularity on the axis at $r=0$ (generally, at some $r_0$); the wavefunction therefore must go to zero at exactly the same points, that is, inside the vortex core. In other words, both real and imaginary parts of $\psi$ go to zero at $r=0$ yielding a zero density at the singularity $r=0$. Therefore, the fact that the velocity diverges as $r \to 0$ is not physically inconsistent: there are no atoms in the vortex core which move with infinity speed. The vortex line is a hole in the superfluid, that therefore renders its bulk a multiply-connected space. 

Mathematically, the flow pattern has the characteristics of a classical vortex line, as can be found in elementary fluid dynamics textbooks: a node at $r=0$ and a velocity field around it which decays as $r \to \infty$ away from the axis.

Note that the integer $n$ of Eq.~\eqref{Eq:WavefunctionII} determines how many $2\pi$ multiples are added to the condensate phase on a $\phi \to \phi + 2\pi$ single rotation and is also called the charge of the vortex. A quantized vortex is a topological excitation of the system with higher energy than the ground state (the condensate without the vortex). It can be shown that the energy of the vortex increases with the square of the vortex charge, $E \propto n^2$. Therefore, for given charge $n$, it is energetically favourable for the system to contain $n$ singly-charged vortices rather than one multiply charged vortex \cite{Kawaguchi2004,Kwon2014}.

\subsection{Quantum turbulence \label{SubSec:QT}}
\begin{figure}
\centering
\includegraphics[width=0.6\textwidth]{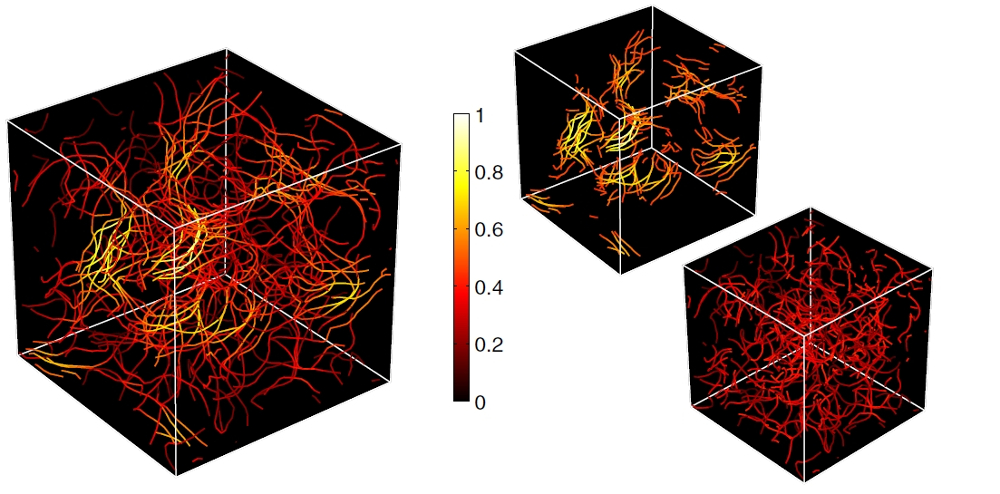}
\caption{Left panel: tangle of vortex lines computed in a periodic domain to model superfuid helium, analyzed in terms of the local vorticity. The vortex lines are locally colored according to the magnitude of the local `coarse-grained' vorticity. Yellow (lighter color) corresponds to large vorticity ($\bom > 1.4 \omega_{rms}$), 
red (darker color) to small vorticity ($\bom<1.4 \omega_{rms}$); in other words, the yellow lines are vortices which are part of a bundle of vortices which are all locally parallel to each other. Right panel: yellow and red lines plotted separately. Source: paper of Baggaley \emph{et al.} \cite{Baggaley2012}.
\label{Fig:Baggaley-Laurie-Barenghi-fig3}
}
\end{figure}

For the sake of simplicity, the vortex lines described in the previous sections were considered straight. We know however that vortex lines can sustain helical deformations away from the straight position called Kelvin waves. These waves rotate at angular velocity $\omega \sim \kappa k^2$, and propagate at phase velocity $\omega/k$, where $\kappa=h/m$ is the quantum of circulation, $k=2 \pi/\lambda$ is the wavenumber and $\lambda$ the wavelength. We also know that two lines which approach each other can reconnect, forming a cusp which then relaxes into Kelvin waves.
%
Very disordered configurations of vortex lines are easily created in liquid helium or atomic BECs and can be manifestation of quantum turbulence. The possibility of such quantum turbulence was already touched upon by Feynman in his seminal work on quantized vorticity \cite{Feynman1955}. The impact of Feynman's -- and also Hall's and Vinen's \cite{Hall1956a,Hall1956b} work -- was such that for years research towards QT has been mainly concerned with $^4$He. The recent interest in atomic BECs has been heavily motivated by progress in understanding quantum turbulence in liquid helium ($^4$He and $^3$He) experiments \cite{Skrbek2012,Barenghi2014b}. A striking discovery has been that, under appropriate forcing, quasiclassical behavior arises in such quantum systems,  displaying statistical properties that characterize ordinary turbulence; an example is the celebrated Kolmogorov $-5/3$ scaling of the energy spectrum \cite{Barenghi2014a} which suggests the existence of a classical energy cascade from large to small length scales. This classical-like scaling defines the \textit{quasiclassical turbulence} or \textit{Kolmogorov turbulence} and can be found for a specific inertial range, provided that the energy flux inside this range is maintained fixed with constant injection of energy in large scales and dissipation at small scales at the same rate. This interpretation requires self-similarity throughout a long range of scales; large bundles of vortices transfer energy to smaller bundles, in a process which goes down to the scale of single vortices. 
Numerical simulations \cite{Baggaley2012} determined that if the energy spectrum of the turbulent system obeys the Kolmogorov $k^{-5/3}$ scaling, the vortex tangle contains transient regions where the vortex lines are oriented in the same direction (vortex bundles); the large scale flows generated by such parallel lines concentrate the energy in the small $k$ region of the energy spectrum. Figure \ref{Fig:Baggaley-Laurie-Barenghi-fig3} clarifies that such vortex bundles also contain many random vortex lines; in other words, the spatial polarization is only partial. However, although Kolmogorov energy spectra have been observed, there is yet no direct experimental observations of these vortex bundles.

Kolmogorov-like turbulence has been seen in helium at both high and low temperatures. Quasiclassical behavior in the limit of high temperature $T \to T_c$ is not surprising, particularly in liquid helium driven by bellows or propellers at temperatures just below $T_c$: in this regime we expect the turbulent normal fluid to dominate the dynamics due to its relative larger density ($\rho_n \rho \approx 1$ and $\rho_s \to 0$, where $\rho_s$, $\rho_n$ and $\rho=\rho_s+\rho_n$ are respectively the normal fluid, superfluid and total density).
Quasiclassical behavior in the low temperature limit is more surprising, but can be understood if we recognize that the energy transfer responsible for the cascade arises from the key nonlinearity of the Euler equation, the $({\bf v} \cdot \nabla) {\bf v}$ term. This result confirms the view that quantum turbulence contains the fundamental mathematical skeleton of classical turbulence.
%

Numerical simulations and, more recently, direct visualisation based on tracer particles, show that quantum turbulence consists of a disordered tangle of vortex lines. In the high temperature range  ($T>1~\rm K$ in $^4$He) the vortex lines remain relatively smooth as friction with the normal fluid damps the Kelvin waves; kinetic energy is thus turned into heat by the normal fluid via viscous forces. In the low temperature range  ($T<1~\rm K$ in $^4$He) the vortex lines are very wiggly, as there is not enough friction to smooth out the Kelvin waves. However, other physical processes dissipate turbulent energy (see Sec.~\ref{Sec:EnergyTransport}). Rapidly rotating Kelvin waves of very short wavelength are created by wave interactions (a process called the Kelvin wave cascade \cite{Vinen2001,Barenghi2014b}), which, if their wavelength is short enough, emit phonons; the sink of turbulent kinetic energy is acoustic rather than viscous.

Under other conditions, a different kind of turbulence (called \textit{ultraquantum turbulence} or \textit{Vinen turbulence}) has also been found both experimentally \cite{Walmsley2008} and numerically \cite{Baggaley2012e}, characterized by random tangles of vortices without large-scale, energy-containing flow structures.

We have no direct experimental judgement as of whether turbulence in atomic BECs is Vinen-like, Kolmogorov-like or both. However, from the present analysis (see Secs.~\ref{Sec:Scales} and \ref{SubSec:Differences}) it results that in a trapped BEC Vinen ultraquantum turbulence is more likely to emerge. It has been recently demonstrated \cite{Zamora2015} that in the case of only two vortices initially in an orthogonal configuration, or even in the case of one doubly-charged vortex precessing and decaying in a spherically symmetric $T=0$ trapped gas, there appear Kolmogorov-like spectra of the energy in the momentum space, but only over a wavenumber range of one decade. Numerical calculations based on the Gross-Pitaevskii equation \cite{Kobayashi2005} suggest that Kolmogorov scaling should emerge from the dynamics of vortices in a trapped Bose gas, but there is no experimental verification of the effect yet. More theoretical work and experiments are necessary to characterize the turbulence observed in BECs, particularly because of the difficulty in experimentally measuring the velocity field within a trapped condensate. What has become clear recently is that excitations and vortices in a trapped BEC display very rich dynamics, which should be fertile hunting ground for theory and experiments, see Fig.~\ref{Fig:NaturalScales}. In the present Review, especially in Sec.~\ref{Sec:Experimental}, we focus on features of the turbulent quantum gas that are experimentally accessible and might constitute Kolmogorov-independent criteria for turbulence.

Last, another distinction which should be made when studying turbulence is its nature regarding forcing. Forced systems can exhibit the so-called \textit{stationary turbulence} that requires a constant injection of energy and also removal of it at the same rate by dissipative mechanisms. In experiments in 3D systems, energy injection happens in larger scales, which then cascades down in a fixed flux to the small scales (direct energy cascade). The opposite happens for the 2D scenario (inverse energy cascade). However, in nature, a stationary turbulent system is destined to decay at a certain point, when the system is ceased to be forced. This frames a very distinct out-of-equilibrium scenario known as \textit{decaying turbulence}. 

\subsection{Length scales} \label{Sec:Scales}
Understanding classical turbulence involves studying processes that happen in many scales which are usually separated by orders of magnitude. In that respect, QT is no different. The ratio $L/\xi$ quantifies the available space for its development; distance $L$ is the typical largest scale of the system (e.g. size of a helium container or extension of the trap for an atomic superfluid) and $\xi$ is the healing length, which sets the approximate size of the vortex core (see also Sec.~\ref{SubSec:GrossPitaevskiiModel}). A large value of $L/\xi$ enables the development of self-similarity throughout several scales, increasing thus the inertial range. For a quantum system to exhibit quasiclassical spectral characteristics (Kolmogorov scaling), a relatively large ratio is necessary and the quantity $\log (L/\xi)$ tells us roughly how many decades are available between $\xi$ and $L$ for this to happen. In that sense, systems with small ratio present few decades [($\log (L/\xi) \approx 1$ or $2$]. This is an intrinsic spatial limitation that forbids formation of large-scale, self-similar structures. On the other hand, in experiments involving superfluid He $\log (L/\xi) \approx 4$ and this number justifies why both quasiclassical and ultraquantum limits could be observed \cite{Maurer1998}. In contrast, current experiments with atomic BECs display $1$ to $2$ decades and the observation of complicated vortex tangle dynamics in such relatively small-ratio systems \cite{Henn2009} shows evidence of ultraquantum turbulence. 

\begin{figure}
\centering
\includegraphics[width=0.9\textwidth]{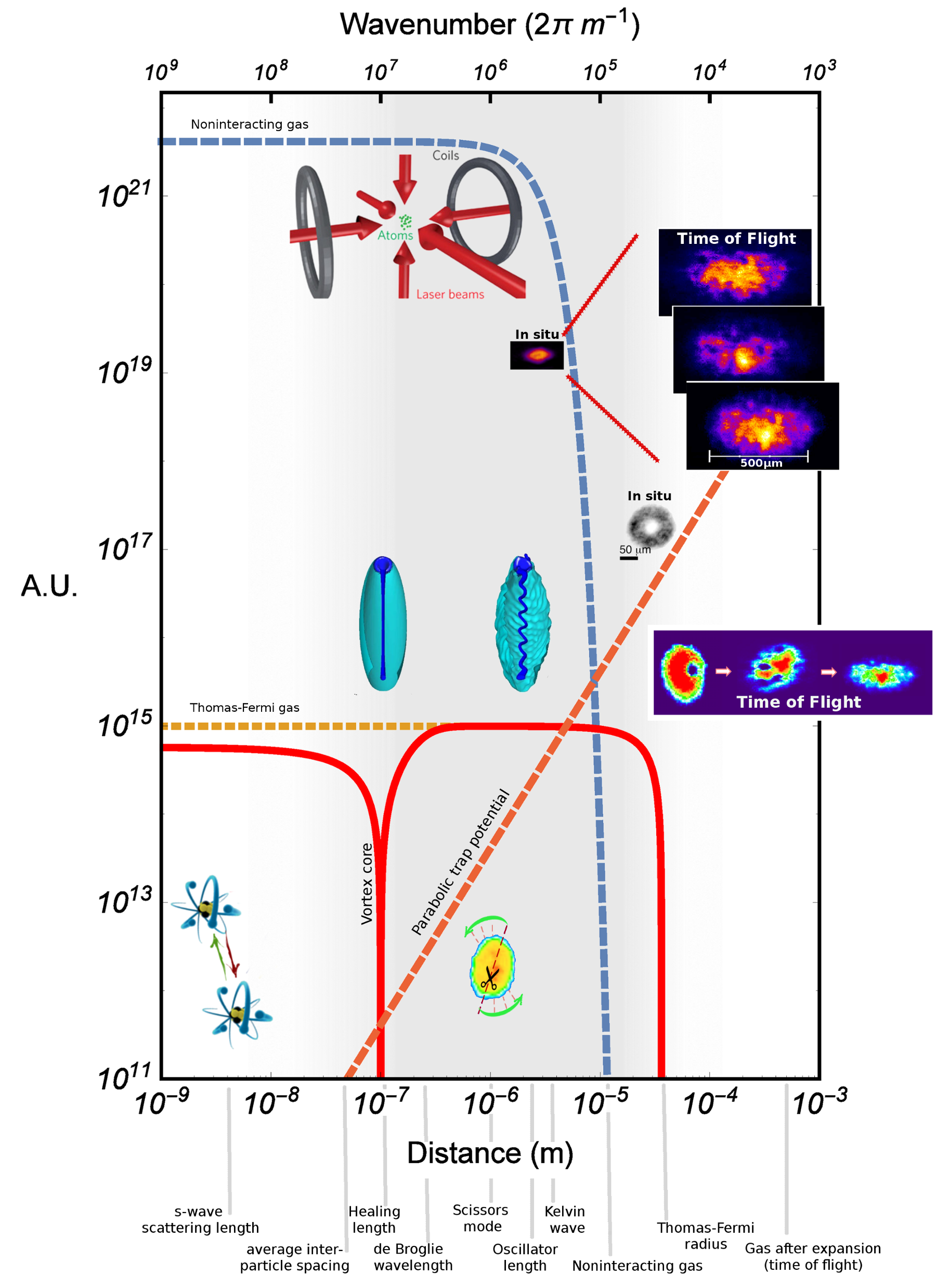}
\caption{Natural scales appearing in a trapped Bose-Einstein condensate of $\sim 10^5$ $^{87}Rb$ atoms. In  such a system there is a variety of scales and physical interactions available, spanning microscopic to meso- and macroscopic scales; from atom-atom interactions to vortex (Kelvin) waves and from quadrupolar and breathing modes to turbulent self-similar time-of-flight expansion of the gas. Notably, turbulence involves the energy transfer over many different scales. However the physical mechanisms (vortex cascades for instance) involved owe to be same in nature. The relevant scales for quantum turbulence would be these of the shaded area. For details on the experimental values used here see \cite{SemanThesis}. Source of the inset images: papers of J. Est{\'e}ve \cite{Esteve2013}, Simula \emph{et al.} 
\cite{Simula2008}, Neely \emph{et al.} \cite{Neely2013}, Seman \emph{et al.}  \cite{Seman2011} and the theses of Henn \cite{HennThesis} and G. Bagnato \cite{BagnatoThesis}.}
\label{Fig:NaturalScales}
\end{figure}

In Fig.~\ref{Fig:NaturalScales} we present the scales that naturally appear in a trapped BEC of $10^5$ rubidium particles at ultralow temperatures: from distances as low as few n$m$ (extension of the scattering length and interatomic spacing) up to the size of $mm$ of an expanded gas. As we can see, there is a plethora of different physical mechanisms and phenomena manifesting over various scales. However, the relevant for turbulence scales span about two decades (shaded region of Fig.~\ref{Fig:NaturalScales}). 

\subsection{Differences between classical, helium and quantum turbulence \label{SubSec:Differences}}
There are significant physical differences between classical and quantum turbulence which affect the nature of the turbulence and the experiments which can be performed.

\begin{enumerate}
\item{\emph{Choice of parameters}}
When studying turbulence in liquid helium, flow properties such as density and viscosity can be varied by changing the pressure and the temperature of the sample. The vortex core size (which is proportional to the superfluid healing length) and the quantum of circulation are fixed and cannot be changed. In atomic BECs, virtually all parameters can be varied by large amounts by tuning the number of atoms, the confining trap, and the strength of the interactions. This freedom however has a drawback, because it makes it difficult to compare the hydrodynamics regimes achieved in different experiments. 
In most experiments, atomic BECs are confined by harmonic external potentials, so the condensate  density and other quantities such as the sound speed and the healing length depend on the position. Again, this feature makes it difficult to infer general properties about the dynamics of turbulence, more so because results are reported in harmonic oscillator units, which, unlike natural units based on the healing length, are unrelated to the vortices. Fortunately, box-like traps have been created recently \cite{Gaunt2013}.
\item{\emph{Vorticity}}
The key difference between classical turbulence and quantum turbulence is the nature of the vorticity, which is continuous and unconstrained in the former and discrete and quantized in the latter. Quantum turbulence thus offers a clear advantage over classical turbulence: the `building blocks' (vortices and eddies) are clearly defined and simpler to detect \cite{TsubotaJoPCM2009}: they are zero-density points (or lines) of the condensed gas, around which the quantum mechanical phase wraps by multiples of $2 \pi$. 
\item{\emph{Range of length scales}}
Another major difference between turbulence in classical fluids, turbulence in liquid helium and turbulence in atomic gases is the range of length scales available. In ordinary fluids, vorticity is a continuous field. The largest vortices are as large as the whole system, $D$, and the size of the smallest eddies is of the order of the Kolmogorov length $\eta \approx D Re^{-3/4}$. Smaller eddies are destroyed by viscous forces.  Here, the Reynolds number $Re=v D/\nu$ is a dimensionless velocity which quantifies the intensity of the turbulence, $v$ is the flow velocity measured at the scale $D$, and $\nu$ is the kinematic viscosity of the fluid. 
Consider, for instance, the experiment of Grant et al. \cite{grant1961} in which a small vessel dragged a submarine probe which measured velocity fluctuations of water in the Seymour Narrows (a tidal channel near Vancouver) and confirmed the predictions of Kolmogorov's turbulence theory. We estimate $D \approx 0.75~{\rm Km}$ and $\eta \approx 0.3~{\rm mm}$, which means that vortices ranged in size over six orders of magnitude.

In superfluid helium (and also atomic BECs) vortices are discrete, not continuous, and the intensity of the turbulence is quantified by the vortex line density $L$ (length of vortex line per unit volume); typical values are $L \approx 10^4$ to $10^6~{\rm cm^{-2}}$, which means that a typical vortex separation is in the range from $\ell \approx L^{-1/2} \approx 10^{-2}~{\rm cm}$ to $10^{-4}~{\rm cm}$. The vortex core size is of the order of the superfluid healing length, which is $\xi \approx 10^{-8}~{\rm cm}$ in $^4$He and $10^{-6}~{\rm cm}$ in $^3$He-B. There is hence a very large separation of length scales between vortex core size, intervortex distance and system size. Considering the experiment of Maurer and Tabeling \cite{Maurer1998}, who first verified Kolmogorov's law in helium, we get: $\xi \approx 10^{-8} \ll \ell \approx 10^{-4} \ll D \approx 10{\rm cm}$. In atomic BECs, where typical atom numbers are $N \approx 10^4$ to $10^8$, the scales are somewhat different (see also Fig.~\ref{Fig:NaturalScales}). As in helium, the vortex core is of the order of the healing length $\xi$, which is now larger (typically $\xi \approx 10^{-5}cm$) and the intervortex distance $\ell$ is only few times the healing length. The system size $D$ is still larger than the latter, typically few dozens to hundreds of ${\rm \mu m}$. Even though the inequality $\xi < \ell < D$ still holds true, these length scales are now comparable, which has implications for the nature of the resulting turbulence. The reason is the following. Traditionally, a classical turbulent state is associated with the nucleation of vorticity (for instance, at the boundaries, by Kelvin-Helmoltz or baroclinic instability) followed by nonlinear vortex interactions. Such interaction usually involves an energy cascade from `parent' to `daughter' eddies (see Fig. \ref{Fig:TurbulenceGeneral}) and the formation of smaller and smaller structures until the appearance of scaling laws (such as Kolmogorov's 5/3 law). At very short length scales, energy is then dissipated. There is evidence that a similar process occurs in superfluid helium \cite{Barenghi2014a,Barenghi2014b};
whether it happens in an atomic BEC gas clearly depends on the range of length scales available, which is currently debated \cite{White2014,Reeves2012}. 
\item{\emph{Two-fluid nature}}
Another important difference between classical and helium turbulence is the two-fluid nature of the latter. Below the superfluid transition, liquid helium is a homogeneous, intimate mixture of two fluid components, the viscous normal fluid (which carries the entropy of the system) and the inviscid superfluid. The relative proportion of superfluid to normal depends on the relative temperature $T/T_c$. The thermal excitations, which make up the normal fluid, interact with the superfluid vortex lines (mutual friction) thus coupling the two fluids. Only at sufficiently low temperature (typically $T < 0.5 T_c$) in both $^4$He and $^3$He-B the normal fluid is effectively negligible and we face pure quantum turbulence. That is to say, turbulence of discrete vortex lines undamped by friction \cite{Walmsley2007,Bradley2006,Bradley2011,Hosio2013}. At higher temperatures ($T>0.5 T_c$) the nature of turbulence is affected by the mutual friction. In $^4$He the normal fluid has a relatively small kinematic viscosity (about sixty times less than that of water) and therefore the normal fluid is turbulent in most experiments. We face therefore the rather difficult problem of two distinct turbulent fluids -- the normal fluid and the superfluid -- which are dynamically coupled. Historically, this was the first problem which was studied; we refer the reader to Refs.~\cite{Donnelly1991Quantized,Nemirovskii2013} and references therein. In $^3$He-B, the kinematic viscosity of the normal fluid is very large, so in most experiments the normal fluid is laminar or effectively at rest. We face therefore the simpler problem of quantum vortices moving under friction forces; indeed, there has been some success in predicting whether such forces can prevent the onset or turbulence or not \cite{Finne2003,Kobayashi2005b,TsubotaJoPSJ2008}. 

The situation in atomic BECs is similar: pure quantum turbulence exists only if $T\ll T_c$. Unlike liquid helium however, in most experiments the thermal gas is in a ballistic rather than fluid regime. Still, the idea of friction should hold, for example it is predicted \cite{Jackson2009} that an off-centre vortex will spiral out of the BEC as predicted by the vortex dynamics in helium. However, the experimental conditions are such that allow for a sufficiently good control of the temperature and hence the relative density of the thermal cloud.
\item{\emph{Dissipation}}
Ordinary fluids are viscous. That is, unless energy is supplied to compensate for viscous losses of kinetic
energy, classical turbulence decays. In a two-fluid system, at high relative temperature the inviscid component can transfer energy to the viscous component via friction. We expect therefore that any sound waves, solitons and vortices in the condensate will be damped by the presence of the thermal cloud. As said above, at sufficiently low temperatures the thermal gas is negligible and quantum fluids are effectively pure superfluid. Surprisingly, turbulence will still decay: the kinetic energy contained in the vortices will not remain constant, but will be turned into sound waves within (and at the surface of) the condensate (see also Sec.~\ref{SubSec:Phonons}).
\end{enumerate}

\subsection{A strict definition of quantum turbulence? \label{Sec:ProbDef}}
Quantum turbulence is a relatively young field of physics research and the question has been raised as to whether there is (or there should be) a proper definition of the term `quantum turbulence'. This question unavoidably triggers a second question: what is `turbulence'?

The term `quantum turbulence' was first used in 1982 in the Ph.D. thesis of one of the present authors \cite{Barenghi1982} and later in 1986 by Donnelly and Swanson \cite{DonnellySwanson1986} in an article about turbulence in superfluid helium. Until 1986, it was customary to refer to turbulence in liquid helium as `superfluid turbulence` (an example is Tough's extensive 1982 review article \cite{Tough1982}). Donnelly and Swanson shifted the attention from the property that the turbulent fluid in question is a  superfluid (i.e. that it has zero viscosity) to the property that the vorticity is quantized (a consequence of the existence of a macroscopic wavefunction).  
This change of point of view was almost prophetic, as later experiments demonstrated that superfluid turbulence decays despite the lack of viscosity. In 2000, Davis et al \cite{Davis2000} used a vibrating grid to excite turbulence in $^4$He at mK temperatures ($T/T_c \approx 10^{-3}$, a temperature regime in which the normal fluid is utterly negligible), and observed that, when left unforced, the turbulence quickly vanishes. The observation raised a puzzle: why does turbulence decay if the fluid has zero viscosity?

Now we know that in $^4$He, at temperatures above $1~\rm K$, the turbulent kinetic energy contained in the quantum vortices (created by stirring helium with grids or propellers) is transferred to the normal fluid by mutual friction, and then is converted into heat by viscous forces, as in ordinary fluids. Despite the normal fluid being negligible, below $1~\rm K$ the kinetic energy of the vortex lines is not conserved: the acceleration of rapidly rotating Kelvin waves along the vortex lines \cite{Leadbeater2003} and the cusps created by vortex reconnections \cite{Leadbeater2001} induce phonon emission. In other words vortices decay into sound. Similar results were obtained later in superfluid $^3$He-B \cite{Bradley2006} and, more recently, also in 2D BECs \cite{Kwon2014}. This scenario was also confirmed by 3D and 2D numerical simulations based on the Gross-Pitaevskii model of a Bose-Einstein condensate \cite{Nore1997b, Stagg2015}.

We conclude that the key property of superfluid turbulence is not the fact that the fluid in question has no viscosity (at temperatures low enough acoustic dissipation plays the same role of viscous dissipation at higher temperatures). It is rather the quantization of vorticity that results in discrete vortex lines. Donnelly and Swanson were indeed correct: the term `quantum turbulence' captures the essence of the problem. Quantum turbulence is therefore the turbulence of a system described by a macroscopic complex wavefunction  - the property which implies the quantization of circulation.

We turn the attention to the second question - what is `turbulence'? Fluid dynamics researchers have never felt the need of a proper definition of turbulence\footnote{Kolmogorov's student A. Yaglom was advising against any attempt to define turbulence. Nonetheless, prominent scientists as Neumann (1949), Bradshaw (1971) and Liepmann (1979) have proposed such definitions. The interested reader is referred to \cite{Tsinober2009} for a conceptual discussion.} . It is generally assumed that turbulence is a state of irregular, spatially and temporally disordered flow, characterized by a very large number of degrees of freedom (over a very large range of length scales and time scales) which are simultaneously excited and interact in a nonlinear fashion. To describe this state of motion, we often seek statistical properties and scaling laws. But which form of nonlinearity is meant by this definition? If we called turbulence any nonlinear system with many degrees of freedom, the term would become too generic to be useful; we must restrict the use of the term to systems whose interaction is described by the same $({\bf v} \cdot \nabla) {\bf v}$ nonlinearity of the classical Euler equation:
\begin{equation}
\frac{\partial {\bf v}}{\partial t}
+ ({\bf v} \cdot \nabla) {\bf v} = -\frac{1}{\rho} \nabla p,
\label{eq:Euler}
\end{equation}
where ${\bf v}$ is the velocity, $p$ the pressure and $\rho$ the density. Mathematically, the Euler equation is the skeleton of the Navier-Stokes equation which governs the motion of ordinary viscous fluids (and which can be further generalized to the presence of magnetic fields, thermal effects, non-Newtonian stresses etc). The nonlinear term  $({\bf v} \cdot \nabla) {\bf v}$ arises simply because Euler equation -- the manifestation of Newton's second law for a continuum -- describes the force (per unit volume) acting on a moving fluid parcel, not at a fixed point in space. It is important to notice that the Euler equation is also the skeleton of the Gross-Pitaevskii equation which models a weakly interacting condensate in the mean-field approximation at sufficiently low temperatures. 

We conclude that the terms normally used in the literature are indeed justified: turbulence is the time-dependent, space-dependent state of irregular motion, characterized by a huge number of degrees of freedom which interact via the fundamental nonlinearity of the Euler equation. Since this equation is mathematically at the core of our descriptions of ordinary fluids, superfluid helium and atomic BEC the term `turbulence' can be applied to all these systems, provided they are disordered enough.
 

Yet simple, this analysis is useful because it suggests how to improve our understanding of quantum turbulence.
As said, turbulence in a classical fluid contains many degrees of freedom, which we can think of as eddies. How many degrees of freedom? Since the vorticity is a continuous field, it makes no sense to count them, but we can still give a quantitative answer to this question (see next Section) by noticing that the usual measure of the intensity of the turbulence, the Reynolds number, represents the ratio of the smallest to the largest eddy. On the contrary, in a quantum fluid the number of vortices (in 2D) or their length (in 3D) can be counted or measured because vortices are discrete entities. Then the question becomes: how many vortices are needed for turbulence? 

We argue that this is a good question (cf. Sec.~\ref{SubSubSec:Distinction}). We know that a system with as little as four point vortices moving in a 2D incompressible, inviscid fluid not restricted in space is chaotic \cite{Aref1983} in the sense that initially close trajectories diverge in time away from each other with a positive Lyapunov exponent. The same holds true for a system of three vortices confined in a 2D BEC \cite{Kyriakopoulos2014}. But turbulence, according to our definition, is more than chaos. Turbulence is also governed by statistical and scaling laws (such as the celebrated Kolmogorov law of homogeneous isotropic turbulence) and metastable coherent structures. We have already connected the relative size of the smallest and largest eddies in the Kolmogorov inertial range to the range of length scales. It is natural to ask how many vortices are required to observe, for example, the emergence of Kolmogorov scaling, over how many orders of magnitude this scaling persists and how can one measure the experimental signatures of it. All of the above constitute challenging open questions in today's research in QT; we will come back in Sec.~\ref{Sec:Discussion}.
\section{Essentials of theory \label{Sec:Essentials}}

In this Section we give a brief account of the theoretical models that are commonly followed in the literature of trapped BECs and summarize the main physics in the study of turbulence. Note that we chose to present the quantum and classical theoretical parts in the same chapter, as this will ease the understanding of the origin of similarities in the behavior of quantum and classical fluids. In particular, we discuss the reduction of mean-field models used for the description of quantum gases to classical equations of motion and thus highlight the success but also the weakness of these models. We stress that we do not attempt to present the theory of classical turbulence in full, but only the principal ideas and characteristics.

\begin{figure}
\begin{center}
\includegraphics[trim={0cm 1.5cm 1cm 8.7cm},clip, width=0.45\textwidth]{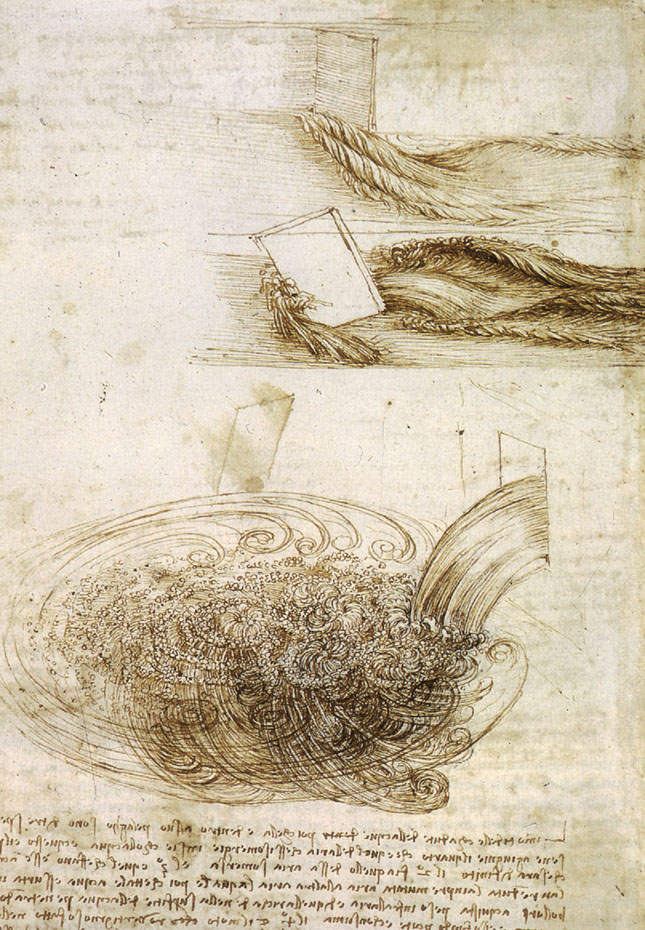}
\includegraphics[width=0.45\textwidth]{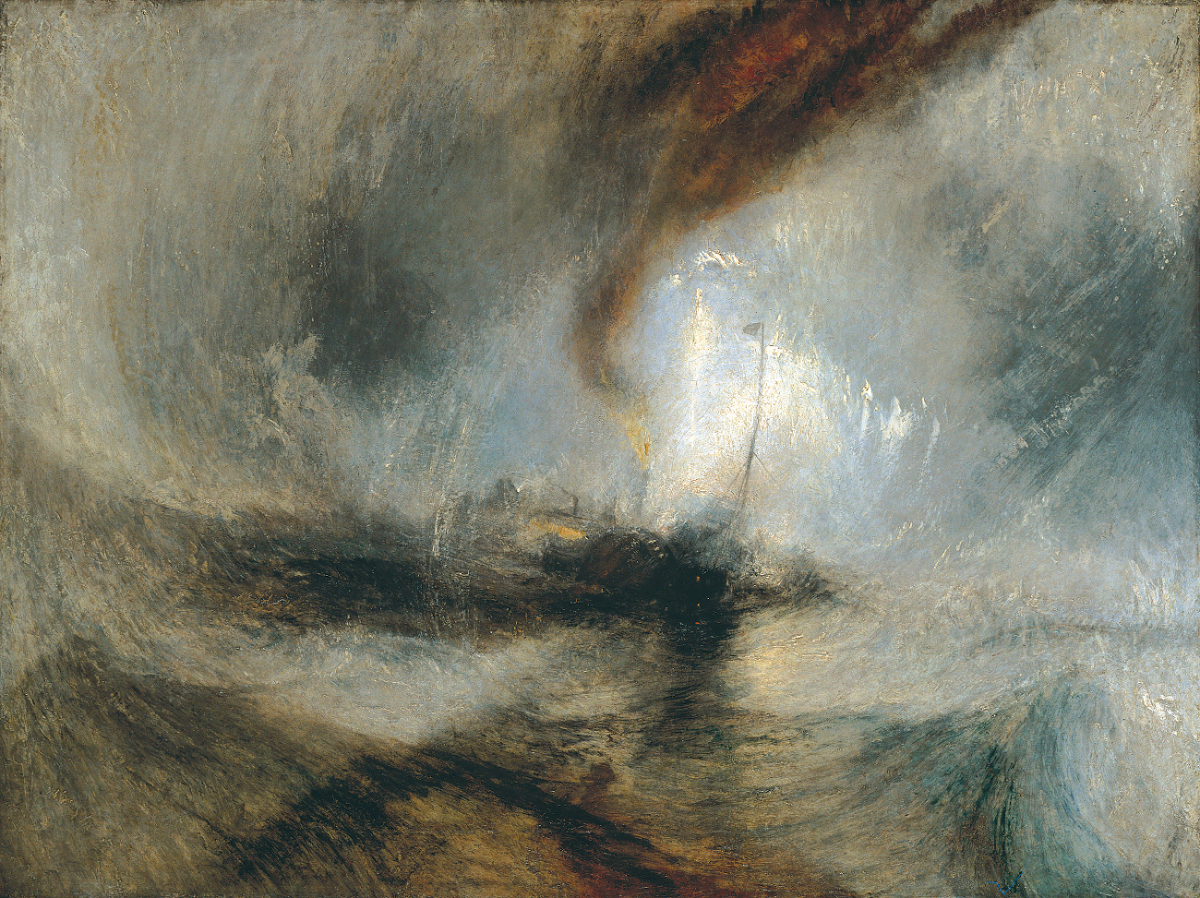}
\end{center}
\caption{Left: Drawing from ``Studies of water passing obstacles and falling'' from Codex Leicester by Leonardo da Vinci (1452-1519).  Right: ``Snow Storm: Steam-boat off a harbour's mouth'', by J. M. W. Turner (1775-1851). Source: \textit{Wikimedia Commons}. \label{Fig:arts}}
\end{figure}

\subsection{Brief account of classical turbulence \label{SubSec:BriefCT}}
Turbulence -- the nonregular motion of fluids, undesired in many everyday situations -- has been known for centuries. The extraordinary properties of turbulent fluids have attracted the attention of great thinkers in the history of science and also arts. The idea that turbulent behavior arises from intertwined eddies in a fluid is a milestone in the historical development of the subject, which dates back to Leonardo da Vinci's drawings of water falling into a pond, see the pattern of vortices at different scales shown in Fig.~\ref{Fig:arts}. Despite this precise observation of the natural world, da Vinci's suggestion was not fertilized; neither scientists (who clearly then lacked the necessary mathematical tools to formalize it), nor artists followed up Leonardo's observation, who continued representing turbulence as a foamy unstructured mess, see for example Turner's ``Snow Storm: Steam-boat off a harbour's mouth'' (Fig.~\ref{Fig:arts}).

That turbulent flow involves eddies exchanging energy has only been given a mathematical justification centuries later. The nonlinear differential equation which generalizes the Euler equation and predicts the motion of a real, viscous fluid was written in the mid 19th century. In the simple case of a fluid of constant density (e.g. water, which is essentially incompressible in typical situations), the Navier-Stokes equation describing a solenoidal field takes on the form
\begin{equation}
\frac{\partial {\mathbf v}}{\partial t} + ({\mathbf v} \cdot \nabla) {\mathbf v} =
    -\frac{1}{\rho} \nabla p +\nu \nabla^2 {\mathbf v} + {\mathbf g}
\label{Eq:NS}
\end{equation}
and
\begin{equation}
\nabla\cdot {\mathbf v}=0
\end{equation}
where $\mathbf v$ is the velocity, $p$ the pressure, $\rho$ the density, $\nu$ is the kinematic viscosity of the fluid (the ratio of viscosity and density) and $\mathbf g$ the external forces. Note the extra term at the right hand size which generalizes Euler's original and idealized fluid equation. 

\begin{figure}
\centering
\includegraphics[width=0.40\textwidth]{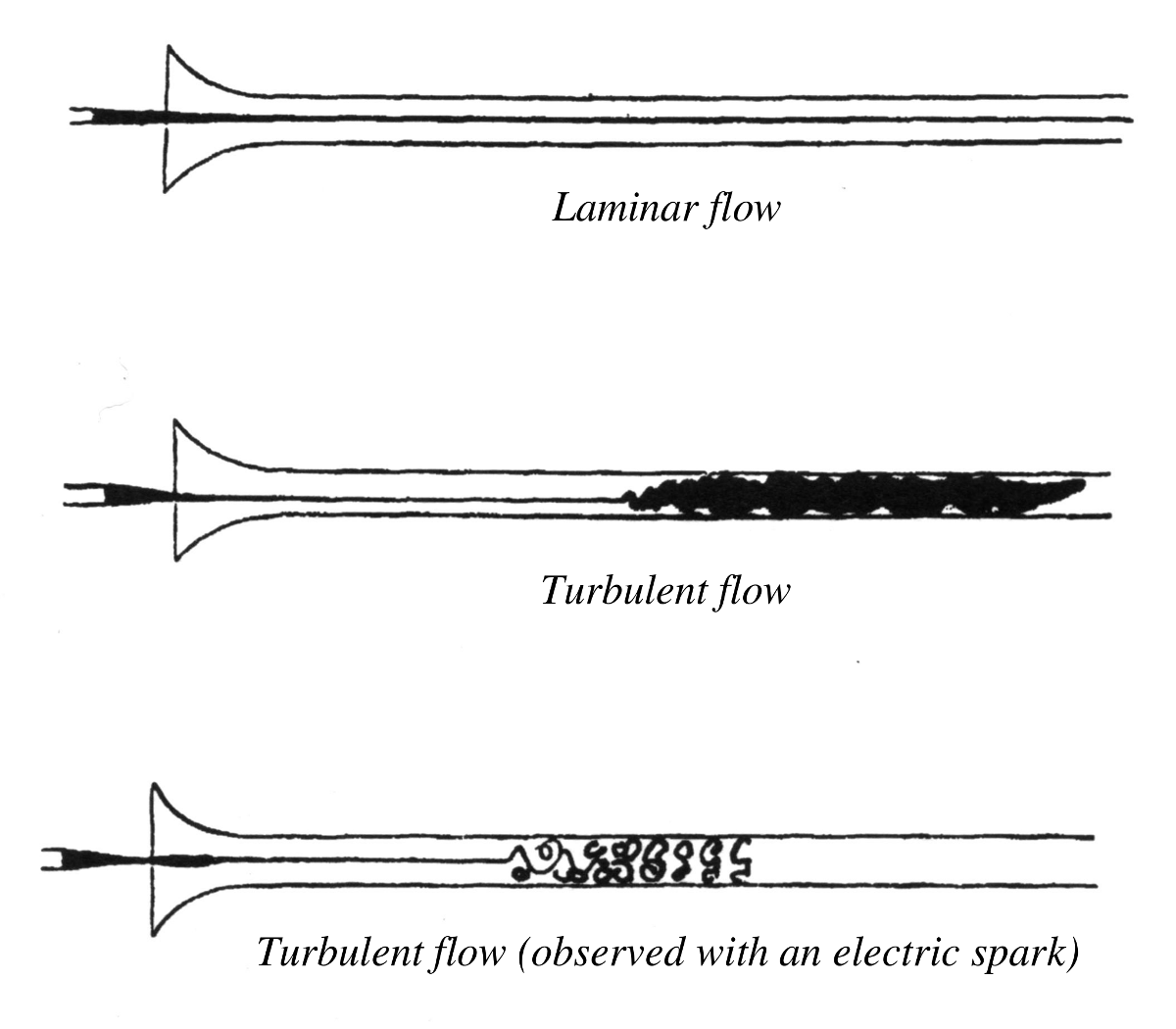}
\caption{Osborne Reynolds' original drawing of the observation on the different types of flow of water through a pipe. Reynolds quantified the different emergent flows of water by using the celebrated eponymous number $Re$  \cite{Reynolds1883} (see text). Source: \textit{Wikimedia Commons}. \label{Fig:ReynoldsObservation}}
\end{figure}

The Navier-Stokes equation is remarkably powerful, and describes a huge range of phenomena, including turbulent flows. In the late 19th century the British physicist Osborne Reynolds studied experimentally the transition from laminar to turbulent flows in a pipe (see Fig.~\ref{Fig:ReynoldsObservation}) and understood  the interplay of viscous and inertial forces. In particular, Reynolds recognized that the transition requires that the parameter
\begin{equation}
Re=\frac{v D}{\nu}
\label{Eq:Reynolds}
\end{equation}
is sufficiently large. Here $v$ is the average velocity of the fluid in the pipe and $D$ is the pipe's diameter.  This parameter, named Reynolds number, is a dimensionless measure of the velocity, and can be interpreted as an estimate of the ratio of inertial to viscous forces. At small $Re$, viscous forces dominate, and the flow is smooth and laminar; at large $Re$, the laminar profile is destabilized, a large number of eddies of all sizes appear, and the flow becomes turbulent.

The next seminal step was Lewis Richardson's concept of the energy cascade in the 1920s. Richardson, a British meteorologist, understood that in the simplest possible case of turbulence away from boundaries (i.e. a case which is much simpler than Reynolds' experiment) a statistical steady state requires a constant input of energy at large length scales and, at the same rate, the dissipation of energy by viscous forces at the small length scales. At intermediate length scales (the so called \emph{inertial range}), the transfer of energy from large to small eddies is independent of viscosity. This transfer is called the \emph{energy cascade}.
The idea of the cascade was formalized in the 1940s by the Soviet mathematician Andrei Kolmogorov, who associated spectral symmetries and a self-similar behavior to turbulent flow patterns. In the idealized case of a homogeneous and isotropic statistically steady-state, energy flows from the largest scales $D$ to the smallest one $\eta$ where it is turned into heat; within this range, the energy dissipation rate $\varepsilon$ is constant. Therefore the Reynolds number, which is the dimensionless measure of the velocity
at the largest scales, represents also the \emph{range} of length scales available for turbulence. More precisely we have 
\begin{equation}
\frac{\eta}{D} \approx Re^{-3/4}
\label{Eq:Re2}
\end{equation}
(a relation which we have already used in Sec.~\ref{SubSec:Differences}). The quantity $\eta$ is called the Kolmogorov length scale, or the dissipation length scale, and effectively represents the smallest eddies. Kolmogorov suggested that the turbulent state has universal characteristics. A key property of turbulence is the distribution of kinetic energy over the length scales, which according to Kolmogorov is the same for all flows and does not depend on the specific details of each.  It is convenient to describe the length scales $r$ as inverse wavenumbers $k=\vert {\bf k} \vert=2 \pi/r$, where $\bf k$ is the three dimensional wavevector, and introduce the concept of energy spectrum $E_k$. Kolmogorov showed that in the inertial range 
\begin{equation}
E_k = C \varepsilon_k^{2/3} k^{-5/3},
\label{Eq:Kolmogorov}
\end{equation}
where $C$ is a dimensionless constant of order one. In other words, in homogeneous isotropic turbulence most of the energy is contained in the largest eddies (small $k$); smaller eddies of size $r=2 \pi/k$ contain proportionally less energy according to Kolmogorov's law, Eq.~\eqref{Eq:Kolmogorov}.

The above argument applies to three dimensional flow. With the exception of soap films \cite{Rivera1998}, most real flows are three dimensional and two-dimensional (2D) flow models are only approximations of situations created in the laboratory. Fluid dynamics researchers are, however, interested in 2D flows as well, due to their important role as a paradigm for anisotropic 3D turbulence and not necessarily due to its physical realizability. The reduced dimensionality may arise from strong anisotropy,  stratification or rotation (via the Taylor-Proudman theorem), which opens route to a new complex physics present in 2D chaotic flows. This focus by itself has essential, practical applications to oceans, planetary atmospheres and astrophysical systems as useful mathematical idealizations which approximate large scale features of geophysical flows (e.g. ocean and atmospheric currents) which have vertical extension (few km) much smaller than the horizontal extension (thousands of km). 

Dynamics in 2D classical turbulence differs dramatically from 3D. In 1967 the American scientist Robert Kraichnan showed that in 2D flows energy does not cascade from large to small scales (as in 3D), but is rather in the opposite direction; from small to large scales \cite{Kraichnan1967,Kraichnan1971,Boffetta2007}. The same scaling 
$k^{-5/3}$ as for the direct cascade however is found [Eq.~\eqref{Eq:Kolmogorov}]. This \emph{inverse energy cascade} (IEC) is accompanied by a direct (forward) cascade of a second inviscid quadratic invariant - the enstrophy [a measure of vorticity variance, defined by the surface integral on the xy plane $\int dS\left(\nabla\times \textbf{v}\right)^2$] - which makes the energy spectrum scale with $k^{-3}$ for large momenta.

\begin{figure}
\centering
\includegraphics[width=0.50\textwidth]{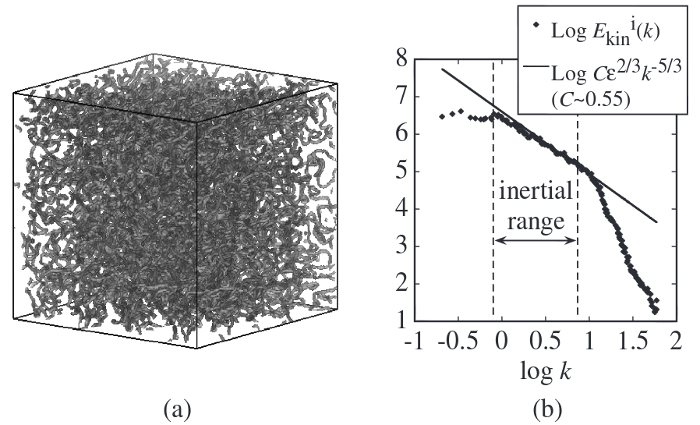}
\caption{(a) A typical vortex tangle. (b) Spectrum of the kinetic energy $E_{\text{kin}}(k,t)$ as a function of the wavenumber $k$ for a quantum turbulent gas. The plotted points are for an ensemble average of 50 randomly selected states. The solid line is the Kolmogorov $-5/3$ scaling. Source: paper of Tsubota \cite{TsubotaJoPSJ2008}.
\label{Fig1}}
\end{figure}

\begin{figure}
\centering
\includegraphics[width=0.50\textwidth]{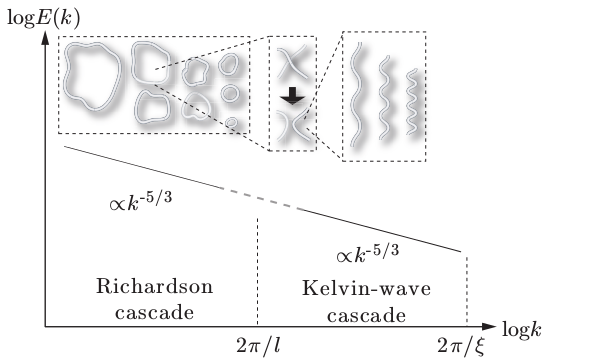}
\caption{Energy spectrum of homogeneous isotropic quantum turbulence at $T=0$ according to current theoretical understanding \cite{Barenghi2014a}. At large length scales ($k \ll k_{\ell}=2 \pi/\ell$ where $\ell$ is the average intervortex distance), the energy spectrum obeys the classical Kolmogorov scaling corresponding to the Richardson cascade:  $E_k \sim k^{-5/3}$. At small length scales ($k \gg k_{\ell}$) the energy spectrum arises from the Kelvin-wave cascade. At very large length scales, $k \approx 2 \pi/(100 \xi)$, energy is dissipated by phonon emission. Source: paper by Tsubota \cite{Tsubota2014}.
\label{Fig7}}
\end{figure}

\subsection{Brief account of quantum turbulence \label{SubSec:BriefQT}}
In the quantum world of superfluid helium and atomic condensates, turbulence acquire new aspects. In Section~\ref{SubSec:Differences} we have already listed the main differences between quantum turbulence and classical turbulence in superfluid helium and atomic condensates: discrete vorticity, quantized circulation, two-fluid nature, different range of length scales available, different energy dissipation. However, there are also similarities and these arise from the same conservation laws of mass and momentum. In particular the latter implies the same vortex interactions predicted by Euler's equation. Moreover, an ``effective''  viscosity which plays a role similar to that of the viscosity in ordinary fluids, can be experimentally and numerically identified \cite{Walmsley2008} in superfluid helium at very low temperatures. The similarities between superfluid helium and atomic condensates include the friction exerted by the normal fluid \cite{Barenghi1983} or by the thermal cloud \cite{Allen2015} which resists the motion of vortices.

We know that quantum turbulence takes the form of a tangle of vortex lines, which interact and reconnect with each other (see Sec.~\ref{SubSec:Vortexreconnections}). However, in trapped condensates visualization or vortex tangles is more difficult and one resides in numerical simulations. In liquid helium it is relatively easy to determine experimentally the length of vortex lines contained in the experimental cell and therefore the vortex line density $L$ (vortex length per unit volume) is a practical measure of the intensity of quantum turbulence. From the knowledge of $L$, one estimates that the typical distance between the vortex lines is $\ell \approx L^{-1/2}$.

Panel (c) of Fig.~\ref{Fig:TurbulenceGeneral} shows a small tangle of vortex lines inside an inhomogeneous, spherically trapped atomic condensate \cite{White2010}; note the oscillations of the surface of the condensate induced by the motion of the vortices. Panel (a) of Fig.~\ref{Fig1} shows a vortex tangle in a homogeneous condensate \cite{TsubotaJoPSJ2008}, computed inside a periodic domain. Panel (b) of the same figure shows the classical $k^{-5/3}$ scaling laws of the energy specrum of the incompressible kinetic energy (see next section). 

\begin{figure}
\centering
\includegraphics[width=0.45\textwidth]{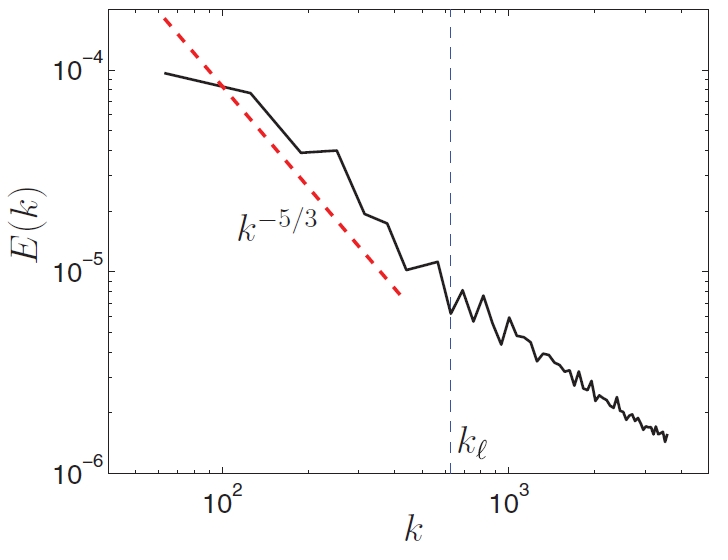}
\includegraphics[width=0.45\textwidth]{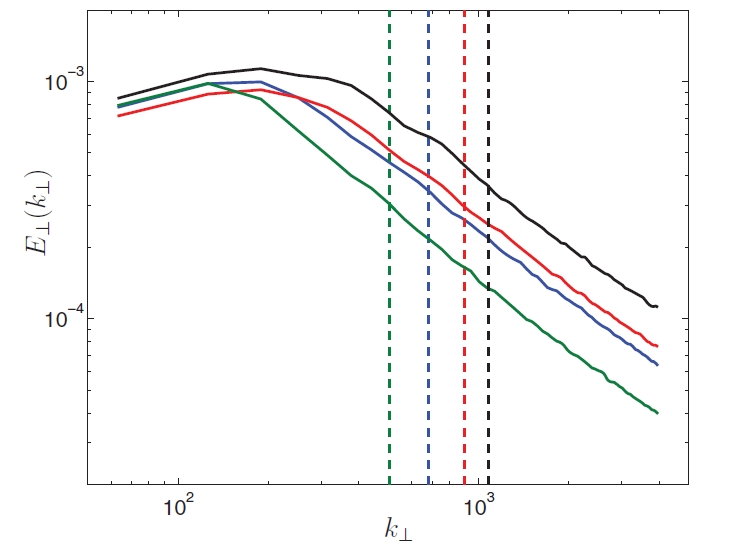}
\caption{Superfluid energy spectrum $E_k$ corresponding to thermally-driven (left) and mechanically-driven (right) helium flows at intermediate temperature $T=1.9~\rm K$. For wavenumbers $k<k_{\ell}=2 \pi/ \ell$, where $\ell$ is the average intervortex spacing, the spectrum of mechanically-driven turbulence (left) is consistent with the classical $k^{-5/3}$ Kolmogorov scaling (most of the energy is concentrated at the largest scales $k \ll k_{\ell}$) whereas the spectrum of thermally-driven turbulence (right) lacks energy at the largest scales and scales as $k^{-1}$ at large wavenumbers. Source: paper by Baggaley \emph{et al.} \cite{Baggaley2012d}. \label{Fig:Sherwin}
}
\end{figure}

It is important to notice that not all turbulent states which have been studied display the classical Kolmogorov scaling. If the superfluid is thermally driven by a small heat flux (the normal fluid being uniform), the superfluid energy spectrum has the shape of a broad `bump', lacks energy at the largest scales \cite{Baggaley2012d} and decays as $k^{-1}$ at large $k$; since this is the spectral signature of individual vortices, we infer that the flow is random in character.  On the contrary, if the superfluid is driven by a turbulent normal fluid, for example by mechanically exciting it with grids or propellers \cite{Maurer1998},  the energy spectrum is consistent with the classical $k^{-5/3}$ Kolmogorov scaling \cite{Baggaley2012d}; the fact that most of the energy is at the largest length scale is consistent with a relative large-scale polarization of the vortex lines, which creates large-scale flows. Fig.~\ref{Fig:Sherwin} compares the two types of spectra: thermally-driven (right) and mechanically-driven (left). It is important to notice that each of the spectra corresponds to numerical simulations with the same turbulence intensity: the vortex line density (length of vortex line per unit volume) is the same ($L=12,000~\rm cm^{-2}$) and so is the size of the computational box ($D=0.1~\rm cm$). The difference between the results -- the slope and the presence/absence of large scale, energetic eddies -- is remarkable. 

Similar spectra, classical and nonclassical, have been observed \cite{Walmsley2008} and computed \cite{Baggaley2012e} in helium at very low temperatures (in the absence of normal fluid): they are usually referred to as quasiclassical (or Kolmogorov) and ultraquantum (or Vinen) turbulence \cite{Volovik2003}.

What happens at very low temperatures at length scales smaller than the intervortex spacing $\ell$ raises theoretical controversies and still lacks experimental verification. Let us call $k_{\ell}=2 \pi/\ell$ the wavenumber which corresponds to the intervortex distance. The previous discussion referred to hydrodynamical scales $k \ll k_{\ell}$. We are now concerned  with length scales in the range $k_{\ell} <k < 2 \pi/\xi$ where $\xi$ is the healing length (vortex lines cannot bend on scales shorter than the vortex core radius).

It is generally understood that, without friction effects, the interaction of Kelvin waves produces shorter and shorter wavelengths, until the angular frequency $\omega$ of the wave is fast enough to efficiently radiate sound away. This process, called the Kelvin-wave cascade, is often described in terms of the one-dimensional wavenumber $k$ and the wave amplitude, or occupation number $A=n(k)$. Two conflicting theories emerged. The first, proposed by Kozik and Svistunov \cite{Kozik2004}, predicted that $n_{KS}(k) \sim k^{-17/5}$. To convert the one-dimensional Kelvin spectrum to the three-dimensional energy spectrum we recall
that $E \sim \hbar \omega(k)$ and that, for Kelvin waves, $\omega(k) \sim k^2$; we obtain that $E\sim k^{-7/5}$ (see also the discussion in Sec.~\ref{SubSec:WT}). The second theory, proosed by L'vov and Nazarenko \cite{LvovNazarenko2010}, results in $n_{LN}(k) \sim k^{-11/3}$, hence $E \sim k^{-5/3}$. The difference between $k^{-17/5}=k^{3.40}$ and $k^{-11/3}=k^{3.67}$
is small, and there have been attempts to determine the exponent numerically; the most recent work \cite{Krstulovic2012,Baggaley2014b,Kondaurova2014} supports the theory of L'vov and Nazarenko, but we still lack direct experimental evidence. 

The following possibility has also been raised: that of a bottleneck between the Kolmogorov cascade and the Kelvin cascade, which may change the shape of the spectrum in the region $k \approx k_{\ell}$ \cite{Lvov2007}. This effect may arise because the rate at which energy flows down the (three dimensional) Kolmogorov cascade is different by the (slower) rate at which energy can progress into the (one-dimensional) Kelvin cascade. A somewhat different approach, suggesting the smooth transition of a Kolmogorov cascade of superfluid turbulence to a Kelvin-wave cascade has been suggested in Refs.~\cite{Kozik2008,Lvov2008}. Again, experimental evidence is lacking.

Last, more recently, there has been discussion on the possible introduction of a `quantum' Reynolds number or a `quantum' Navier-Stokes equation \cite{Finne2003,Tavares2013,Reeves2015} to distinguish between laminar and turbulent superfluid helium flows and to characterize turbulence in a quantum gas. These are discussed in Sec.~\ref{SubSec:Reduction}.

\subsection{Theoretical models for BECs \label{SubSec:ThModels}}

\subsubsection{Biot-Savart model} \label{SubSec:Biot-SavartModel}
The foundations of a quantitative description of superfluid helium under rotation were set by Hall and Vinen in the mid 1950's \cite{Hall1956a,Hall1956b}. Some decades later, computational simulations in quantum turbulence were pioneered by Schwarz \cite{Schwarz1985}. Schwarz observed that the velocity flow field in a quantum gas can be described by the vortex motion which is represented by a curve $s=s(\xi,t)$ in three-dimensional space, where $\xi$ is the arclength and $t$ is time.  
With these computations the final distribution of the vorticity is determined entirely by the geometry of the cell, the global flow parameters, and the initial vortex distribution. The Biot-Savart (BS) formulation gives the field a self-induced velocity along the curve $s$ in complete analogy to a magnetic field. In this approximation the vortex cores are approximated as infinitesimal, one dimensional objects driven in an inviscid background. The spatial coordinate along each vortex is given by the parametric arclength $s=s(\xi)$.

We can write the velocity field as the rotation of a vector potential, $\mathbf{v}=\nabla\times\mathbf{A}$.  The vorticity is therefore defined by the Poisson equation, 
$\nabla^2\mathbf{A}=-\pmb{\omega}$, 
which has a Green function solution given by
\begin{equation}
\mathbf{A}(\mathbf{r})=\frac{1}{4\pi}\int{\frac{\pmb{\omega}(\mathbf{r^{\prime}})}{|\mathbf{r}-\mathbf{r}^{\prime}|}} d\mathbf{r}^{\prime},
\label{Eq:BS-A}
\end{equation}
where $\mathbf{r}$ is the location of the vortex core.  
Since in the cartesian plane, the vorticity has a constant intensity and is concentrated in the core, a change of variables allows for the rewriting $\pmb{\omega}(\mathbf{r^{\prime}})d\mathbf r^{\prime}=\kappa d\pmb{\ell}$. 
Taking the curl of $\mathbf{A}$ and inserting the above parametrizations the well known BS velocity form is deduced,
\begin{equation}
\mathbf{v_s}(\mathbf{s},t)=\frac{\kappa}{4\pi}\int_\mathcal{L}\frac{\mathbf{r-s}}{|\mathbf{r-s}|^3}\times d\mathbf{r}.
\label{Eq:Biot-SavartLaw}
\end{equation}
Note that, by construction, the above solution $\mathbf{A(r)}$ is an incompressible field. Indeed, as the divergence of a curl, the identity
\begin{equation}
\nabla \cdot \mathbf v = \nabla \cdot \nabla \times \mathbf A = 0
\end{equation}
holds. However, a general (quantum) velocity field $\mathbf{v}$, as we shall see in the following Section, needs not be incompressible but rather of the form:
\begin{equation}
\mathbf v^\prime = \mathbf v_i + \mathbf v_c,
\label{Eq:HelmholtzDecom}
\end{equation}
where $\mathbf v_i\equiv \mathbf v=\nabla\times \mathbf A$ is the field that is incompressible (i.e. has zero divergence) and $\mathbf v_c=\nabla \phi$ is the irrotational (and compressible) part of the field. 
Hence, the velocity field described by the BS model gives \emph{only} the incompressible part of $\mathbf v^\prime$. Physically, this suggests that $\mathbf v_i$ captures the vortex dynamics, as described by the BS model. 
The form of $\mathbf v^\prime$ in Eq.~\eqref{Eq:HelmholtzDecom} is known as Helmholtz decomposition and its importance will become obvious in the analysis that follows (see Sec.~\ref{Sec:Hydro}).

Except for very simple cases, complete analytical solutions with this formulation are not possible. Additionally, BS calculations are computationally expensive, since $N^2$ operations are required over a grid of $N$ points \cite{Barenghi2001}. To reduce the computational ef{}fort, the BS integral is often replaced by
\begin{eqnarray}
\mathbf v_{\text{LIA}} &\approx& \beta(\pmb s^{\prime}\times \pmb s^{\prime\prime}),
\label{Eq:BSExpansion}
\end{eqnarray}
where $\beta=\frac{\kappa}{4\pi R}\ln\left(R/\xi\right)$ and the derivatives are taken with respect to the arclength $s^{\prime}=ds/d\xi$ \cite{Barenghi2001}. In that way the nonlocal contributions of the integral of Eq.~\eqref{Eq:BSExpansion} are neglected and thus this simplification is called the \emph{local induction approximation (LIA)}. LIA is commonly used both for analytical intuition and for computational tractability.  With this simplification it is apparent that the vortex lines move in a direction perpendicular to both the line curvature ($R=1/s^{\prime\prime}$) and tangent ($s^{\prime}$) with a velocity inversely proportional to the curvature.

The BS law describes individual vortices in the zero temperature limit.  However, high temperature viscous effects and vortex-vortex interactions are not naturally included -- in neither of the two approaches -- and need to be inserted \emph{ad hoc} in each simulation. A quantum vortex reconnection (QVR) occurs when two vortices approach each other and connect at one point. During a reconnection the two lines topologically change into a different configuration, as shown in Fig.~\ref{Fig:Reconnections}. QVRs have been observed both in He II \cite{Bewley2006,Bewley2008} and BECs \cite{Serafini2015}. To circumvent this problem, one needs to artificially insert the occurrence of QVR in the BS or LIA models \cite{Schwarz1985,Samuels1992}. To achieve this a cut-off vortex-vortex distance is introduced, beyond which the vortices are assumed to reconnect (see for instance \cite{Kivotides2000} and references therein). The absence of QVR is a critical failing of the BS simulations as it has been shown that reconnection processes are fundamental to the dynamics of QT.

The BS approach, even extremely simple can capture several of the features of QT and give statistical insight on the behavior of driven quantum fluids, especially when studying distribution of momenta. However, the lack of reconnections in the model (or the \emph{ad-hoc} insertion of reconnection events) together with the assumption of infinitesimally small vortex cores are drawbacks that one cannot ignore. Infinitesimal vortex cores, that the BS model admits, do not translate well in gaseous dilute BECs: a vortex core can have a diameter comparable even to the extension of the entire condensate! Hence a formal and more precise description of the fluid alternate methods should be employed.

The mean-field Gross-Pitaevskii (GP) theory, instead, can describe the above situations. There, hydrodynamic and quantum behaviors naturally coexist and reconnections arise, as we shall see, from its solutions.

\subsubsection{Gross-Pitaevskii model} \label{SubSec:GrossPitaevskiiModel}
The GP approximation is today the most used theory in the BEC community and also beyond. It was formulated originally by Eugene Gross and independently by Lev Pitaevski in 1961 in order to describe quantized vorticity in a Bose gas \cite{Gross1961,Pitaevskii1961}. Within a semi-classical approach they derived the celebrated Gross-Pitaevskii (GP) -- also known as nonlinear Schr\"odinger -- equation. The defining assumption of this mean-field theory is that all bosons of the system reside in the same quantum state (total condensation) and hence the whole system is approximated by one single-particle state $\psi(r,t)$. The particle density $\rho$ of the system is then given by $|\psi|^2$. In other words, for a system of $N$ particles of mass $m$, according to the Gross-Pitaevskii theory, the knowledge of one them is enough to characterize the collective properties of the whole system.

The wavefunction of the relaxed state is found by solving the eigenvalue problem:
\begin{equation}
\mu~\psi(\mathbf r) = \left(-\frac{\hbar^2}{2m}\nabla^2 + V_{\text{trap}}+g|\psi(\mathbf r)|^2 \right)\psi(\mathbf r),
\label{Eq:TIGP}
\end{equation}
i.e. the time-independent GP equation. The eigenvalue $\mu$ is the chemical potential, $\nabla$ the usual nabla operator and $\hbar$ the Planck constant. The time evolution is given from the solution of the time-dependent GP equation:
\begin{equation}
i \hbar \frac{\partial \psi(\mathbf r,t)}{\partial t} = \left(-\frac{\hbar^2}{2m}\nabla^2 + V_{\text{trap}} + g|\psi(\mathbf r,t)|^2 \right)\psi(\mathbf r,t).
\label{Eq:TDGP}
\end{equation}
At ultralow temperatures and densities the particles are assumed to interact only weakly via an $s$-wave scattering and this is modelled by a contact (two-body) pseudo-potential of Dirac delta form $\delta(|\mathbf r - \mathbf r^\prime|)$. This assumption gives rise to the `self-interaction' term $\propto |\psi|^4$ in the energy of the system. The parameter $g$ measures the strength of this interaction and is proportional to the s-wave scattering length: $g=4\pi a_s\hbar^2/m$. The parameter $g$ can be either positive or negative, depending on the sign of the s-wave scattering length, $a_s$. The $V_{\text{trap}}$ is the confining external potential and is -- in general -- function of space and time, i.e. $V_{\text{trap}}=V_{\text{trap}}(\mathbf r,t)$. The wavefunction $\psi$ is normalized by the total number of atoms in the system,
\begin{equation}
\int{|\psi(\mathbf{r})|^2d^3\mathbf{r}}=N.
\label{Eq:GPNormalization}
\end{equation}

Keeping in mind that $\psi$ is a complex-valued function, one can perform the following transformation:
\begin{equation}
\psi(\mathbf r,t) = f(\mathbf r,t) e^{i S(\mathbf r,t)},
\label{Eq:Madelung}
\end{equation}
known as \emph{Madelung transformation}. $f$ and $S$ are real functions and give the (square root of) the density $\rho$ and phase of the condensate accordingly.
Inserting this in Eq.~\eqref{Eq:TDGP} and separating real and imaginary parts we get the two equations \cite{Pethick2008}
\begin{equation}
\frac{\partial(f^2)}{\partial t} = - \frac{\hbar}{m}\nabla\cdot(f^2\nabla S)
\label{Eq:GPpolar1}
\end{equation}
and
\begin{equation}
-\hbar\frac{\partial S}{\partial t} = -\frac{\hbar^2}{2mf}\nabla^2f + \frac 1 2 mv^2 + V(\mathbf r)+ gf^2,
\label{Eq:GPpolar2}
\end{equation}
where $\mathbf{v}=\frac{\hbar}{m}\nabla S$ is the field velocity, as defined in Sec.~\ref{SubSec:Superfluidity}, and $v$ its norm.
Equation \eqref{Eq:GPpolar1} is the continuity equation [see Eq.~\eqref{Eq:continuityQuant}]. This becomes clear if one identifies the term  $f^2\frac{\hbar}{m}\nabla S$ with the flux $\mathbf j=\rho\mathbf v$. In other words, the inclusion of contact interactions in the GP model does not affect the conservation of the probability $|\psi|^2$.

The GP equation is the mean-field approach to calculate the properties of turbulent BECs. In the limit of a ground state condensate at zero temperature and infinite particle number, the GP equation is exact \cite{Lieb2000,Lieb2002}. One feature of importance is that the lines of quantized vorticity and vortex reconnections appear as solutions of the GP model, rather than \textit{ad hoc} added as in the BS approach \cite{Barenghi2001}.

For an infinite, time-independent system, the vortex core diameter can be estimated as follows. Like a soliton, a vortex is a topological defect whose extension at equilibrium is expected to be of the order of the healing length. The healing length $\xi$ of a (uniform) condensate is defined as the length scale over which the interaction energy becomes equal (or comparable) to the kinetic one. Thus,
\begin{equation}
\frac{\hbar^2}{2m}|\nabla\psi|^2\sim g|\psi|^4
\label{Eq:BalanceHealingLength}
\end{equation}
or
\begin{equation}
\frac{\hbar^2}{2m}\frac{|\psi|^2}{\xi^2} = g|\psi|^4  \Rightarrow ~ \xi = \frac{\hbar}{\sqrt{2mg\rho_0}},
\label{Eq:HealingLength}
\end{equation}
(see also \cite{Pethick2008}). For a condensate that is locally perturbed at some point $r_0$, $\xi$ gives the length after which the density reassumes its initial value $\rho_0$. In other words, it is the distance that it takes for the gas to `heal' the disturbance. A quantum vortex, being a local node of the density, can also be thought to have an opening of diameter $\xi$.

Last, in the regime of the parameters that condensation of trapped rubidium atoms normally takes place (that is, temperature in the $\rm{nK}$ regime, particle number of the order of around $10^5$ and $s$-wave scattering length of about $100~\alpha_s$) the wavefunction of the gas at equilibrium varies smoothly and slowly. Hence the kinetic contribution to the energy (scaling as the second spatial derivative) can be neglected. This is known as the \emph{Thomas-Fermi} (TF) approximation and with it the spatial extension of the cloud of the condensed atoms can be easily determined. If we neglect the kinetic energy operator in Eq.~\eqref{Eq:TIGP} we obtain for the density of the ground state:
\begin{equation}
\rho(\mathbf r) = \frac{\mu-V_{\text{trap}}(\mathbf r)}{g}.
\label{Eq:TFdens}
\end{equation}
We see that the density is merely the inverse of the confining potential $V_{\text{trap}}$, scaled and shifted ($\mu$ is the chemical potential and is fixed for some given particle number $N$). Naturally, this formula does not hold for $\mu<V_{\text{trap}}$. Otherwise, it gives a satisfactory approximation to the expansion of the background density of gas, over which quantum vortices with expansion $\xi$ are `seeded'.

When in the TF regime, the gas has a well-defined extension, known as the Thomas-Fermi radius $R_{\mathrm{TF}}$, that is found from the boundary condition $V_{\text{trap}}(R_{\mathrm{TF,x}},R_{\mathrm{TF,y}},R_{\mathrm{TF,z}})=\mu$. In the case of a harmonic potential $V_{\text{trap}}=m/2(\omega_x^2 x^2+\omega_y^2 y^2+\omega_z^2 z^2)$ we obtain:
\begin{equation}
R_{\mathrm{TF,i}} = \sqrt{\frac{2\mu}{m\omega_i^2}}, 
\label{Eq:TFRadius}
\end{equation}
for each direction $i=x,y,z$. 

\subsubsection{Reduction of mean-field to classical hydrodynamic equations \label{SubSec:Reduction}}
An interesting feature of the quantum gas is the resemblance of its dynamical equation to that of a classical incompressible gas. Indeed, recasting the GP equation [Eq.~\eqref{Eq:TDGP}] to the form of Eqs.~\eqref{Eq:GPpolar1}-\eqref{Eq:GPpolar2} and taking the gradient of Eq.~\eqref{Eq:GPpolar2} we obtain:
\begin{equation}
\frac{\partial\mathbf{v}}{\partial t} = -\frac{1}{m \rho} \nabla{}p - \nabla\left(\frac{v^2}{2}\right) + \frac{1}{m} \nabla\left(\frac{\hbar^2}{2m\sqrt{\rho}}\nabla^2\sqrt{\rho}\right) - \frac{1}{m}\nabla V_{\text{trap}},
\label{Eq:QuantumEuler}
\end{equation}
where $p=\rho^2 g/2$ and $\rho=f^2$ (see also \cite{Pethick2008}). The term $\nabla v^2$ is kinetic energy while the term $\nabla p$ is a pressure gradient. Note that the only term of Eq.~\eqref{Eq:QuantumEuler} that depends on $\hbar$ is the left term of the right-hand side. This term, also known as \emph{quantum pressure}, originates from the kinetic energy as well. However, it corresponds to the `zero-point' quantum mechanical motion and is significant only when the density of the gas varies `rapidly' in space and negligible otherwise. More specifically, 
let us imagine that the wavefunction changes significantly in space over a distance $l$. Then the classical pressure is of order $\rho g/ml$ while quantum pressure is of order $\hbar^2/m^3 l^3$. Hence, only for distances  less than $\hbar/(m\rho g)^{1/2}$, which is of order of the healing length [Eq.~\eqref{Eq:HealingLength}], the quantum pressure dominates. 

By taking the classical limit $\hbar\rightarrow 0$ the above equation becomes identical to the Euler hydrodynamic equation for a classical irrotational fluid ($\nabla \times \mathbf v = 0$), which is the dissipation-free ($\nu=0$) form the of the NS equation [Eq.~\eqref{Eq:NS}]. Then, the role of external forces $\mathbf g$ plays the potential $\nabla V_{\text{trap}}$. That the Schr\"odinger equation can be reshaped into a \emph{quantum Euler} equation [Eq.~\eqref{Eq:QuantumEuler}] and the identification of the quantum pressure term was already noted by Erwin Madelung in 1926. The German scientist showed, in two letters \cite{Madelung1926,Madelung1927}, how one can derive both the continuity and the quantum Euler equations, sometime collectively called \emph{Madelung equations}.

Note that, an effective dissipation can be included in the GP equation (see, for instance \cite{Tsubota2002} and references therein). The phenomelogical modification is the addition of an imaginary term in the GP equation, making its time evolution complex: $i \partial_t \rightarrow ( i- \gamma)\partial_t $, accounting thus for thermal effects and damping. The hydrodynamic form of the dissipative GP equation now, here derived for an incompressible quantum fluid, corresponds to a \emph{quantum Navier-Stokes} equation:
\begin{equation}
\frac{\partial\mathbf{v}}{\partial t} = -\frac{1}{m\rho} \nabla{}p - \nabla\left(\frac{v^2}{2}\right) + \frac{1}{m} \nabla\left(\frac{\hbar^2}{2m\sqrt{\rho}}\nabla^2\sqrt{\rho}\right) - \frac{1}{m}\nabla V_{\text{trap}} -\nu_q \nabla^2\mathbf v, 
\label{Eq:QuantumNS}
\end{equation}
where $\nu_q\equiv \hbar\gamma/2m$ \cite{Bradley2012}.

\subsubsection{Beyond mean-field theories}
The GP theory is a widely used and successful mean-field description for the statics and dynamics of interacting bosonic gases. It has been applied to various scenarios involving vortices, solitons and other collective excitations and the agreement to experiment has been found to be satisfactory (see for instance Ref.~\cite{Dalfovo1999} for a review). Thus, the GP approach is a natural choice as a theoretical tool to study quantum turbulence in ultracold gases as well.

However, the turbulent gas is a largely perturbed gas, i.e. a system with an average excitation energy many times higher than the relaxed-state energy. In a typical experiment the gas is brought to turbulence by shaking and oscillating the confining trap for various different times (see Sec.~\ref{Sec:Experimental}). It could well be that the coherence and condensation of the system is lost throughout the dynamics. To check the consistency of a mean-field approach one needs to go beyond the limits of mean-field and relax the restriction that the system stays condensed throughout the dynamics. In other words, there should be no \emph{a priori} reason why a single one orbital describes \emph{all} of the system bosons at all times. A consistent way to go beyond the GP picture is the multi-configurational time-dependent Hartree for bosons (MCTDHB) and the equations bearing the same name \cite{Streltsov2006,Streltsov2007,Alon2008}. This approach is capable of describing fragmentation and loss of coherence; it includes many single-particle states that are time-dependent, the non-zero occupations of which give rise to non trivial correlations. Correlations and loss of coherence are important aspects that are experimentally measurable and also shed light into the structure of complicated and highly excited perturbed systems, such as the turbulent gas prepared in the laboratory and discussed in the following Sections.

Besides a recently published work \cite{Wells2015}, and despite the numerical techniques being openly available \cite{ultracold.org} literature is still lacking consistent studies of the many-body (i.e. beyond mean-field) phenomena in turbulent systems.

\subsection{Hydrodynamic turbulence in BECs \label{Sec:Hydro}}
Turbulence, as said, is generally associated with names such as Kolmogorov and Richardson and ideas such as the local transfer of energy between eddies of similar sizes giving rise to the so-called energy cascade. Despite simple, it is important to explicitly demonstrate the dependence of the energy in the wavenumber (momentum) $k$ and the emergence of power laws. Turbulence in BECs -- both the quasiclassical and ultraquantum types discussed earlier -- involves the nonlinear propagation of the velocity field, as is obvious from the governing equations (Schr\"odinger or quantum Euler). This manifestation of turbulence is hydrodynamic in nature in the sense that it requires a nontrivial velocity field, usually containing quantized vortices, i.e. line singularities. However, this is not the only possibility for turbulence in a BEC: see the upcoming Sec.~\ref{SubSec:WT}.


Due to the statistical nature of turbulence it is convenient to define the averaged energy distribution in momentum space
\begin{eqnarray}
E^{(3D)}\left(\mathbf{k}\right) & = & \frac{1}{2(2\pi)^{3}}\int d\mathbf r\left\langle \mathbf{v}(\mathbf{x})\cdot\mathbf{v}(\mathbf{x}+\mathbf r)\right\rangle e^{i\mathbf{k}\cdot\mathbf{r}},\label{eq:energyK} \\
E_{\text{tot}} & = & \int d\mathbf k E^{(3D)}\left(\mathbf{k}\right) = \frac{1}{2}\left\langle \mathbf{v}(\mathbf{x})\cdot\mathbf{v}(\mathbf{x})\right\rangle ,
\end{eqnarray}
where $\left\langle \cdots\right\rangle $ stands for the statistical average. Note that the above holds for an isotropic velocity field $E^{(3D)}\left(\mathbf{k}\right)$. In that case, the latter depends only on the magnitude $k=|\mathbf{k}|$ and can be expressed in terms of a one-dimensional distribution
\begin{eqnarray}
E^{(1D)}\left(k\right) & = & 4\pi k^{2}E^{(3D)}\left(k\right),\\
E_{\text{tot}} & = & \int_{0}^{\infty}dk\, E^{(1D)}\left(k\right).
\end{eqnarray}

In general, as discussed in Sec.~\ref{SubSec:Biot-SavartModel}, a velocity field can be separated into an incompressible (or solenoidal) part with $\nabla\cdot\mathbf{v}_{{\rm i}}=0$ and an irrotational field with $\nabla\times\mathbf{v}_{c}=0$. This means that the Richardson picture of big vortices decaying into smaller vortices should be looked for in the field with non-vanishing vorticity, i.e. the incompressible part of the velocity field.

A similar separation has also to be applied to a generic compressible superfluid system. In contrast to liquid helium systems, atomic Bose-Einstein condensates, due to the weak interparticle interaction in these systems, have the advantage of allowing a theoretical modelling of its dynamics in terms of the GP equation [Eq.~\eqref{Eq:TDGP}] which has total energy given by the functional 
\begin{equation}
E=\int d^{3}r\,\left(\frac{\hbar^{2}}{2m}\left|\nabla\psi(\mathbf{r},t)\right|^{2}+V_{{\rm trap}}\left|\psi(\mathbf{r},t)\right|^{2}+\frac{g}{2}\left|\psi(\mathbf{r},t)\right|^{4}\right).
\end{equation}
In this case the total energy $E=E_{{\rm K}}+E_{{\rm V}}+E_{{\rm I}}+E_{{\rm Q}}$ can be decomposed in its kinetic $E_{\rm K}$, potential $E_{V}$, interaction $E_{I}$, and quantum $E_Q$ parts \cite{Bradley2012,Horng2009}  according to 
\begin{eqnarray}
E_{{\rm K}} & = & \frac{m}{2}\int d^{3}r\rho(\mathbf{r},t)\left|\mathbf{v}(\mathbf{r},t)\right|^{2},\label{eq:EK}\\
E_{{\rm V}} & = & \int d^{3}r\rho(\mathbf{r},t)V_{\rm{trap}}(\mathbf{r},t),\\
E_{{\rm I}} & = & \frac{g}{2}\int d^{3}r\rho(\mathbf{r},t)^{2},\\
E_{{\rm Q}} & = & \frac{\hbar^{2}}{2m}\int d^{3}r\left|\nabla\sqrt{\rho(\mathbf{r},t)}\right|^{2},\label{eq:EQ}
\end{eqnarray}
Note that, $E_Q$ is the energy owing to the zero-point motion (see also Sec.~\ref{SubSec:Reduction}) and shall go to zero in the classical limit, when, for instance, examining the gas at scales much larger than the healing length $\xi$.

It is also convenient to define the so-called \emph{density averaged velocity field} $\mathbf{w}(\mathbf{r},t)=\sqrt{\rho(\mathbf{r},t)}\mathbf{v}(\mathbf{r},t)$ which can be then separated (again, according to the Helmholtz theorem) in its solenoidal $\nabla\cdot\mathbf{w}^{(i)}(\mathbf{r},t)=0$ and irrotational $\nabla\times\mathbf{w}^{(c)}(\mathbf{r},t)=0$ parts. This, in turn, allows the kinetic energy $E_{\rm K}$ to be separated in its compressible and incompressible parts \cite{Bradley2012}
\begin{eqnarray}
E_{\rm K}^{(i)} & = & \frac{m}{2}\int d^{3}r\left|\mathbf{w}^{(i)}(\mathbf{r},t)\right|^{2},\\
E_{\rm K}^{(c)} & = & \frac{m}{2}\int d^{3}r\left|\mathbf{w}^{(c)}(\mathbf{r},t)\right|^{2},\\
E_{{\rm K}} & = & E_{\rm K}^{(i)}+E_{\rm K}^{(c)},
\end{eqnarray}
which also have a simple representation in Fourier space 
\begin{eqnarray}
E_{\rm K}^{(i,c)} & = & \frac{m}{2}\int d^{3}k\left|\tilde{\mathbf{w}}^{(i,c)}(\mathbf{k},t)\right|^{2},\\
\tilde{\mathbf{w}}^{(i,c)}(\mathbf{k},t) & = & \frac{1}{\left(2\pi\right)^{3/2}}\int d^{3}k\, e^{i\mathbf{k}\cdot\mathbf{r}}\mathbf{w}^{(i,c)}(\mathbf{r},t).
\end{eqnarray}
Analogously to the classical case a one-dimensional energy distribution can also be defined for isotropically distributed $\mathbf{w}^{(i)}(\mathbf{k},t)$
\begin{eqnarray}
E^{(3D)}\left(k\right) & = & \frac{m}{2}\left\langle \left|\tilde{\mathbf{w}}^{(i)}(\mathbf{k},t)\right|^{2}\right\rangle, \\
E^{(1D)}\left(k\right) & = & 4\pi k^{2}E^{(3D)}\left(k\right),
\end{eqnarray}
in three and one dimension respectively. 
Since the dynamics and interactions of quantum vortices are strongly nonlinear problems, analytical approaches are very restricted and almost all the literature in this subject deals with numerical simulations. In Refs. \cite{Kobayashi2005,Kobayashi2005b}, high-accuracy numerical simulations of the GP equation for initial states with random phase profile were performed showing the existence of a Kolmogorov-like $-5/3$ law for $E^{1D}(k)$, suggesting thus some similarity between the turbulent dynamics of $\mathbf{w}^{(i)}(\mathbf{r},t)$ and the velocity field in incompressible Navier-Stokes equation. 

The analogy between the classical and quantum fluid can be carried even further by considering the possibility of Richardson cascades involving the big structures with many quantized vortices forming bundles of vorticity and decaying into smaller structures in a self-similar fashion. This process is believed to generate direct energy cascades in 3D as in the classical case \cite{Baggaley2012c}. Several authors also claim the existence of inverse energy cascades, while direct enstrophy cascades are still object of dispute \cite{Reeves2012}.

\subsection{Wave turbulence in BECs \label{SubSec:WT}}
In addition to the turbulence associated with the motion and interaction of vortices, there are some processes that can occur in BECs which are classified under the umbrella of \emph{wave turbulence} (WT). In that case the turbulence emerges due to interacting dispersive waves -- in analogy to eddies in hydrodynamic turbulence. WT, the interaction of waves in nonlinear media, appears as a cascade with power laws in the wave energy spectrum (see Ref.~\cite{Fujimoto2015} and references therein). In BECs the motion of lines of quantized vorticity and also acoustic waves are both nonlinear and fall in this classification. The system is said to exhibit \textit{weak wave turbulence}, when its turbulence is described by weakly nonlinear dispersive equations of motions. The weak nonlinearities allow an analytic description as described in Refs.~\cite{Nazarenko2011,Zakharov1992}.

In this description, nearly all the properties associated to the stationary turbulent states can be obtained from the knowledge of the wave dispersion relations and the form of their lowest order nonlinear term. In many applications of this type, the system can be described by a complex field whose equation of motion in Fourier representation can be written up to its lowest nonlinear term.

\paragraph{Phonons}
Sound waves are small amplitude excitations $\delta\psi(\mathbf{r},t)$ over the background macroscopic wavefunction $\psi_{0}(\mathbf{r},t)$ \cite{Kevrekidis2008}. For a homogeneous background one may consider $\psi_{0}(\mathbf{r},t)=\sqrt{\rho_{0}} e^{-\frac{ig}{\hbar}\rho_{0}t}$. By substituting the wavefunction $\psi(\mathbf{r},t)=\left[\sqrt{\rho_{0}}+\delta\psi(\mathbf{r},t)\right]e^{-\frac{ig}{\hbar}\rho_{0}t}$
and keeping only the smallest nonlinearity, i.e. neglecting any third-order
terms in $\delta\psi$, into Eq.~\eqref{Eq:TDGP}, we get
\begin{equation}
i\hbar\frac{\partial\delta\psi}{\partial t}=-\frac{\hbar^{2}}{2m}\nabla^{2}\delta\psi+g\rho_{0}\left(\delta\psi+\delta\psi^{\ast}\right)+g\sqrt{\rho_{0}}\left(2\delta\psi^{\ast}\delta\psi+\delta\psi^{2}\right),
\end{equation}
In Fourier representation, the above equation becomes
\begin{eqnarray}
i\hbar\frac{d\delta\psi\left(\mathbf{k}\right)}{dt} & = & \frac{\hbar^{2}k^{2}}{2m}\delta\psi\left(\mathbf{k}\right)+g\rho_{0}\left[\delta\psi\left(\mathbf{k}\right)+\delta\psi^{\ast}\left(\mathbf{k}\right)\right]\label{eq:motion1}\\
 &  & +\frac{g\rho_{0}}{\left(2\pi\right)^{3/2}}\int d\mathbf{k}_{1}\int d\mathbf{k}_{2}\left[\delta\left(\mathbf{k}-\mathbf{k}_{1}-\mathbf{k}_{2}\right)\delta\psi\left(\mathbf{k}_{1}\right)\delta\psi\left(\mathbf{k}_{2}\right)+2\delta\left(\mathbf{k}+\mathbf{k}_{1}-\mathbf{k}_{2}\right)\delta\psi^{\ast}\left(\mathbf{k}_{1}\right)\delta\psi\left(\mathbf{k}_{2}\right)\right].\nonumber 
\end{eqnarray}
In order to analyze the plane-wave solutions and their nonlinear corrections it is convenient to work with the Bogoliubov transformed field \cite{LvovPhysicaD2005,Dyachenko1991}
\begin{equation}
a(\mathbf{k})=\frac{1}{2}\left(\sqrt{\frac{\omega(k)}{\beta(k)}}+\sqrt{\frac{\beta(k)}{\omega(k)}}\right)\delta\psi\left(\mathbf{k}\right)+\frac{1}{2}\left(\sqrt{\frac{\omega(k)}{\beta(k)}}-\sqrt{\frac{\beta(k)}{\omega(k)}}\right)\delta\psi^{\ast}\left(-\mathbf{k}\right),
\end{equation}
which has inverse the transformation 
\begin{equation}
\delta\psi\left(\mathbf{k}\right)=\frac{1}{2}\left(\sqrt{\frac{\omega(k)}{\beta(k)}}+\sqrt{\frac{\beta(k)}{\omega(k)}}\right)a\left(\mathbf{k}\right)-\frac{1}{2}\left(\sqrt{\frac{\omega(k)}{\beta(k)}}-\sqrt{\frac{\beta(k)}{\omega(k)}}\right)a^{\ast}\left(-\mathbf{k}\right),
\end{equation}
with $\beta(k)=k^{2}\hbar^{2} / 2m$ and $\omega(k)=\sqrt{\beta(k)^{2}+2g\rho_{0}\beta(k)} .$
In this way the equation of motion for $a(\mathbf{k})$ can be obtained
\begin{eqnarray}
i\hbar\frac{da\left(\mathbf{k}\right)}{dt} & = & \omega(k)a\left(\mathbf{k}\right)\\
 & + & \int d\mathbf{k}_{1}\int d\mathbf{k}_{2}V\left(\mathbf{k};\mathbf{k}_{1},\mathbf{k}_{2}\right)\left[\delta\left(\mathbf{k}-\mathbf{k}_{1}-\mathbf{k}_{2}\right)a\left(\mathbf{k}_{1}\right)a\left(\mathbf{k}_{2}\right)+2\delta\left(\mathbf{k}+\mathbf{k}_{1}-\mathbf{k}_{2}\right)a^{\ast}\left(\mathbf{k}_{1}\right)a\left(\mathbf{k}_{2}\right)\right],\label{eq:motion2}
\end{eqnarray}
where the quantity $V(\dots)$ is the interaction coefficient and depends on the interaction parameter $g$. The latter, in the long wavelength regime scales as \cite{Nazarenko2011}
$$V(k_1,k_2,k_3) \sim \frac{\sqrt{k_{1}k_{2}k_{3}}}{\rho_{0}^{1/4}},$$ 
while the dispersion coefficient behaves as $\omega(k)\approx\hbar c_{s}k$, with the sound velocity being $c_{s}=\sqrt{g\rho_{0}/2m}$.
The $V$-dependent terms inside the integral in Eq.~\eqref{eq:motion2} describe the three-wave interactions responsible for triggering the cascade process. The first term correspond to the decay of two waves into one ($k_{1}+k_{2}\rightarrow k$) while the second term corresponds to the opposed process ($k+k_{2}\rightarrow k_{1}$).

Similarly to the hydrodynamic case, the energy is also a measure of central importance and in weak interacting case is mainly given by its kinetic part 
\begin{equation}
E=\int d^{3}k\,\omega_{k}n_{k}=\int_{0}^{\infty}dk\, E^{(1D)}(k),
\end{equation}
where $n_{k}=a_{k}^{*}a_{k}$ and $E^{(1D)}(k)$ is the so-called one-dimensional energy distribution.

According to the statistical theory of out-of-equilibrium waves, presented in \cite{Nazarenko2011,Zakharov1992}, such three-wave processes allow the existence of a steady state characterized by a direct energy cascade. Indeed the power-law distribution associated with the energy spectrum of the steady-state turbulent system can be entirely determined from the scaling properties of the dispersion relation and the interaction coefficient. For the special case in Eq.~\eqref{eq:motion2}, where dispersion and interaction coefficients scale as $\omega(\lambda\mathbf{k})=\lambda\omega(\mathbf{k})$
and $V(\lambda\mathbf{k}_{1};\lambda\mathbf{k}_{2},\lambda\mathbf{k}_{3})=\lambda^{3/2}V(\mathbf{k}_{1};\mathbf{k}_{2},\mathbf{k}_{3})$, the energy spectrum turns out to be of the Zakharov-Sagdeev type \cite{Nazarenko2011}:
\begin{equation}
E^{(1D)}\sim k^{-3/2}.
\end{equation}

\paragraph{Kelvin waves}
Another WT process that can occur in BEC is associated with the vibratory motion of vortex lines and vortex loops, the so-called  Kelvin waves (see also Sec.~\ref{SubSec:QT}). The first WT theory for this system was formulated by Kozik and Svistunov in \cite{Kozik2004}. In such a formulation, the relevant interaction term involves 6-wave processes. In his book \cite{Nazarenko2011}, Nazarenko points out that two different cascade processes can simultaneously occur within a WT system lead by $N$-wave interactions, with $N$ being an even number. A direct energy cascade is present, where the energy density in Fourier space ($\omega_{k}n_{k}$) flows from large to small length scales. In addition to this, there can also be an inverse \textit{wave action} ($n_{k}$) cascade. Such dual cascades have close analogies with the energy/enstrophy dual cascade in hydrodynamic turbulence.

In the Kozik-Svistunov theory, the dispersion relation, neglecting logarithm factors, is $\omega_{k}\sim k^{2}$, which leads to the direct energy cascade spectrum 
\begin{equation}
E^{(1D)}\sim k^{-7/5},
\end{equation}
while the inverse wave action spectrum is given by \cite{Kozik2004} 
\begin{equation}
E^{(1D)}\sim k^{-1}.
\end{equation}

Interestingly and in contrast to hydrodynamic turbulence, WT relies only in the nonlinearity induced by the two-body interactions that the $\sim g |\psi|^2$ (and consequently $V$) term captures. Note that, the $g$-dependence of the GP interaction is not present in the quantum Euler equation [Eq.~\eqref{Eq:QuantumEuler}] and hence hydrodynamic turbulence does not explicitly depend on the particle-particle interaction of the bosons of the cloud.

\subsection{What all this has to do with the experiments? \label{SubSec:WhatExp}}
In time-of-flight experiments, indirect measurements of the kinetic energy density are possible. This stems from the fact that, after a long enough time of expansion $\tau_f$, the observed density of the gas in a standard absorption image process has the form of the distribution of the momenta $k=m x/\hbar\tau_f$ \emph{in situ}, i.e. inside the trap, before the ballistic expansion of the gas:
\begin{equation}
\rho(x;\tau_f) \propto \tilde\rho(k).
\end{equation}
However and since, as we saw, this energy is composed from several component parts [Eqs.~\eqref{eq:EK}--\eqref{eq:EQ}], it is a quite challenging task to associate any observed power law spectrum $E^{(1D)}\sim k^{-\nu}$ with specific cascade processes. That is, an experimental absorption imaging method cannot decide what type of turbulence is there. As shown above, we have for the sound wave cascade $\nu=3/2=1.5$ and for the direct energy Kelvin wave cascade $\nu=7/5=1.4$. These exponents are very close to the Kolmogorov $\nu=5/3\approx1.67$ from hydrodynamic turbulence which complicates even further the determination of which process is actually responsible for any observed power laws.

Vorticity in superfluids and quantum systems is probably the most important ingredient of quantum turbulence. In recent experiments (see Sec.~\ref{Sec:Experimental}) a $^{87}$Rb Bose condensate is used to observe and investigate quantum turbulence, by means of a weak off-axis magnetic field gradient, which perturbs the BEC and injects kinetic energy onto it. This nucleates on the condensed-thermal interface of the BEC and sets up experimental conditions to the emergence of turbulence. It is important for the development of turbulence that the nucleated vortices are not coplanar, not ordered and let to proliferate and interact. Once the turbulent regime is set, the condensate is then released and expands under free fall. The atomic density profile is acquired using resonant absorption imaging, some milliseconds (typically $15~ms$) after the gas has been released from its trap (time of flight). The calibrated density images are used to determine the \emph{in situ} momentum distribution of the BEC. Power laws in the experimentally measured momentum and energy distributions have been observed, making these states deviate significantly from the unperturbed gas. We present these studies in Sec.~\ref{SubSec:MomDist}. Additional characteristics of the system, such as the finite number, size and temperature of the condensate, shall play a role in the energy injection mechanisms and are further discussed throughout Sec.~\ref{Sec:Experimental}.

\section{Energy transport and dynamics \label{Sec:EnergyTransport}}
\subsection{Vortex reconnections \label{SubSec:Vortexreconnections}}

As mentioned in Sec.~\ref{Sec:Background}, the quantization of vorticity is a unique feature of superfluids and condensates. The study of vortex propagation and interactions are central in the study of turbulence, at least in its hydrodynamic manifestation. The experimental settings, so far, that brought BECs to turbulence have generated a large number of vortices that eventually proliferate, reconnect and tangle with each other \cite{Henn2009,White2014a}. A quantum vortex reconnection (QVR) is a general phenomenon in which two vortex lines interact, approach each other, connect at a given point for some (short) time and exchange tails. This process happens repeatedly in highly excited gases, like the turbulent systems. Hence, the study of QVR is central in turbulent dynamics as reconnections are, to a large extent, responsible for the energy transfer between different scales. As such, QVR can be considered as a fundamental mechanism of QT.

Quantum vortices were first observed experimentally in trapped BECs in 1999-2000 in two different laboratories, in JILA and Paris \cite{Matthews1999,Madison2000}. The vortices were nucleated with different mechanisms (phase imprint and stirring with a laser, respectively) in the two experiments. Since then, vortices have been extensively studied in gaseous BEC and nowadays they can be routinely created (see \cite{Fetter2009,Fetter2010} for reviews on this topic). Interestingly, the existence of vortices in superfluids has, decades earlier, stimulated the formulation of the Gross-Pitaevskii theory \cite{Gross1961,Pitaevskii1961}  (see also Sec.~\ref{SubSec:GrossPitaevskiiModel}). In liquid helium QVRs have been, with impressive detail, monitored by the group of Lathrop in Ref.~\cite{Bewley2006,Bewley2008} and can be seen in the video of Ref.~\cite{LathropVideo1}.

Even though, the present experimental work pertains to trapped gases, we begin the exposition with theoretical results for homogeneous systems. This happens in order to, firstly, give a brief historic account of the principal results and, secondly, to emphasize on the fact that QVR is a behavior that can be seen across many different physical settings. 
Vortex reconnection is a universal phenomenon appearing in different classical and quantum systems, and also arises within different theories. In Fig.~\ref{Fig:Reconnections} we present reconnections that emerge as solutions of the (a) NS equation, (b,c) the GP equation and also (d) the many-body MCTDHB equations (see Sec.~\ref{Sec:Essentials}). 
As already mentioned, the vortex filament models of the Biot-Savart theory (presented in Sec.~\ref{SubSec:Biot-SavartModel}) cannot describe the reconnection \emph{per se}; an \emph{ad hoc} inclusion of it is, however, possible. 

\subsubsection{In homogeneous space}
A line vortex moving in the velocity field produced by another vortex will exert a force to the latter, and vice versa. This is the mutual Magnus force:
\begin{equation}
\mathbf F \propto - \rho_s s^\prime \times \mathbf{v}_i(s),
\label{Eq:VortexForce}
\end{equation}
where $\rho_s$ is the superfluid density and $\mathbf{v}_i(s)$ is the sum of the velocity fields of both vortex lines, (see Eq.~\eqref{Eq:Biot-SavartLaw}). Note that the vortex line -- unless completely straight -- will feel a force on it induced by its own velocity field due to its curvature. Hence, vortex lines in the three-dimensional space will feel a force by the two velocity fields combined, resulting in a mutual attraction or repulsion. depending on their relative orientation. For a generic initial configuration of the vortex lines in space this will result in parts of the lines mutually attracting and approaching until reconnection. It is only the extreme case, where the two filaments lie totally straight in free space and parallel to each other, that the force (say, attraction) does not depend on $z$; then the two lines will approach each other keeping their initial parallel shape until the collide. However, boundary conditions, density inhomogeneities or local random fluctuations along the vortex line will break this symmetry and reconnection at one point -- i.e., the familiar tail-exchange event -- will eventually occur.

\begin{figure}
\centering
\includegraphics[width=0.16\textwidth]{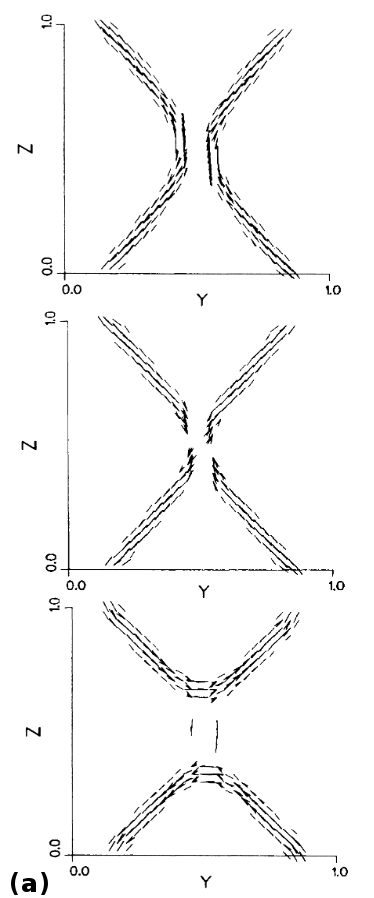}  ~~~~
\includegraphics[width=0.16\textwidth]{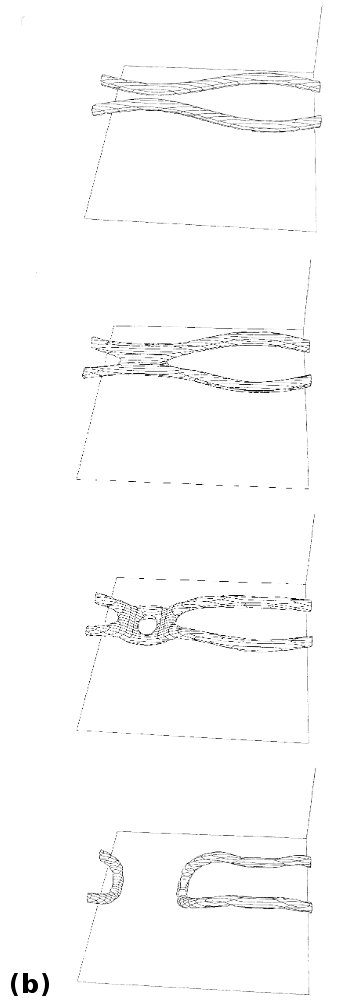}   ~~~~
\includegraphics[width=0.16\textwidth]{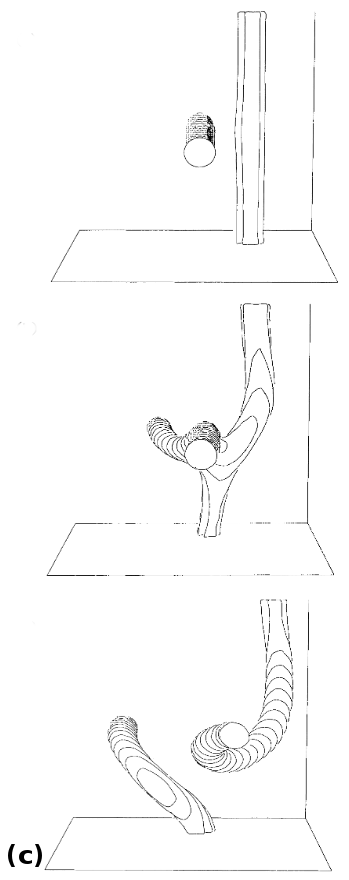}
\includegraphics[width=0.17\textwidth]{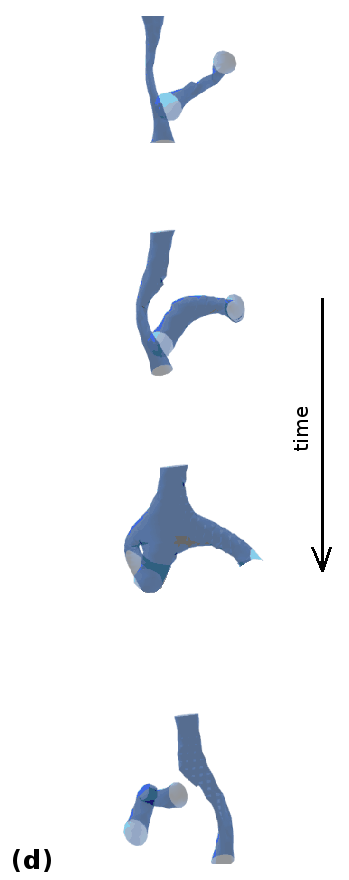}
\caption{Universality in the route of vortex reconnection. Examples of vortex reconnections as calculated within different theories, with the time running from top to bottom in all cases. In (a) is shown the vorticity field of the Navier-Stokes equation for an initial condition of two vortex tubes (source: paper of Ashurst and Meiron \cite{Ashurst1987}). In (b) [(c)] quantum vortices are plot initially in an anti-parallel (orthogonal) configuration as found by solving the Gross-Pitaevskii equation in homogeneous space. Source: paper of Koplik and Levine \cite{Koplik1993}. In (d) a vortex reconnection is shown for initially orthogonal vortices as computed within a many-body theory (MCTDHB) for a parabolically and isotropically trapped gas. Source: paper of Wells \emph{et al.} \cite{Wells2015}.}
\label{Fig:Reconnections}
\end{figure}

Since decades, reconnections have been studied extensively both in the context of quantum and classical theories. In 1987 Ashurst and Meiron simulated reconnections using a techique that combines the BS model and NS equation \cite{Ashurst1987}, as seen in Fig.~\ref{Fig:Reconnections} (a). Later on, in 1993, Koplik and Levine showed that quantum vortex reconnections appear in the dynamics of the GP equation \cite{Koplik1993}, shown in Figs.~\ref{Fig:Reconnections} (b) and (c). In 1994 Waele and Aarts with a BS model claimed a universal route to reconnection for all kinds of initial vortex-antivortex arrangements \cite{deWaele1994}. More recent relevant work includes the study of vortex reconnections in untrapped superfluids \cite{Tebbs2010}, the numerical computation of the minimum distance between two approaching vortices as function of time \cite{Zuccher2012} and the calculation of the energy spectra of gases with reconnecting vortices \cite{Nemirovskii2013,Nemirovskii2014}.

\subsubsection{In a trap}
\begin{figure}
\centering
\includegraphics[width=0.6\linewidth]{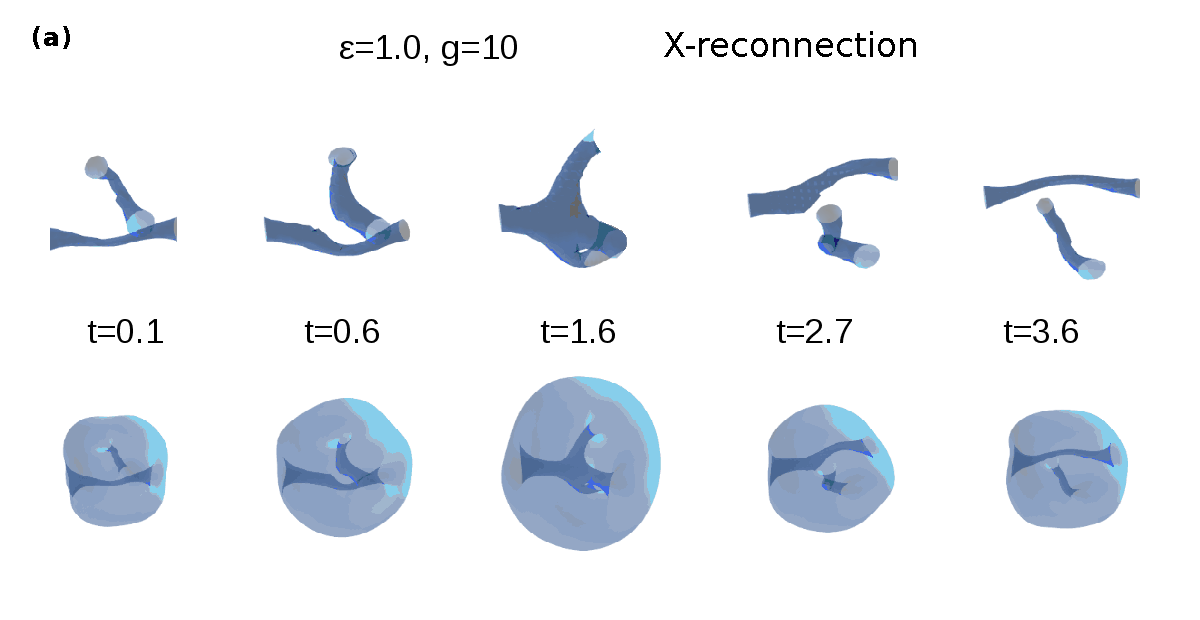}
\includegraphics[width=0.6\linewidth]{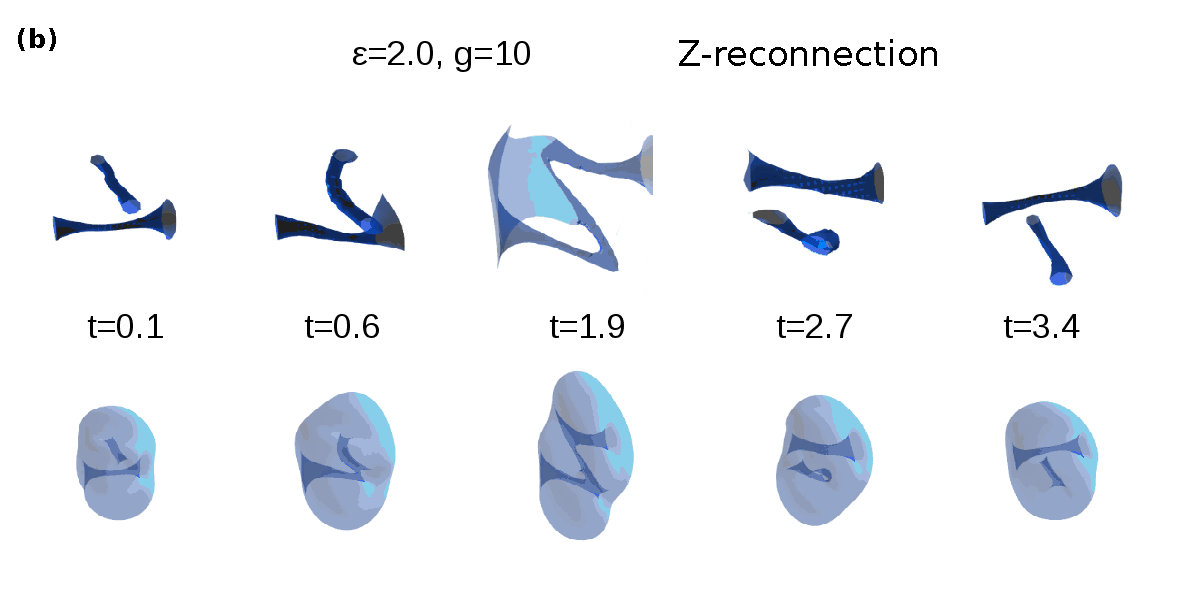}
\label{Fig3}
\caption{Different paths to quantum vortex reconnection depending on the trap anisotropy. (a) X-reconnection: In a symmetrically trapped BEC the vortices follow the expected path and reconnect at the centre, where the density is maximum. (b) Z-reconnection: When the aspect ratio of the $xy$-to-$z$ trapping frequencies is changed to $2:1$ the vortices follow a different path and they reconnect at two points in the outer regions of the gas, where the density is minimum. In both panels shown are the vortex tubes only (upper raw) and the vortices within the bulk of the gas (lower raw). Source: paper of Wells \emph{et al.} \cite{Wells2015}.}
\label{Fig:ReconnectionsAnis}
\end{figure}

As said, experiments that involve vortices are performed in parabolically trapped gases and, hence, studies of vortex interactions in non-homogeneous systems are necessary. Lately, the interest has shifted to simulate realistic situation, where vortices are seeded in a trapped gas and let proliferate. Qualitatively, the QVR process does not change dramatically when the system density is no more homogeneous. The nonuniformity of the density induces an extra force on the vortex line. For instance, a single vortex isotropically confined will precess around the centre of the trap. The tail-exchange process between two vortex lines described above still occurs, interestingly, only as long as the trapping anisotropy does not exceed a critical value \cite{Wells2015}. That is, for isotropic and close-to-isotropic traps the QVR happens in the same familiar way, at the center of the condensate where the density is higher. For larger anisotropies the QVR tends to happen at the margins of the gas, where the density is lower (see Fig.~\ref{Fig:ReconnectionsAnis}). 

The impact of finite temperatures on the QVR in trapped BECs has been studied in \cite{Allen2014} using the GP approach coupled to a thermal cloud. It is found that thermal effects do not inhibit or alter the reconnection process, extending thus the universality of the QVR to both zero and finite-temperature systems.

Simulations of QVR have also been obtained with methods that go beyond the common mean-field approach (as the GP model) \cite{Wells2015}. In specific, it is possible to solve the many-body time-dependent Schr\"odinger equation as an initial value problem in three dimensions, for a state of the trapped gas containing two perpendicular vortices. This initial state is usually created by enforcing a desired phase field in the initial wavefunction and a density node as well, accounting for the vortex core. In Ref.~\cite{Wells2015} where QVR is studied in anisotropic traps, the time-dependent Schr\"odinger equation is solved with the \textit{Multi-Configurational Time-Dependent Hartree for Bosons} (MCTDHB) method \cite{Streltsov2006,Streltsov2007,Alon2008} and its numerical implementation (MCTDH-X)\cite{ultracold.org}. There, it is found that the `walls' of the trap will heavily interfere with the reconnection process, as long as the trap anisotropy exceeds some critical value. This impacts the `shape' of the reconnection (see Fig.~\ref{Fig:ReconnectionsAnis}) as well as the time which the vortices spend during the reconnection. Furthermore, this process might also alter the familiar empirical law $d\sim \sqrt{t_0-t}$ that gives the minimum distance $d$ of two approaching vortices as a function of time.

\begin{figure}
\centering
\includegraphics[width=0.43\textwidth]{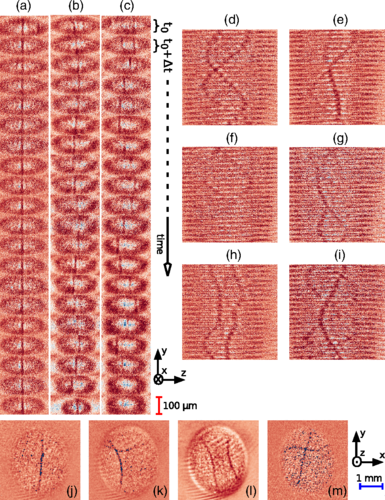}
\caption{Vortex dynamics as seen in the laboratory. Observed oscillations and possibly reconnections of individual vortices created in a cigar-shaped BEC of $10^7$ sodium atoms. Source: paper of Serafini \emph{et al.} \cite{Serafini2015}.}
\label{Fig:ReconnectionsExp}
\end{figure}

QVR has been explored in the laboratory in liquid helium systems \cite{Bewley2008,Paoletti2008,Paoletti2010} and only very recently experimental evidence for it in a trapped BEC has been reported \cite{Serafini2015} (see Fig.~\ref{Fig:ReconnectionsExp}).

\subsubsection{Kelvin waves}
The QVR is a fundamental process of energy transport in quantum fluids with vorticity with important consequences on the turbulent dynamics. Not only does each individual reconnection release kinetic energy, but it nucleates vibrations on individual vortex cores. The energy release can take the form of phonon radiation and vibrations on the vortex core that are carried by helically rotating Kelvin waves. The excited by the reconnection phonons and waves can offer an explanation in one of the fundamental questions of QT: what the nature of the dissipation is, at zero-temperature frictionless fluids \cite{Fonda2014}. Indeed, Vinen postulated that high frequency oscillations of a vortex core can efficiently produce phonon radiation, allowing for an eventual dissipations of energy in an inviscid fluid \cite{Vinen2000}. It is natural to assume that turbulent vortex tangles decay with mechanisms like phonon radiation and helical waves excitations.

Kelvin waves are long scale perturbation along the vortex line that can exist in both classical and quantum fluids \cite{Kozik2004,Nazarenko2005}. In superfluids it has been suggested that cascades of Kelvin waves can transfer energy from the scale of the intervortex spacing down to smaller scales, as small as the vortex core \cite{Kivotides2000,TsubotaJoPCM2009,Fonda2014}. Therefore, Kelvin modes become an important mechanism in the decay of turbulence \cite{Kondaurova2014} as energy is directed to the smaller scales of the system, where it is eventually dissipated.

In Ref.~\cite{Kivotides2000} the authors, solve numerically the BS model for an initial configuration of four vortex rings and report on a Kelvin-wave cascade owing to ``individual vortex reconnection events which transfers energy to higher and higher wave numbers $k$''. Recalling that, simulations with the BS approach involve no compressible energy at all, the mechanisms of energy transfer studied in the works of Refs.~\cite{Kivotides2000, Kondaurova2014} and others, pertain to vortex energy exclusively. This suggests, that cascades of energy mediated by Kelvin waves assist the decay of turbulence, independently of the existence of other collective modes as phonons or others. In Ref.~\cite{Simula2008} simulations confirmed that such cascades can serve as a mechanism for dissipation of energy within the GP theory too. In specific, a gas in an axisymmetric elongated trap possessing a single vortex is studied as an initial condition problem. A time-dependent rotating potential perturbs the initial state and its time-evolution shows the emergence of helical Kelvin waves along the vortex line. The Kelvin waves eventually decay to longer wavelength excitations via emission of phonons, thus making this process a zero-temperature energy decay mechanism.

In Ref.~\cite{Bretin2003} Kelvin modes in a $^{87}$Rb BEC have been indirectly seen, through the decay of quadrupole modes. In liquid helium Kelvin waves have been visualized in \cite{Fonda2014}.

\subsection{In two dimensions \label{SubSec:2DQT}}
\subsubsection{General characteristics of 2DQT}
In subsection \ref{SubSec:BriefCT} we already mentioned that 2D classical turbulence is very different from the 3D case and presents interesting features, such as the inverse energy and enstrophy cascades. The search for similar startling characteristics of the 2D classical counterpart in quantum fluids has been intense. While the $-5/3$ scaling (also expected for the 2D classical turbulent fluids, as predicted by Kraichnan \cite{Kraichnan1980}) has been found in simulations of the GP equation for homogeneous systems, several points concerning energy cascade direction is still an open debate.

In contrast with the classical analogue, where assuming two-dimensionality is in most cases an approximation, in quantum flows two-dimensionality can be achieved by exploiting the controllability that experimental gaseous atomic condensates offer. Using suitable trapping potentials, atomic BECs can be easily shaped so that vortex dynamics is 2D rather than 3D making them ideal systems to study 2D turbulence \cite{White2014}. In these settings, one direction is so tightly confined that the dynamics along it is practically frozen and the system is  quasi-2D. Precisely, the anisotropic optical trapping makes the Thomas-Fermi radius $R_{\mathrm{TF}}$ [Eq.~\eqref{Eq:TFRadius}] on the transverse direction much smaller (factor of $10$ or more) than the $R_{\mathrm{TF}}$ of the radial direction, rendering thus transverse excitations energetically costly. The quantized nature of quantum vorticity together with this reduction of dimensionality implies that the properties of the vortex line are no longer relevant for dynamics, as the one-dimensional vortex now becomes a zero-dimensional point in the plane. Kelvin waves and their cascade, that is the fundamental process for QT decay in  three-dimensional turbulence, are not applicable in two dimensions. Vortex-antivortex annihilation is now a salient feature and, moreover, clusters of vortices appear as the large-scale structures that play a role analogue to the 3D vortex bundle. Another practical advantage of using atomic condensates to explore 2DQT is that, unlike liquid helium, 2D quantum vortices can be directly imaged and, unlike classical systems, the motion of such 2D vortices is not hindered by viscous effects or friction with the substrate.

The 2D problem has been investigated in both homogeneous and trapped systems, and are being discussed in the following.

\paragraph{Homogeneous Systems}

\begin{figure}[t!]
\begin{center}
\includegraphics[width=0.5\textwidth]{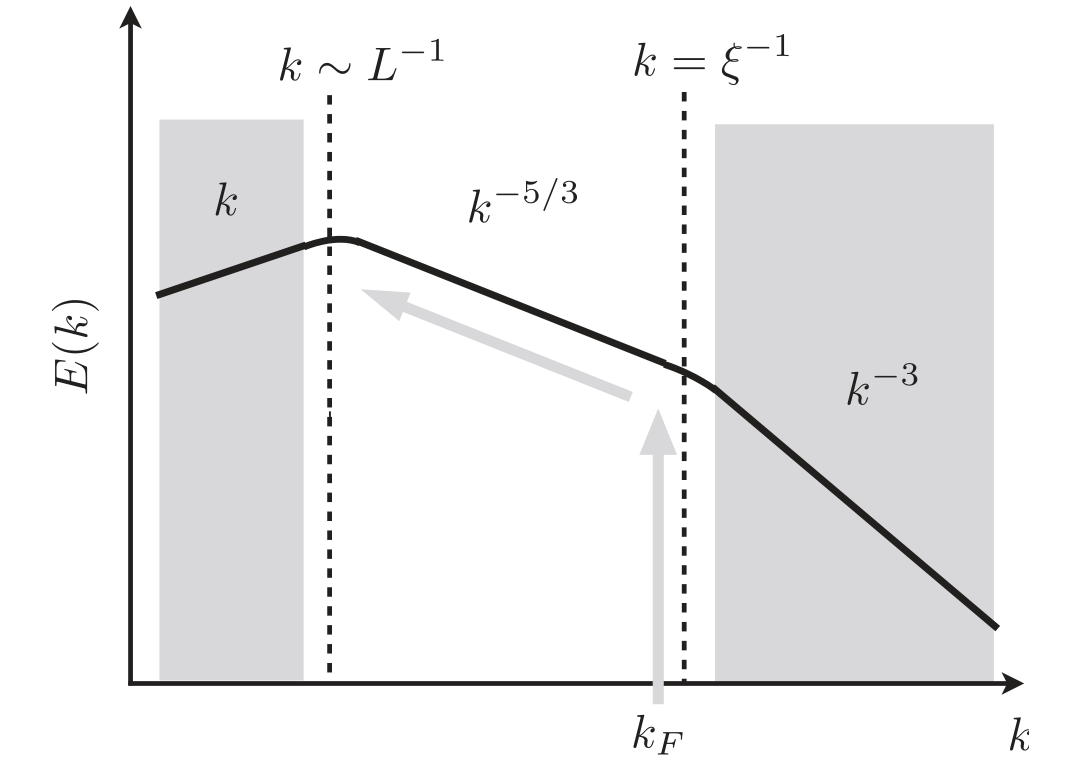} 
\caption{Qualitative picture of the incompressible kinetic energy spectra for a 2D system. The $k^{-3}$ part of the spectrum appears due to the structure of the vortex core while the $k$ region pertains to distances larger than the largest intervortex distance $L$ and has no net vorticity. The nonshaded region is the inertial range where Kolmogorov scaling $k^{-5/3}$ manifests. Source: paper of Bradley \emph{et al.} \cite{Bradley2012}.}
\label{2Dspectra}
\end{center}
\end{figure}

Several numerical studies have explored the generation of turbulence in 2D homogeneous condensates. The investigation of the evolution of the system towards turbulence leads to important questions regarding vortex spatial distributions. As we have discussed, in order to show the Kolmogorov scaling, a turbulent system must also display self-similarity throughout several scales in an inertial range, which requires particular vortex distributions. For the appropriate quasiclassical (with relatively large ratio $L/\xi$) 2D turbulent regime, the incompressible kinetic energy is found \cite{Bradley2012} to follow a spectrum of
\begin{eqnarray}
E^{(2D)}_i \propto k^{-3}, &\text{for}& ~~ k\gg \xi^{-1},  \label{2Dspectrum-a}\\
E^{(2D)}_i \propto k^{-5/3}, &\text{for}& ~~ k < \xi^{-1} \label{2Dspectrum-b},
\end{eqnarray}
which is qualitatively illustrated in Fig.~\ref{2Dspectra} for system where forcing takes place in small scales $k_\mathrm{F}\sim \xi^{-1}$. In 2D classical turbulence, a direct enstrophy cascade accompanies the inverse energy cascade (IEC), which scales as $k^{-3}$ and so analogues in the quantum system have been sought. As Eq.~(\ref{2Dspectrum-a}) shows, the same scaling is found in 2D quantum systems for a completely different reason, however; the problem of what the quantum analogue of enstrophy should be remains. In contrast to its classical counterpart, the velocity field profile of a quantized vortex is responsible for the $k^{-3}$ scaling, therefore representing solely the internal structure of the vortex core. This same quantized nature of quantum vortices makes enstrophy to be proportional to the number of vortices \cite{Bradley2012,White2014} and the possibility of vortex-antivortex annihilations forces enstrophy not to be an inviscid quantity in quantum fluids. A different interpretation is typically adopted in this case: the approximate conservation of vortex number is taken to be the analogous of the enstrophy conservation in classical systems, helped by long-lived clustering of like-signed vortices. In the context of decaying turbulence, the authors in Ref.~\cite{Numasato2010} created initial states from random phases and evolved them using the GP equation, highlighting the effect of non-conservation of vortex number. Although their system also developed a statistical distribution of vorticity consistent with Kolmogorov scaling, the compressible nature of the superfluid and the absence of forcing make enstrophy in these systems a non-conserved quantity; the vortex number always decays due to pair annihilation processes and no dynamical mechanism keeps injecting pairs at the same rate of their removal. This fact prohibits the development of an IEC and the calculated energy flux becomes direct, as in the 3D case. A detailed study of energy transfer mechanism in 2D systems using the GP equation can be very challenging, since determining the energy flux depends on knowing how compressible and incompressible energies couple. A way to go around this problem is proposed in Ref.~ \cite{Billam2015}, where a modified point-vortex model is introduced to mimic the effects of the GP equation and dissipation in a purely incompressible fluid. Their results suggest that the IEC due to vortex interactions appears only for large systems with moderate dissipation. A transient dual energy-enstrophy cascade is also verified, validating the quasiclassical behavior for such systems.

In Ref.~\cite{Bradley2012} the authors investigated the characteristics of the 2D incompressible kinetic energy spectrum and discussed its several attributes. Particularly, after sampling both randomized and clustered vortex distributions they consistently verified the emergence of the $-5/3$ scaling in the latter cases, where the distribution of vortex separation is imposed to follow a specific power-law. While Ref.~\cite{Bradley2012} investigates the spectral characteristics by sampling over, reportedly turbulent, quasiclassical states, in Ref.~\cite{Reeves2013} a dynamical simulation of an actual forced homogeneous system suggests an IEC. A superfluid flow past a grid of obstacles simulated using the dissipative GP equation displays clear features of a forced turbulent state. Alongside the Kolmogorov -5/3 scaling, the creation of clusters of like-signed vortices and the calculation of the energy flux have provided evidence for the existence of the IEC. The 2D energy spectrum and scaling laws of quasiclassical 2DQT have been further computed in numerical simulations (see Refs.~\cite{Numasato2010,Nowak2011}) and are in accordance with Eqs.~(\ref{2Dspectrum-a}) and (\ref{2Dspectrum-b}).

\paragraph{Trapped Systems}

\begin{figure}
\centering

\includegraphics[width=0.65\textwidth]{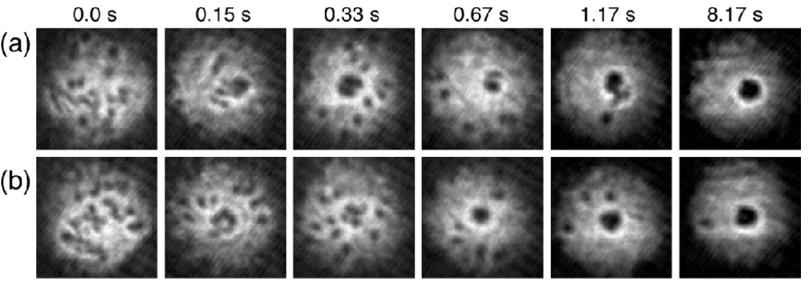}
\caption{Turbulent flow in a 2D quantum gas. Energy is injected by stirring a gas of $^{87}Rb$ atoms trapped in a toroidal potential and gradually transferred to larger scales. Vortices that are created from the stirring motion coalesce to form a persistent current (centre of the trap). The images shown here are for different waiting times, after the stirring has been performed. In all images the trap potentials has been switched off and the gas is ballistically expanding. Source: paper of Neely \emph{et al.} \cite{Neely2013}. 
}\label{Fig:Neely}
\end{figure}

Despite the first investigations into QT in trapped BEC having focused on looking for quasiclassical characteristics \cite{Parker2005}, in trapped systems we are usually dealing with systems that typically exhibit only the ultraquantum limit of turbulence or, at most, an underdeveloped large-scale self-similarity. As opposed to the homogeneous case, the finite size of the trapped system typically hinders the development of large-scale motion. In Ref.~\cite{White2012}, for instance, the authors performed simulations of a paddle potential (of an approximate elliptical shape) that stirred an atomic condensate with dimensions of current 2D experiments in BECs. The degree of vortex clustering was measured using statistical methods of nearest-neighbors and first-order point correlations (known as Besag's function). Although the stirring process happened at small-scales (forcing scale few times that of the healing length $\xi$), the system size prevented the growth of vortex clustering for several different stirring parameters; such growth would be an indication of inverse incompressible energy transfer. Another example of such a finite-size effect can be found in Ref.~\cite{Reeves2012}, where a simulation of a trapped BEC stirred by a Gaussian potential of width $\sim 4\xi$ is performed. Different stirring regimes were found based on the vortex shedding process and vortex distribution, which depend on the barrier amplitude and stirring velocity. The incompressible energy spectrum was studied for several cases. Particularly, the infrared region of the spectrum ($k\xi <1$) displays a close resemblance to Kolmogorov spectrum when stirring produces clustering of like-signed vortices. However, the $-5/3$ scaling can only be seen for a region of $2\pi/30<k\xi<1$ (less than a decade), highlighting the lack of scales available (as discussed in Secs.~\ref{Sec:Scales} and \ref{SubSec:Differences}). Similarly, in the context of both forced and decaying turbulence, Ref.~\cite{Neely2013} showed that the incompressible energy spectrum also mimics Kolmogorov for roughly a decade in $k$-space when clustering of like-signed vortices is present. In this latter case, vortices were nucleated in a condensate trapped in a toroidal geometry by small-scale stirring using a laser beam. For certain driving and waiting times the authors observed the formation of a persistent current around the torus, which persists after a ballistic expansion. This behavior was verified both experimentally and through numerical simulations and, by using a qualitative argument, they concluded that energy is being transferred from small scales ($\sim \xi$) by the initial stirrer to large ones ($\sim L$) by forming a persistent current along the periphery of the gas (see Fig.~\ref{Fig:Neely}). Despite the same lack of available space in the above discussed systems, this is the first demonstration of an inverse energy transfer in trapped BECs, the analogous behavior of classical 2D incompressible turbulence.

Similar configurations were used to explore vortex shedding and annihilation processes in both experiment \cite{Kwon2014,Kwon2015} and simulations \cite{Stagg2015}, focusing on the decay of 2DQT itself. Specifically, the work of Ref.~\cite{Kwon2014} studied a sodium gas confined in a quasi-2D trap that is brought to a turbulent state by sweeping a repulsive laser beam of Gaussian shape through its center. After producing $\sim 60$ vortices, the repulsive laser beam was turned off. The time evolution of decay of the vortices was then observed and a nonexponential behavior for the total number of vortices $N_v$ in time was verified. In specific, a phenomenological description was proposed by means of a rate equation given by
\begin{equation}
\frac{dN_v}{d t} = -\Gamma_1 N_v - \Gamma_2 N^2_v,
\label{Eq:DecayKwon}
\end{equation}
where $\Gamma_1,\Gamma_2$ are real, positive parameters, attributed to one-body and two-body losses and are determined by the experiment close to $0.87$ and $1.88$ respectively. The decrease in the number of the vortices is seen in the first panel of Fig.~\ref{Fig:Kwon}. The crescent shapes (see lower panel of Fig.~\ref{Fig:Kwon}) observed in the experiments are visualizations of vortex-antivortex collisions, culminating in annihilations; such processes give rise to the two-body term in Eq.~(\ref{Eq:DecayKwon}).

The effect of stirring laser beams with different shapes or along different paths was investigated in Refs.~\cite{White2012,Reeves2012,White2014a,Neely2010,Tsatsos2015}. However, in all of the cited cases, the number of positive and negative vortices generated is approximately the same; all vortex configurations investigated had approximately zero polarization. Since irrotational flow is a hallmark property of superfluidity, the polarization of the vortex configuration (i.e. the relative proportion of positive and negative vortices) plays the role of net rotational angular velocity $\Omega$ of a classical fluid. A recent study of the effect of polarization on these quantum systems can be found in \cite{Cidrim2015}. There the authors found that the phenomenological description given by Eq.~(\ref{Eq:DecayKwon}) does not properly describe the vortex decay when a strong dependence on polarization is present and proposed a different approach. 

Successive vortex nucleations as the above are candidate mechanisms for turbulent energy transfer and are reminiscent of a B{\'e}rnand-von K{\'a}rm{\'a}n vortex street, that theoretically has been shown to be possible in a trapped BEC \cite{Sasaki2010}. However, the experimental verification of the latter still needs to be carried out.

\subsubsection{Vortex annihilation}
An extreme case of vortex reconnection is the annihilation of two vortices. This can happen only in 2D (or quasi-2D) geometries, where a vortex is considered to be a point in the plane. A vortex has thus only one orientation (perpendicular to the plane) and its velocity only two possible configurations: it either rotates clockwise or anticlockwise. Conventionally, one assigns a `positive charge' in the former and a `negative charge' in the latter. Indeed, a simple (classical) theoretical model that considers vortices as charged point particles has been developed and applied to various scenarios (see the works of Refs.~\cite{Torres2011,Navarro2013} and references therein). According to it, point vortices in trapped symmetric 2D systems move and mutually interact with $\sim 1/r$ forces. Even though this model takes into account no vortex annihilation -- analogous to the absence of vortex reconnection in the BS model -- it captures nicely the complicated and chaotic dynamics of systems of few co-rotating or counter-rotating vortices \cite{Koukouloyannis2014,Kyriakopoulos2014}.

The creation of pairs of vortices, again in a quasi-2D gas, has been studied and experimentally demonstrated \cite{Neely2010} in the case of an obstacle moving through (sweeping) the superfluid. It was shown that vortices of opposite charge (vortex dipoles) can be formed, are stable and proliferate, as long as the relative velocity of the moving obstacle exceeds a critical value. For even higher velocities vortex dipoles whose vortices are multiply charged can appear. Periodic and controlled generation of vortex dipoles has also been studied in Ref.~\cite{Kwon2015a,Kwon2015b}, where vortex dipole generation is favored for a penetrable moving potential, again, for a moving velocity above some critical value $v_c$. However, in all of the above cases, the number density of vortices $\rho_v$ is relatively small (less than $10$). In contrast, in situations where $\rho_v$ is sufficiently large, vortex collisions are highly probable. 

\begin{figure}
\centering
\includegraphics[width=0.70\textwidth]{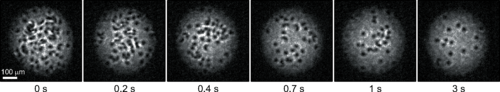}
\includegraphics[width=0.30\textwidth]{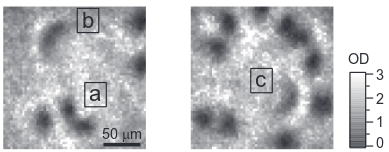} 
\caption{Vortex annihilation as a mechanism of turbulence relaxation. In a recent experiment \cite{Kwon2014} a large number of vortices has been created in an oblate condensate of $10^6$ sodium atoms by rapidly sweeping a transverse laser across the gas. Upper panel: Sequence of absorption images where the number of vortices is seen to decrease in time according to Eq.~\eqref{Eq:DecayKwon}. Lower panel: detail of vortex-antivortex annihilation that assists the relaxation of the turbulent gas. Source: paper of Kwon \emph{et al.} \cite{Kwon2014} 
\label{Fig:Kwon} }
\end{figure}

In vortex-antivortex collision events, two vortex cores can coalesce and annihilate [Fig.~\ref{Fig:Kwon}(a)], evolve into a solitary wave [Fig.~\ref{Fig:Kwon}(b)] and eventually disappear as atoms fill up the empty space [Fig.~\ref{Fig:Kwon}(c)]. These are the fundamental processes of 2DQT decay, that transfers incompressible to compressible energy in the form of sound. Below we discuss the consequences of this energy transfer to 2DQT and its partial suppression by the interesting process of \emph{vortex clustering}.


\subsubsection{Vortex clustering}
\begin{figure}[h!]
\begin{center}
\includegraphics[width=0.5\textwidth]{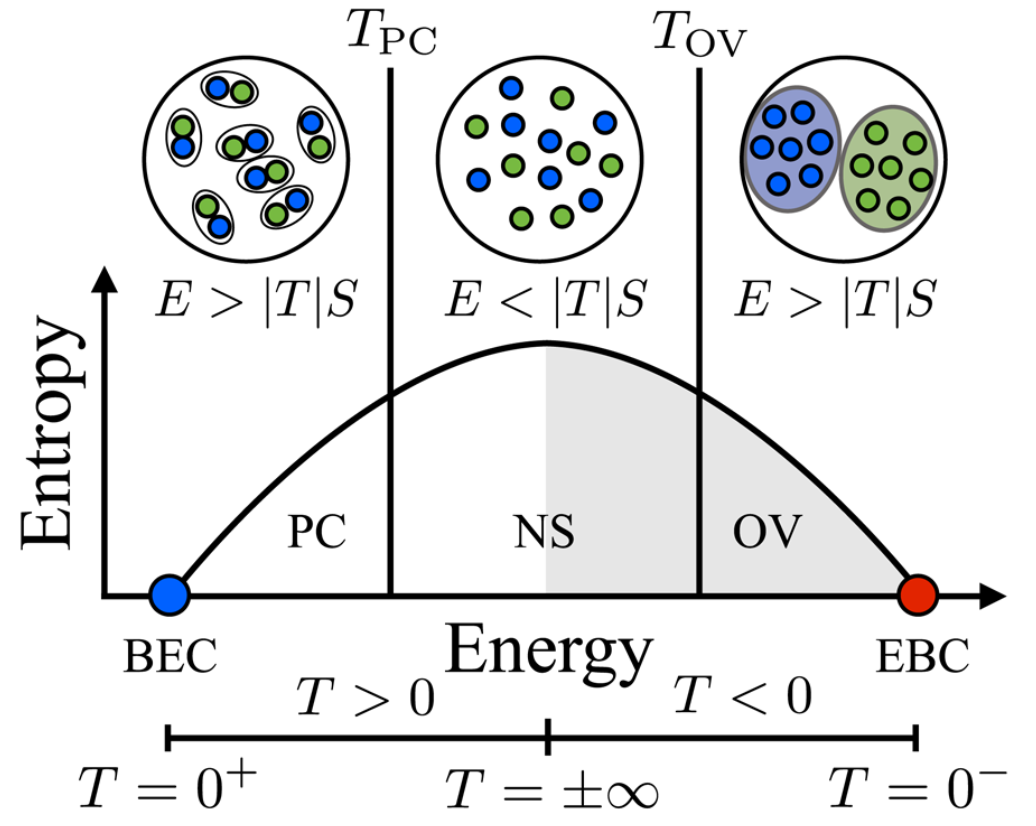} 
\caption{Negative-temperature and Onsager condensation. The scheme shows the behavior of entropy for a point vortex model as a function of temperature. As temperature decreases and reaches negative values, the system passes through a phase-transition where vortex clustering, a coherent Onsager vortex (OV) state, becomes favorable against vortex dipoles and unbound vortices. For $T=0^+$ a zero-entropy Bose-Einstein condensate (BEC) is present. Its negative temperature counterpart ($T=0^-$) is an Einstein-Bose condensate (EBC). At $T = \pm \infty$ entropy is a maximum and the vortex distribution is said to be in a entropy dominated normal state (NS). The vortex binding-unbinding phase transition separates this normal state from the pair collapse (PC) state at positive temperature. Source: paper of Simula \emph{et al.} \cite{Simula2014}.}
\label{negT}
\end{center}
\end{figure}

The clustering of like-signed vortices was predicted to appear in 2D forced turbulent systems by the `vortex gas' theory of Onsager and was applied to quantum fluids in \cite{Simula2014}. An interesting interpretation provided by such theory is that clustering represents a phase transition for a certain distribution of vortices associated with a negative-temperature state (see Fig.~\ref{negT}). An effective temperature $T$ can be defined in terms of the entropy $S$ generated by a certain vortex configuration with (incompressible kinetic) energy $E$, i.e. $S\equiv E/T$. The so-called Onsager condensation happens when it is energetically favorable for the system to organize itself in clusters, as the right part of the Fig.~\ref{negT} shows. As opposed to a configuration where  vortex dipoles form and annihilations are highly probable (left region of the schematic plot in Fig.~\ref{negT}), these states tend to suppress vortex annihilations. The effective temperature of such configuration is negative and is the key signature of an inverse energy cascade (IEC) in turbulent systems, since clustering process represents transfer of incompressible energy from small to larger scales.  

%

\subsection{Phonons and dissipation \label{SubSec:Phonons}}
Even though a superfluid is an inviscid, frictionless medium, effective drag forces can arise when an obstacle moves in its bulk  \cite{Frisch1992,Winiecki1999,Nore2000,Stiessberger2000}. In that case a Landau criterion applies, according to which the flow is dissipationless as long as the velocity of the obstacle of size $R$ stays below a critical value $u_s\sim\hbar/mR$ \cite{Stiessberger2000}. Beyond this value, collective excitations are favored, energy is being absorbed from the obstacle and so an effective dissipation is felt by the external drive. Experimentally investigated in \cite{Raman1999}, this simple idea sheds light in a fundamental mechanism, that plays an important role in turbulent energy transfer as well.

A phonon, i.e., a wave-like collective excitation, is one of the simplest form of excitations in superfluids and condensates, omnipresent in complex processes such as QT generation and decay. Their signature in the energy spectrum is seen in the compressible part of the kinetic energy. Incompressible energy $E_{\rm K}^i$ can be converted to compressible $E_{\rm K}^c$ in a trapped gas in the various turbulent processes, allowing thus a partial dissipation of energy from the incompressible part. Since the former is responsible for the vortex dynamics (the analogue to the vortex energy in CT) 
the diminishing of $E_{\rm K}^i$ in favor of $E_{\rm K}^c$ can be seen as turbulent energy dissipation. Physically, this means that energy is taken away from vortex modes towards phonons, surface modes and sound propagating throughout the cloud. However, energy conversion can be bidirectional and even oscillations between compressible and incompressible energies are possible \cite{Horng2009}.

Phonons in a BEC can be triggered by numerous processes and excitations \cite{Stamper-Kurn1999}. Focusing in a turbulent decaying state, one shall emphasize the role of vortices and their interaction with phonons. Oscillating vortices can emit phonons reducing thus their (incompressible) energy from the scale of the vortex-core size to larger system-wide scales \cite{Parker2003}. Similar phonon emissions can happen in processes of vortex reconnections or Kelvin waves propagating on a vortex \cite{Leadbeater2001}.
The energy loss due to phonon radiation was calculated in the above-mentioned work by solving the GP equation in homogeneous space. Interestingly, the authors found that the amount of the lost vortex line length is a function of the angle at which the vortex lines reconnect. 

This emergent effective dissipation mechanism -- present in conservative systems as well -- is not to be confused with the dissipation caused by the presence of thermal particles. In the latter case, one includes a phenomenological imaginary term $i\gamma$ in the Hamiltonian to include damping effects. The non norm-conserving dynamics that such a theory describes is considered to account for thermal effects and energy or particle loss of the superfluid towards the thermal cloud. 
\section{Experimental emergence and characterization \label{Sec:Experimental}}

Nowadays, BECs are being routinely produced by various groups around the world. The atomic samples used vary and include hydrogen, sodium, rubidium, lithium, potassium, cesium, calcium, strontium, chromium, dysprosium, erbium and more. The methods used to trap and cool down an atomic sample can also be very different, but they most commonly include the following stages \cite{Ketterle1999enrico}: 
\begin{enumerate}
\item capture and cool the hot sample in a magneto-optical trap (MOT),
\item transfer the atoms into a conservative trap potential (typically, magnetic or optical),
\item cool down the temperature using evaporative or sympathetic cooling techniques to reach the quantum degeneracy. 
\end{enumerate}
Typically, the trapping potentials are parabolic isotropic or anisotropic. However, uniform box-like potentials \cite{Gaunt2013} or other arbitrary geometries have been reported \cite{Henderson2009}.

Vortices in atomic BECs have been produced using a variety of methods: phase imprinting, stirring with a blue-detuned laser, rotating the trap potential and others (see \cite{Fetter2001,Fetter2009} and references therein). Recently, in the laboratory of S\~ao Carlos, random arrays of vortices were produced in a 3D BEC by a combined rotation and shaking of the magnetic trap potential, raising thus the possibility to produce configurations of tangled vortices which characterize a 3D quantum turbulent state \cite{Henn2009,Henn2009a,Caracanhas2015}. In Refs.~\cite{Neely2013,Kwon2014} vortices and turbulence were also generated in atomic gases in quasi-2D traps. The BECs produced there were confined in a highly oblate magnetic and optical trap \cite{Neely2010}. In specific, in Ref.~\cite{Neely2013} an additional blue-detuned Gaussian laser beam was directed axially through the trap, creating an annular barrier and a magnetic bias field was applied in cycles to move the center of the harmonic trap, but not the barrier. This motion induced a number of vortices in a disordered distribution, and the authors identified quantum turbulence states. The principal differences with the S\~ao Carlos experiment are the system dimensionality, the methods used to generate vortices and turbulence and, most importantly, the energy flow mechanisms of the system. In this section we mainly describe the experimental work of the S\~ao Carlos group and give a brief account of the principal accomplishments together with some supporting theoretical models.

Quantum turbulence was created in a $^{87}$Rb Bose condensed gas at $100\rm{nK}$, by means of a weak, off-axis, magnetic field gradient that perturbs the cloud and injects kinetic energy to it (see \cite{Henn2008} for a detailed presentation). This excitation nucleates vortices on the condensed-thermal interface of the cloud, which move towards the center of the cloud and set up the experimental conditions for the emergence of turbulence. Once the turbulent regime is reached, the condensate is released and let expand while free-falling. The atomic density profiles were acquired using resonant absorption imaging, after $15$~ms of time of flight (TOF) and calibrated images were used to determine the \textit{in situ} momentum distribution of the BEC. It was observed that, the density profiles of the perturbed gas drastically differ from the nonperturbed ones, offering, moreover, strong evidence of power laws in the measured momentum and energy distributions. Additional characteristics of the system, such as the condensate's finite number, size and temperature, may play a role on the energy injection mechanism and are discussed.


\subsection{Experimental set-up and time sequence \label{SubSec:ExpSetup}}
\begin{figure}[!ht]
\begin{center}
\includegraphics[width=0.90\linewidth]{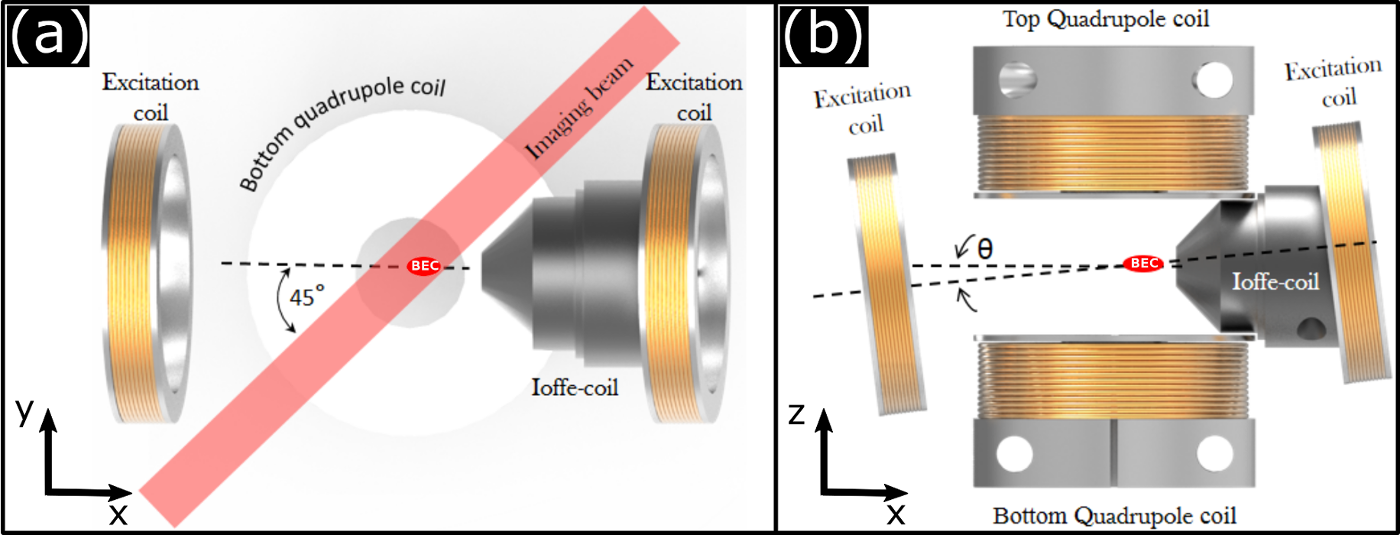}
\caption{Schematic draw of the experimental system. (a) Top and (b) side views of the BEC trapping region. The quadrupole and Ioffe coils correspond to the QUIC trap and the two excitation coils are showing the tilt between the trap and the excitation axis. In (a) the direction of the imaging beam is also shown.}
\label{Fig:BECsystem}
\end{center}
\end{figure}

To achieve Bose-Einstein condensation, an atomic cloud with $2\times 10^8$ atoms of  $^{87}$Rb at $250~\mu\text{K}$ was trapped in a magnetic quadrupole--Ioffe configuration (QUIC trap) \cite{Esslinger1998} characterized by a radial frequency $\omega_r = 2\pi\times210$Hz and an axial frequency $\omega_x = 2\pi\times 23$Hz. Linear ramps of RF-forced evaporative cooling during 20~s were performed to produce a BEC sample containing about $1\times10^5$ atoms with a small thermal fraction ($\sfrac{N_0}{N} \approx 0.9$). After reaching the BEC limit and with the cloud still held inside the trap, a temporally oscillatory magnetic field was superimposed on the magnetic trapping field. This extra field was produced by a fixed pair of anti-Helmholtz coils (called \emph{excitation coils}) whose center (defined by the zero field position) is displaced from the QUIC trap minimum and its axis is tilted in a angle $\theta \approx 5^{\circ}$ from the trap axis. This experimental system is shown in Fig.~\ref{Fig:BECsystem}.

A temporally oscillatory current of the form 
\begin{equation}
I(t) = I_0[1-\cos(2\pi f_{\text{{exc}}}t_{\text{exc}})],
\label{Eq:OscCurrent}
\end{equation}
was applied in the excitation coils, where $f_{\text{exc}}$ is the excitation frequency -- typically $200$~Hz -- and $t_{\text{exc}}$ is the time interval during which this current is on -- typically $5$ to $60$~ms. The maximum current $I_0$ is proportional to the excitation amplitude $A_{\text{exc}}$ of the additional field gradient, that varied from $0$ to $190$~mG/cm. 
The net result of this excitation on the trap potential is:
$i)$ an oscillatory displacement of the trap minimum, an $ii)$ oscillatory deformation of the trap frequencies and $iii)$ an oscillatory inclination of the trap axes \cite{Henn2009a}.

Finishing the oscillatory excitation, the atoms were left trapped for a hold time $t_{\text{hold}}$ before being released from the trap and the measurements were made by analyzing time-of-flight (TOF) absorption images. Further information can be found in Refs.~\cite{Henn2008,Henn2009,Henn2009a,Henn2010,Seman2011}.
Different dynamical regimes of the BEC were observed depending on the excitation time, the excitation amplitude, the hold time and the TOF, which are described in the next sections.
Thus, we were able to observe and distinguish four different such regimes of the excited system:
\begin{enumerate}
\item oscillation of the gas longitudinal axis (scissors mode)
\item nucleation of vortices in a regular configuration
\item dense vortex tangles and turbulence
\item overlapped vortices and granular structures.
\end{enumerate}
In what follows we examine and discuss each of the various structures as well as their interrelation.

\subsection{Regular vortex nucleation and proliferation}\label{SubSec:vortex}
The excitation frequency was the first investigated parameter. It was observed that excitations on the BEC were produced for frequencies closer to these of the trapping frequencies. For $f_{\text{exc}} = 200$~Hz some nontrivial structures in the cloud were observed and, for this reason, the frequency was fixed at this value. The experiments were performed by changing the amplitude and the excitation time while observing the effects of the perturbation to the cloud. The duration of the excitation was explored in the range $0 \le t_{\text{exc}} \le 55$~ms and the excitation amplitude in $0 \le A_{\text{exc}} \le 170$~mG/cm.

\begin{figure}[!ht]
\begin{center}
\includegraphics[width=0.95\textwidth]{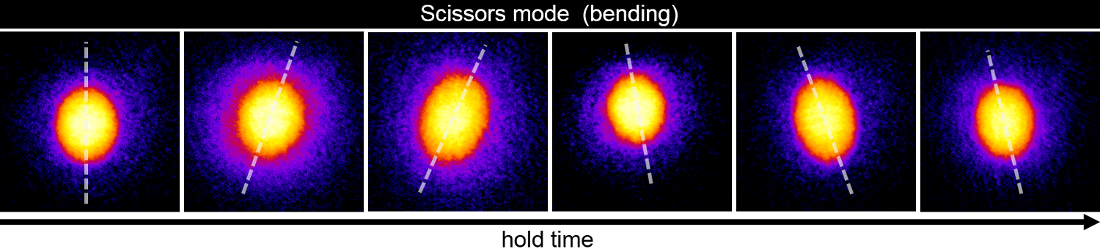}
\caption{Scissors mode. Images of an excited BEC for small excitation amplitudes ($A_{\text{exc}} < 40$~mG/cm) as a function of the hold time, at TOF = 21~ms. The dashed line indicates the incline of the major cloud axis. The hold time increases from left to right.}
\label{fig:scissorsmode}
\end{center}
\end{figure}	

Starting from lower excitation amplitudes, $A_{\text{exc}} < 40$~mG/cm, a bending of the major cloud axis was observed \cite{Henn2009a, Henn2010}, regardless of the excitation time. The observed excitation is the \emph{scissors mode} \cite{Marago2000, Modugno2003} and can be seen in Fig.~\ref{fig:scissorsmode} that shows an image sequence of the cloud bending as a function of the hold time, taken after 21~ms of time of flight. As the hold time increases, the major cloud axis (dashed line) oscillates in a pendulum-like motion.  This observation shows that the external perturbation is able to transfer angular momentum to the atomic cloud.

\begin{figure}[!ht]
\begin{center}
\includegraphics[width=0.95\linewidth]{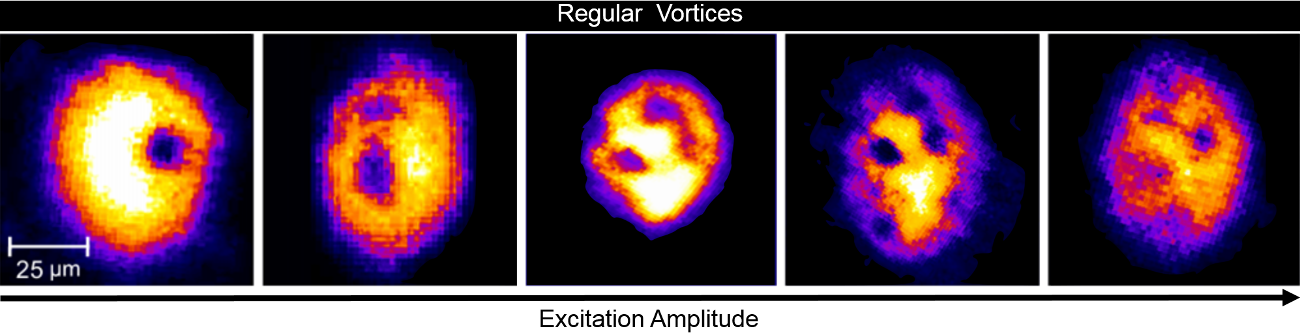}
\caption{Regular vortices in the BEC. Absorption images of an atomic BEC excited with amplitude in the range $40 < A_{\text{exc}}<170$~mG/cm and $20$~ms of excitation time. It shows an atomic BEC with (a) one, (b) two, (c) three and (d)--(e) many nucleated vortices as the excitation amplitude increases. All images were taken after $15$~ms of TOF. Source: Master's thesis of Bagnato \cite{BagnatoThesis}.}
\label{fig:regular_vortices}
\end{center}
\end{figure}

Increasing the excitation amplitude more angular momentum was injected into the cloud. For amplitudes higher than $40$~mG/cm quantized vortices were observed in the cloud density profile \cite{Henn2009a, Henn2010, Seman2010}. At this point, the dependence of the vortex nucleation on the excitation amplitude and time were investigated. By increasing either of these parameters, the average number of vortices grows up. Figure~\ref{fig:regular_vortices} presents an image sequence of the system with different number of vortices. All images were taken after $15$~ms of free expansion.

\begin{figure}[htb]
\begin{center}
\includegraphics[width=0.8\linewidth]{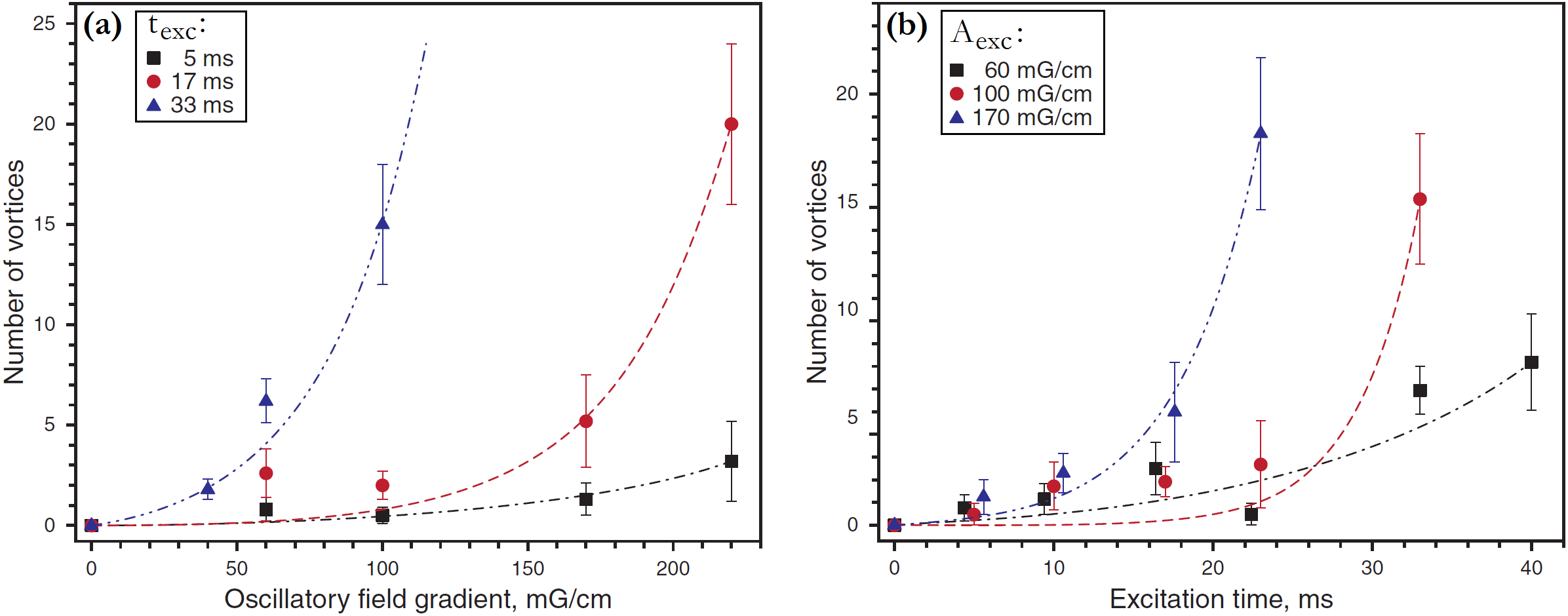}
\caption{Number of vortices as the excitation magnitude is increased. (a) Average number of observed vortices as a function of the excitation amplitude ($A_{\text{exc}}$) for the three fixed excitation times ($t_{\text{exc}}$) shown in the legend of the graph. (b) Average number of observed vortices as a function of  $t_{\text{exc}}$ for three fixed values of $A_{\text{exc}}$ specified in the legend. Source: paper of Seman \emph{et al.} ~\cite{Seman2011a}.}
\label{fig:Nvortices}
\end{center}
\end{figure}

Fig.~\ref{fig:Nvortices}(a) shows the number of vortices as the excitation amplitude increases for three different fixed excitation times. Equivalently, Fig.~\ref{fig:Nvortices}(b) shows the number of vortices when the excitation time was varied for three different amplitudes \cite{Seman2011}. These results show that by increasing the magnitude of the excitation (i.e. increasing the angular momentum transferred to the cloud) a greater number of vortices is produced in the gas.

\begin{figure}[htb]
\begin{center}
\includegraphics[width=0.55\textwidth]{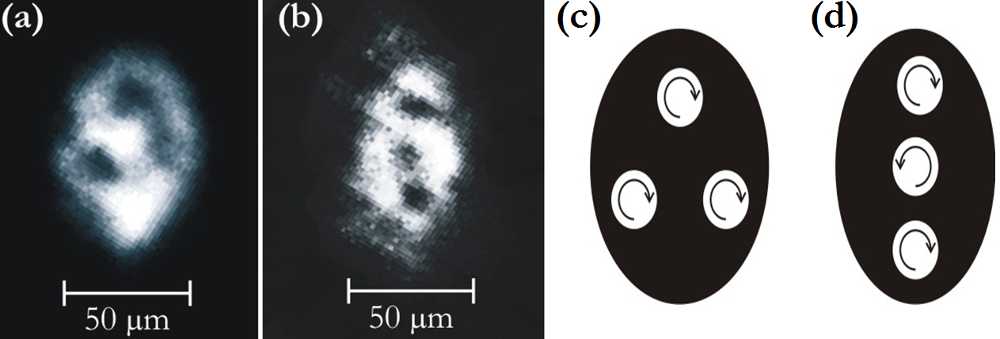}
\caption{Three-vortex configuration produced in the BEC cloud. Absorption images showing configurations of vortices forming (a) an equilateral triangle, or (b) a linear array. Images were taken after 15 ms of free expansion. Panels (c) and (d) show sketches of the BEC with the three vortices of figures (a) and (b), respectively; the arrows show the vortex circulation direction. Source: paper of Seman \emph{et al.}~\cite{Seman2010}.}
\label{fig:3vortex}
\end{center}
\end{figure}

As we can see in Fig.~\ref{fig:regular_vortices}, the vortex distribution in the BEC does not have a regular pattern, like the Abrikosov vortex lattice \cite{Madison2000, Abo-Shaeer2001}. Instead, they are randomly distributed. This is a consequence of formation of vortices and antivortices (vortices with opposite circulation sign), confirmed by observations of three-vortex configurations \cite{Seman2010}. In a three-vortex configuration, two types of vortex distributions have been identified: in the first the vortices are found close to the edges of an equilateral triangle while in the second, the vortices are colinearly arranged. These configurations are shown in Figs.~\ref{fig:3vortex}(a) and (b), respectively.

\subsection{Mechanism of vortex formation} \label{Sec:Counterflow}
Numerous mechanisms are responsible for nucleating vortices in a BEC, with many of them producing parallel vortices with the same topological charge and orientation. As discussed in the previous section, the oscillatory excitation technique is able to nucleate vortices in various orientations and also parallel ones of opposite charge. In superfluid helium, different circulation vortices were created through a counterflow of the normal and superfluid components \cite{Skrbek2011}. In a similar way, a counterflow mechanism that nucleates vortices was observed as a relative motion between the thermal and condensate fractions \cite{Tavares2013}. This motion corresponds to an out-of-phase oscillation mode, which was investigated in Refs.~\cite{Stamper-Kurn1998,Meppelink2009}.

\begin{figure}[htb]
\begin{center}
\includegraphics[width=0.8\linewidth]{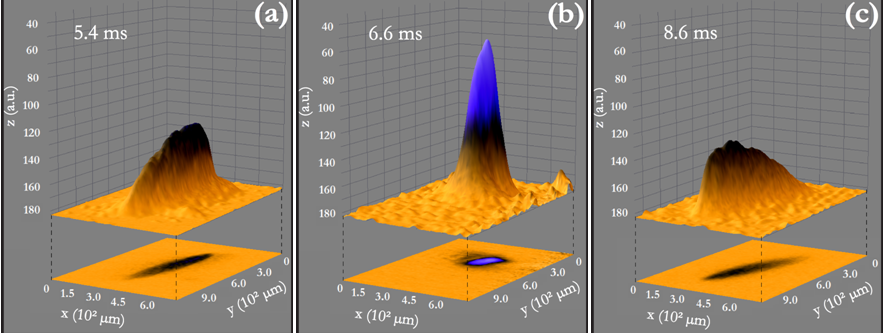}
\caption{Column density profile showing the relative motion between the condensate and the thermal cloud for three different hold times ($t_{\text{hold}}$). The asymmetry in these density profiles as $t_\text{{hold}}$ increases reveals the relative motion between the two components. Source: paper of Tavares \emph{et al.}~\cite{Tavares2013}.}
\label{fig:counterflow3d}
\end{center}
\end{figure}

To analyze this motion, an experiment was performed with a $60\%$ condensate fraction, and introduced the above oscillatory excitation [Eq.~\eqref{Eq:OscCurrent}] at a fixed amplitude ($A_{\text{exc}} = 40$~mG/cm), frequency ($f_{\text{exc}} = 170$~Hz) and excitation time ($t_{\text{exc}} = 100$~ms). The counterflow was analyzed by scanning $t_{\text{hold}}$ and the excited condensate was observed after $23$~ms of time of flight. Figure~\ref{fig:counterflow3d} presents density profiles of the cloud (thermal and BEC cloud) for three different hold times $t_{\text{hold}}$. A closer look to this figure reveals the asymmetry in the density profile; this suggests a displacement between the center-of-mass of the thermal cloud ($\text{CM}_{\text{th}}$) and that of the condensed gas ($\text{CM}_\text{{BEC}}$). Moreover, the displacement changes with $t_\text{{hold}}$, which indicates a relative motion between the two components.

\begin{figure}[htb]
\begin{center}
\includegraphics[width=0.4\linewidth]{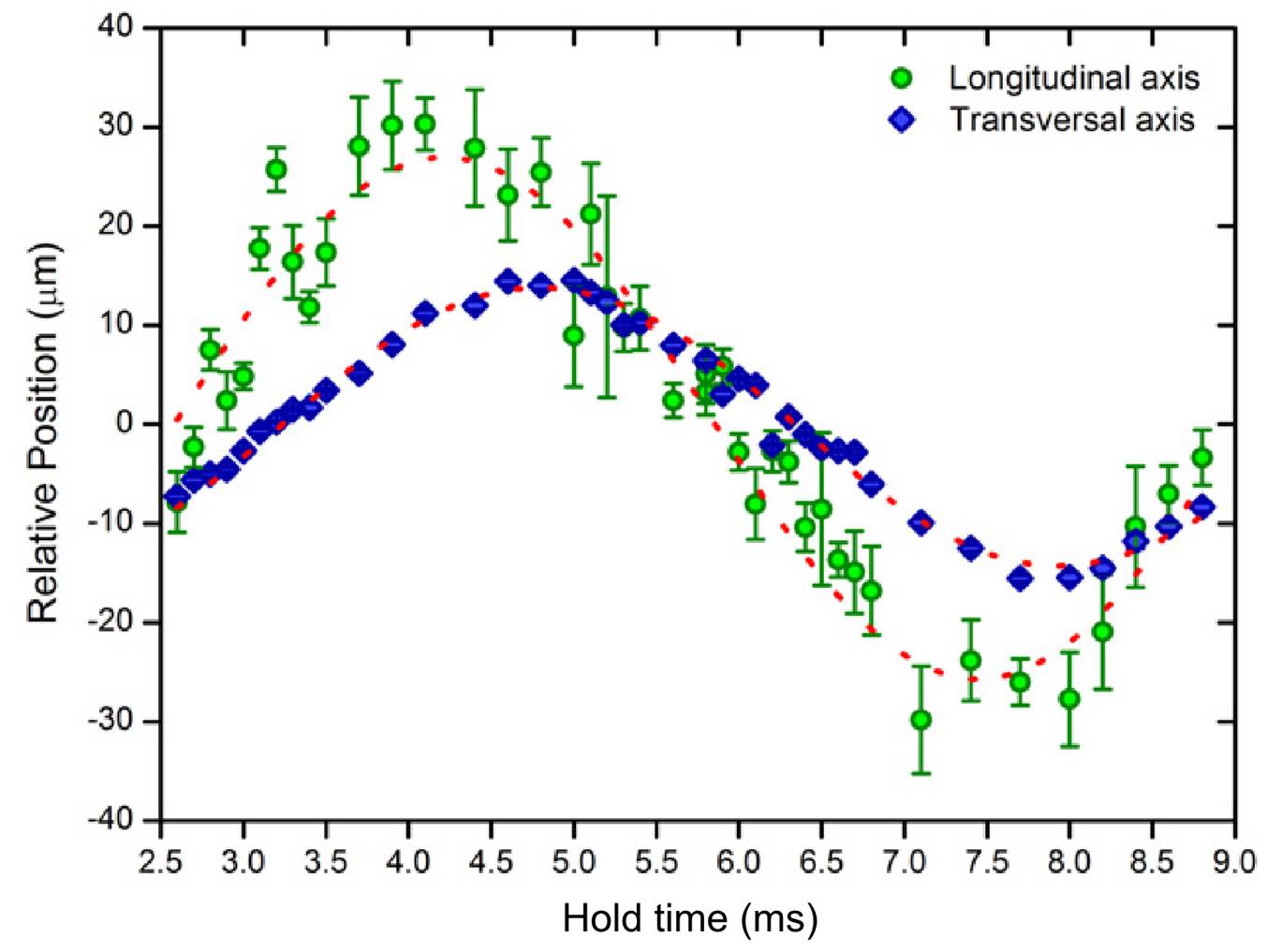}
\caption{Relative position between the condensate and the thermal centers of mass, in the longitudinal and transverse axis, as a function of $t_\text{{hold}}$. The red dotted lines are sinusoidal fittings. Source: paper of Tavares \emph{et al.}~\cite{Tavares2013}.}
\label{fig:counterflow_thold}
\end{center}
\end{figure}

This relative motion was measured by fitting a bimodal profile through the longitudinal ($y$) and transverse ($x$) axes of the absorption images, paying attention to the fact that the center of the Gaussian profile is different from the center of the Thomas-Fermi profile. The difference between the two provides the relative position between each cloud. The relative position as a function of $t_{\text{hold}}$ is plotted in Fig.~\ref{fig:counterflow_thold}, in which sinusoidal fitting indicates an oscillating out-of-phase motion in both axis. The smallest displacement was observed for holding times corresponding to the turning points in the dipolar motion. At these points the counterflow velocity is maximum and has a value of $v_{ns}= 3$~cm/s.
%
%
An important quantity that can be calculated with this velocity is the superfluid Reynold number ~\cite{Finne2003}, 
\begin{equation}
Re_s = \frac{m\xi v}{2\pi \hbar},
\end{equation}
where $\xi$ is the healing length, $m$ is the atomic mass, and $v$ is the flow velocity. By considering $v \equiv v_{ns}$, it was found -- for the above experiment -- that $Re_s$ ranges from $0$ to $1$, This value is, according to~\cite{Finne2003} not sufficiently large for the onset of a turbulent flow.

\subsection{Turbulence}
Turbulence in superfluids is characterized by tangles of vortices with recurring reconnection events that lead to a cascade of energy flow from large to small scales. At $T=0$ the trapped gas contains a relatively small number of atoms (few tens to hundred thousands), much smaller than that in liquid helium. Hence, trapped BECs cannot sustain the same number of quantum vortices as helium superfluids. This brings to surface the issue of length scales and the appearance (or not) of a Kolmogorov-type scaling (see also discussion in Secs~\ref{SubSec:QT}, \ref{Sec:Scales} and \ref{SubSec:BriefQT}). However, experimental signatures of turbulence are evident.

\begin{figure}[htb]
\begin{center}
\includegraphics[width=0.75\linewidth]{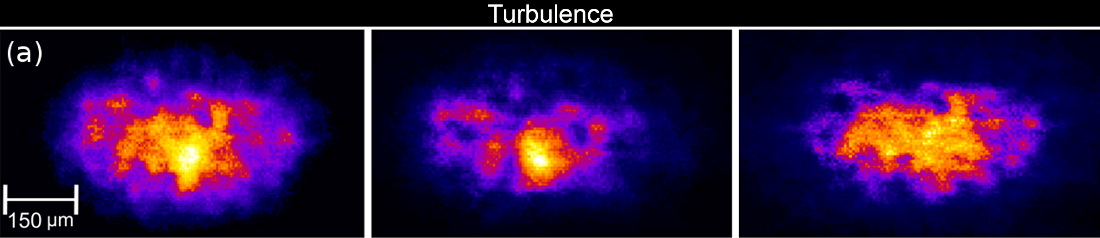}
\includegraphics[width=0.2\textwidth,trim={9cm 0 0 0},clip]{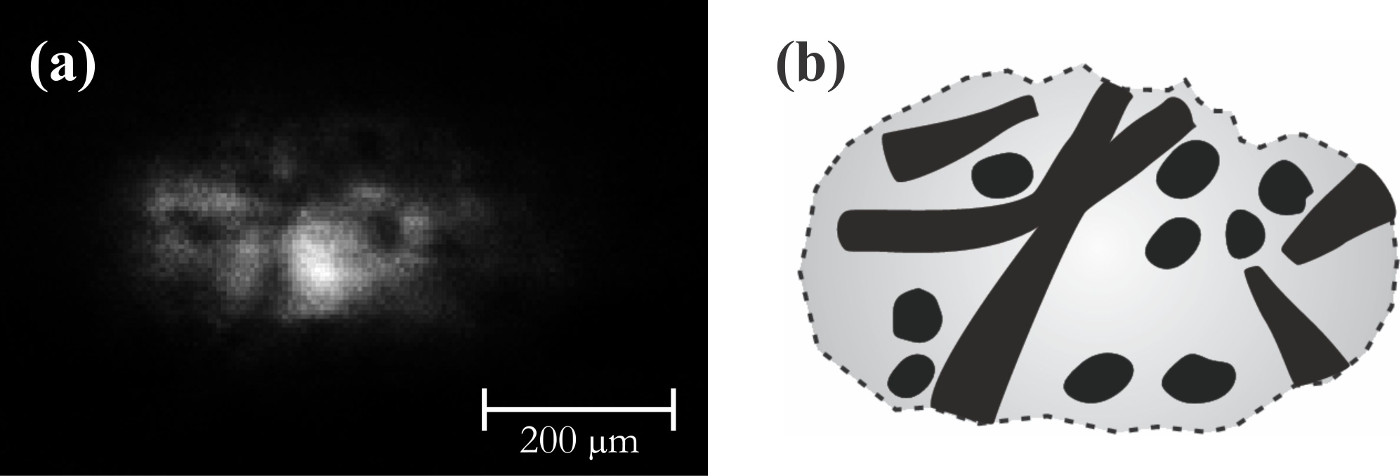}
\caption{Turbulent BEC clouds. (a) Absorption images at $15$ms of TOF showing three different vortex tangle configurations, which characterize the quantum turbulence regime. (b) Sketch of the vortex distribution in the central panel of (a). Source: Master's thesis of G. Bagnato \cite{BagnatoThesis} and paper of Henn \emph{et al.}~\cite{Henn2009}.}
\label{fig:QT_clouds}
\end{center}
\end{figure}

A dramatic change in the cloud behavior was observed as the excitation parameters were further increased. The absorption images showed that the vortices are found along several directions and the vortex lines seem to have a curved profile \cite{Henn2009,Henn2010,Seman2011a,Seman2011}. This irregular configurations suggest the emergence of a configuration of the vortices in tangles, which hints the emergence of quantum turbulence in the sample. Three typical images of the turbulent gas are shown in Fig.~\ref{fig:QT_clouds}(a) and a sketch of the vortex distribution in Fig.~\ref{fig:QT_clouds}(b).

It is worth noticing that vortices in Fig.~\ref{fig:QT_clouds} are distributed in a nonregular way in all directions in the cloud; situation that characterizes a turbulent dynamical regime of the quantum gas. Important to achieve such a configuration was the method followed: the cloud underwent translational and rotational motions that occured differently in the three distinct planes of the cloud.  This approach is closely related to the proposal of M. Kobayashi and M. Tsubota~\cite{Kobayashi2007} where rotations along distinct axes are combined to generate quantum turbulence in a trapped BEC. Another indicative for the emergence of turbulence is the observed counterflow between the thermal fluid and the superfluid \cite{Tavares2013}, as already discussed in Sec.~\ref{Sec:Counterflow}. In liquid helium this causes a viscosity between the two fluids (normal and superfluid) and is the mechanism responsible for the vortex and turbulence generation \cite{Ladner1976,Skrbek2011}.

A novel feature of the turbulent cloud is the anomalous expansion after the trap has been switched off, as later discussed in detail (see Sec.~\ref{SubSec:self-similar}). In contrast to the thermal fluid and the ordinary quantum fluid, the BEC in a turbulent regime expands by keeping its aspect ratio fixed during the whole expansion time. That is, it expands in a self-similar fashion. Figure~\ref{fig:AspRatio} shows the three different clouds (thermal fluid, superfluid and turbulent) expanding and their aspect ratios, during a time of flight. Section \ref{SubSec:self-similar} gives a theoretical account of this property.

Concerning turbulent phenomena, in classical fluids the injected energy is transferred from large scales to smaller one and this process is quantified by the Kolmogorov energy spectrum. Such a power law was also observed in superfluid $^4$He \cite{Maurer1998}. In the case of atomic BECs, an energy spectrum with power-law behavior (like Kolmogov law) was demonstrated theoretically only \cite{Kobayashi2005b,Kobayashi2007}. The observation of the anomalous expansion in turbulent BECs has opened the possibility to look for power laws in the experiments too. Indeed, such results were recently obtained: it appears that the transition to turbulence is followed by power-law-like distributions of the momenta of the gas inside the trap. The details are presented in Sec.~\ref{SubSec:MomDist}.

\subsection{Granular phase \label{SubSec:Granular}}
\begin{figure}[!htb]
\begin{center}
\includegraphics[width=0.55\linewidth]{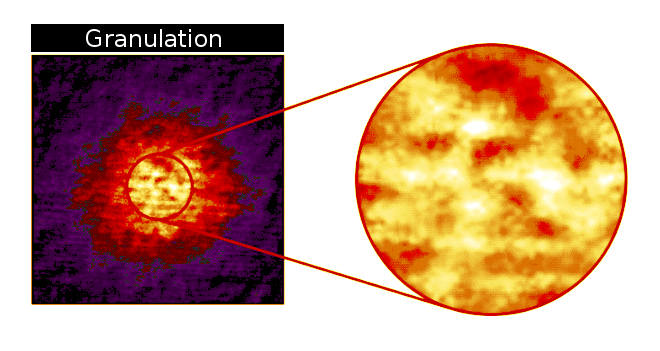} 
\caption{BEC in the granular phase. Absorption image -- and detail -- revealing a density profile showing grains of BEC surrounded by a large thermal cloud. Source: book chapter of \cite{Bagnato2013}.}
\label{fig:granulation}
\end{center}
\end{figure}

For continuing energy pumping into the cloud more vortices want to be created. At some point however the system cannot host more vortices as a maximum number of these is reached. The vortices start then to overlap and collide, destroying thus the regime of vortex turbulence. The system passes to another dynamical regime, that of a granular phase. For higher excitation amplitudes (between 100 and 170 mG/cm) and excitation times greater than 45 ms, the granular phase was observed in the experiments. In this phase a non-equilibrium state appears at which the cloud is composed by localized grains surrounded by a large thermal cloud \cite{Yukalov2009b,Yukalov2009a}. The density of the grains is about a hundred times larger than the surrounding density \cite{Yukalov2015b}. Figure~\ref{fig:granulation} shows a typical absorption image were one can observe the granulation in the core and an almost homogeneous thermal cloud at the borders. 

Among the particular aspects of the granular phase we can distinguish the following: $(i)$ it appears as a heterogeneous phase, with grains of high density surrounded by a rarefied gas -- both components however are formed by the same atoms; $(ii)$ grains are randomly distributed in space; $(iii)$ the grains have well defined phases despite the lack of vortices (zero circulation), which means that the system is overall coherent but \emph{not} locally, since the phases are not the same among the grains. In the work of Refs.~\cite{Yukalov2014,Yukalov2015b} the authors applied the same time-dependent modulation of the trap potential as in the experiment and solved the GP equation for these choices. It was also shown that the grains are metastable structures, since their lifetime is much higher than the local thermodynamic relaxation time.

Interestingly, the granular regime was also experimentally achieved by modulating the atomic scattering length in the experiment performed in the laboratory of R. Hulet using $^{7}$Li atoms \cite{PollackPRA2010}. There -- and also more recent unpublished results -- the gas was confined in a quasi-1D trap and -- by utilizing Feshbach resonances -- the scattering length oscillated in time. For specific such oscillation frequencies the formation of little `islands' where the density is concentrated was observed, bearing thus similarities with the granular phase presented here.

\subsection{Diagram of  excited structures \label{SubSec:Diagram}}
\begin{figure}[htb]
\begin{center}
\includegraphics[width=0.45\linewidth]{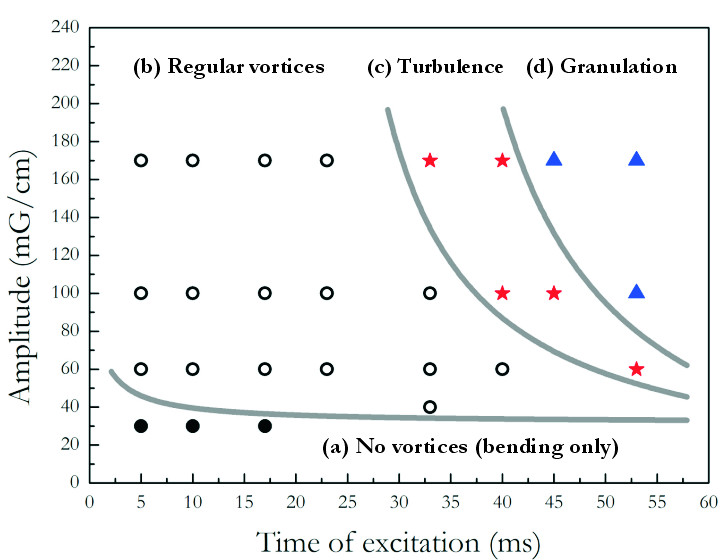}
\includegraphics[width=0.47\linewidth]{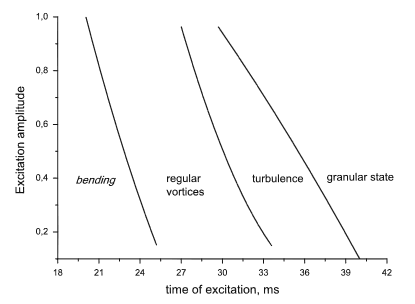}
\caption{Diagram of the various excitation structures. In specific, the amplitude $A_{\text{exc}}$ versus its duration $t_{\text{exc}}$ is shown, presenting regions where four different effects were observed. The left diagram was built from the experimental results and the right one from numerical simulations. Border lines separating different regimes are just guide to the eyes. Source: papers of Seman \emph{et al.}~\cite{Seman2011a} and Yukalov \emph{et al.}~\cite{Yukalov2015b}.}
\label{fig:diagram}
\end{center}
\end{figure}

The oscillatory excitation experiment revealed the possibility to produce four different dynamical regimes as the excitation parameters (amplitude and time) increases. This observations is summarized in a diagram of amplitude against time of excitation, presented in Fig.~\ref{fig:diagram}(a). There, one distinguishes four domains, which corresponds to the different dynamical regimes of the BEC. The purpose of this diagram is to present the direction to which the excitation parameters must be varied to achieve a specific regime \cite{Seman2011}. 

This diagram reveals a very important property of atomic condensates; that of its finite size. Since the total condensate size is only one to three orders of magnitude larger than the healing length, the number of vortices produced in the BEC is limited. This means there is a finite number of vortices that can be created in the BEC and if any further energy is pumped then vortex interactions and reconnections will be promoted \cite{TsubotaJLTP2000}, which can change drastically the superfluid dynamics. In fact, Fig.~\ref{fig:Nvortices} reveals that $\sim 20$ vortices are nucleated before the observed QT in our system.

In the numerical simulations of Ref.~\cite{Yukalov2014}, by solving the nonlinear Schr\"odinger equation, the experimental diagram was reproduced by means of a trap oscillation. Increasing the amount of the injected -- by the oscillating trap -- energy the authors observed the following stages: $(i)$ a slightly perturbed regular BEC in a bending mode; $(ii)$ regular vortices with the presence of both vortices and antivortices; $(iii)$ turbulence formed by random tangle vortices; and $(iv)$ a granular state with a random distribution of grains of dense BEC droplets.
The existence of regular vortices, nonstrongly interacting, implies an energy mainly distributed to the interaction and the velocity field associated with the vortex lines. As soon as reconnections happen and rings start to be formed, the energy begins to flow so that it can accommodate the geometry of the new vortices. 
As energy is accumulated and the vortex density increases, large variations are present, not only of the density but also of the phase, resulting in the final granular state.

\subsubsection{Distinction of the regular vortex and turbulent regimes}
\label{SubSubSec:Distinction}
We can now describe the crossover region from the regime of regular vortices to that of turbulence. A simple model establishes a critical, for the transition, total number of vortices, which is limited by the size of the cloud. Figure~\ref{fig:Nvortices} shows the number of vortices as a function of the amplitude and duration of the oscillatory excitation. The transition from regular vortices to turbulence appears when more energy is pumped to the cloud that makes the latter densely filled with vortices. In this situation, the additional energy pumped in the system cannot nucleate more vortices but instead it induces their motion, resulting in an tangled distribution.

\begin{figure}[htb]
\begin{center}
\includegraphics[width=0.45\linewidth]{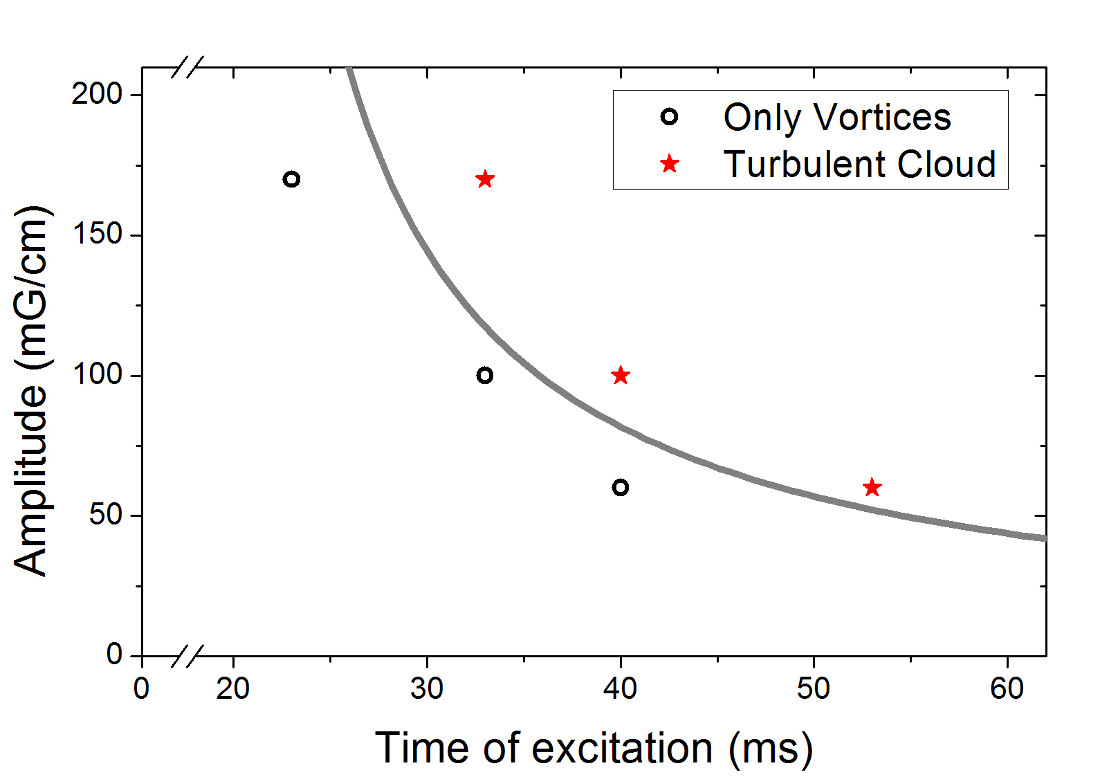}
\caption{Diagram of the excitation amplitude $A_{\text{exc}}$ versus its duration $t_{\text{exc}}$ presenting the transition from regular vortices to a turbulent regime. In contrast to the guide lines drawn in Fig.~\ref{fig:diagram}, the transition line here is a fitting of Eq.~\eqref{eq:Acrit} through critical points considered to be halfway points with the same amplitude. The circles and stars are experimental data. Source: paper of Shiozaki \emph{et al.}~\cite{ShiozakiLPL2011}.}
\label{fig:diagram_finite}
\end{center}
\end{figure}

The transition from the region with regular vortices to that of turbulence is shown in Fig.~\ref{fig:diagram_finite}. In order to distinguish between the two regimes two criteria were adopted. First, the appearance (or not) of regular vortices in the absorption images of the cloud density after TOF, as already discussed. Second, the behavior of the expansion of the cloud in TOF.  

The simple model proposed to understand the transition between the two regimes is based on energy-balance arguments \cite{ShiozakiLPL2011}. We infer that there should exist a minimum energy necessary to be pumped into the superfluid in order to produce a vortex. According to \cite{Pethick2008} the energy needed is:
\begin{equation}\label{eq:vortex energy}
E_{\text{vort}} \approx \frac{{\hbar}^2\pi}{m\ell_{0}^2}ln\frac{\ell_{0}}{\xi},
\end{equation}
where $\xi$ is the healing length, $\ell_{0}$ is the vortex line length and $m$ is the atomic mass. In this model it is assumed that $l_{0}$ is approximately equal to the 	harmonic oscillator length, $a_{\text{HO}}$,
\begin{equation}\label{eq:harm-oscill-length}
\ell_{0} \approx a_{\text{HO}} = \sqrt{\frac{\hbar}{m(\omega_r^2 \omega_z)^{1/3}}} ~.
\end{equation}

Another assumption is that the oscillatory excitation pumps energy into the sample at a rate $R_{\text{pump}}$, which depends linearly on the excitation amplitude $A_{\text{exc}}$, and after $t_{0}$ a fraction $\eta$ of the pumped energy is converted into vortices. The total energy needed for the formation of vortices becomes $E_{\text{pump}}=\eta R_{\text{pump}} (t-t_0)$, where $t$ is the elapsed excitation time. If after a pumping time $t$ a number of vortices $N_{\text{vort}}$ is formed, then the energy balance implies that $\eta R_{\text{pump}} (t-t_0) = N_{\text{vort}} E_{\text{vort}}$ or
\begin{equation}
N_{\text{vort}} = \frac{\eta R_{\text{pump}}}{E_{\text{vort}}} (t-t_0),
\label{Eq:Nvort}
\end{equation}
for the number of vortices.

From the above approximative expression for the number of vortices observed as a function of the excitation time we see that, after the first vortex has been nucleated, the number of vortices should grow linearly with the amplitude and time of excitation. This is not the exact behavior of some points in Fig.~\ref{fig:diagram_finite}. However, the transition between regular vortices and turbulence can occur when the cloud is heavily populated by vortices and this is captured by the model. We can hence define a critical number of vortices as
\begin{equation}\label{eq:number-vort-crit}
N_{\text{vort}}^{\text{(crit)}} \approx \frac{\ell_{0}}{\xi}.
\end{equation}
As the healing length characterizes the size of the vortex core, this expression reflects the maximum number of vortices (or vortex number density) inside the cloud of size $a_{\text{HO}}$ without the vortices overlapping.  In this way, this condition represents the minimum number of vortices needed for quantum turbulence.

Finally, the energy pump rate $R_{\text{pump}}$ of Eq.~\eqref{Eq:Nvort} is proportional to the excitation amplitude $A_{\text{exc}}$. Therefore, the relation
\begin{equation}\label{eq:Acrit}
A_c(t) = \frac{C}{t-t_0}
\end{equation}
that gives the excitation amplitude as a function of time at which the energy is pumped, defines a critical line in the diagram of structures of Fig.~\ref{fig:diagram_finite}. By fitting Eq.~\eqref{eq:Acrit}, it was possible to determine the experimental transition line. The fitting parameters were found to be $C \approx 1.6$~G${\cdot}$ms/cm and $t_0 \approx 17$~ms. For the discussed experiment $\xi \approx 0.06~\mu$m and $a_{\text{HO}} \approx 1.08~\mu$m, which results in a critical number of vortices $N_{\text{vort}}^{\text{(crit)}}\approx 20$ for the onset of quantum turbulence. As seen in Fig.~\ref{fig:Nvortices}, parameters found for $t_0$ and $N_{\text{vort}}^{\text{(crit)}}$ are in good agreement with the experimental results, indicating that this model provides a sufficiently good estimation to the quantum turbulence onset in trapped gases.

\subsubsection{Emergence of QT as an inverse Kibble-Zurek mechanism}
According to the Kibble-Zurek mechanism (KZM), a system that is being quenched or driven past a phase transition at a finite rate, i.e. nonadiabatically, develops domains with distinct long-range parameters. These long-range parameters, during the phase transition, can grow and eventually join, forming thus a larger domain that encompasses topological defects. These topological defects, left-overs of such a nonadiabatic process, are essential ingredients of the KZM. Indeed, KZM predicts the number of the defects as a function of the quench rate. Over the past decades KZM has been successfully applied to numerous classical and quantum systems \cite{Zurek2005}, that vary from cosmological to condensed matter and atomic systems. In such processes, the system ends up in a symmetry-broken state with defects, resulting from the out-of-equilibrium finite-timescale process (for a review see \cite{delCampo2014}.

\begin{figure}
\centering
\includegraphics[width=0.5\textwidth]{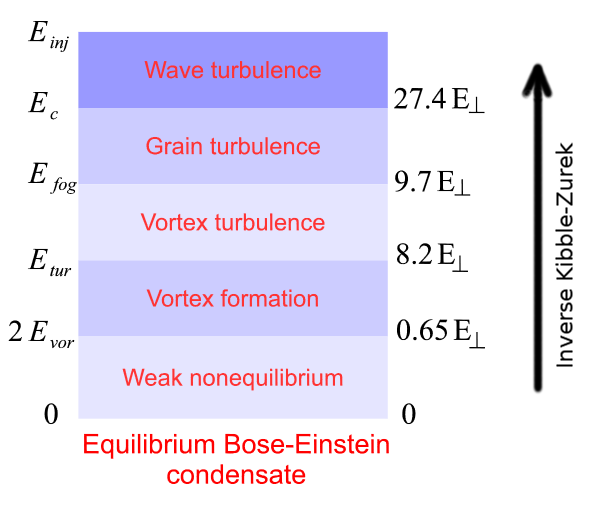}
\caption{Schematic representation of the inverse Kibble-Zurek scenario in a Bose-Einstein condensate. The gas is being excited by absorbing some amount of energy (multiples of the transverse trap energy $E_\perp\equiv \hbar\omega$ in the plot) and passes through different phases. The system begins at an equilibrium state of broken symmetry and is driven to a granular and strong (wave) turbulent phase, where domains of distinct phases coexist. Source: paper of Yukalov \emph{et al.} \cite{Yukalov2015b}.}
\label{Fig:IKZs}
\end{figure}

An interesting qualitative similarity of the KZM to the process that brings a BEC from its ground state to the turbulent or granulated state has been suggested in Ref.~\cite{Yukalov2015}. Specifically, the authors suggest that the sequence of excitations, described above, that brings the gas to the turbulent state constitute an \emph{inverse Kibble-Zurek scenario}. The granular phase, at which little islands of superfluid density form, is thus seen as a collection of distinct regions, reminiscent of the domains of the KZM. An inverse Kibble-Zurek scenario includes the `destruction' of the superfluidity and the transition, from a symmetry-broken state (BEC) to that of wave turbulence, with intermediate stages being vortex states, vortex (hydrodynamic) turbulence and a granular phase (see Fig.~\ref{Fig:IKZs}).

\subsection{Self-similar expansion}\label{SubSec:self-similar}
\begin{figure}[htb]
\begin{center}
\includegraphics[width=0.92\textwidth]{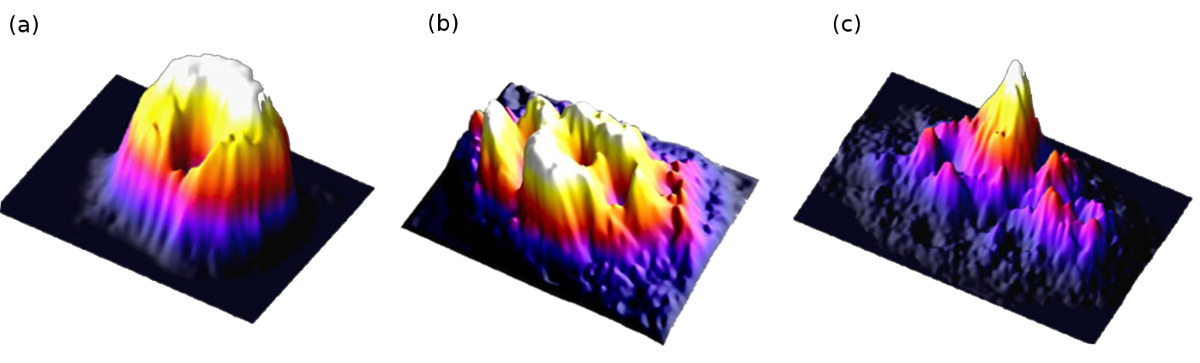}
\caption{Measured density profiles: (a-b) non--turbulent clouds with well--defined vortices, and (c) turbulent cloud showing partial absorption of a configuration of tangled vortices. Source of (a) and (c): paper by Caracanhas \emph{et al.} \cite{Caracanhas2012}. Source of (b): paper by Shiozaki \emph{et al.} \cite{ShiozakiLPL2011}.
\label{fig:clouds3d}}
\end{center}
\end{figure}

As mentioned, a sharp distinction of a turbulent sample from a gas with regular vortices is in the observation of their density profiles. Figure \ref{fig:clouds3d} shows the density profiles for both cases, in a $3D$ perspective. In a sample with regular vortices, the low light absorption of the vortex core appears as a node inside the bulk of the density, whereas in a turbulent cloud the presence of a vortex-tangle configuration no longer preserves this clear (node-bulk) distinction. The turbulent condensate has randomly increasing and decreasing densities along the radial direction due to the out-of-equilibrium irregular vortex dynamics that the (high) excitation energy induced.

Besides the evidence of vortex tangles in the density profile, seen in the absorption images, dynamic measurements of the evolution of a freely expanding condensate also indicate significant dif{}ferences, among the dif{}ferent phases. It is well known that a thermal cloud has an isotropic expansion: regardless of the initial trap potential, after some expansion time the resulting thermal cloud will be spherical. In contrast, a BEC expands anisotropically: the direction that is initially tightly confined in the trap has a faster expansion and the final state has an inverted profile. Heuristically, this is explained from the fact that the \emph{in situ} quantum state has along the direction of the tightest confinement, say $\omega_x$, a wider momentum uncertainty (hence, distribution) $\Delta p_x$ than that of the direction of the loose confinement $\omega_y$: $\Delta{}p_x\sim\hbar/\Delta x > \Delta{}p_y\sim\hbar/\Delta y$, as long as $\Delta x < \Delta y$. The presence of interactions among the atoms makes this effect more pronounce: the energy density $\int|\psi(\mathbf r)|^4dy$ along $x$ pushes the atoms further and faster apart, than the density along $y$.

\begin{figure}[htb]
\begin{center}
\includegraphics[width=0.90\textwidth]{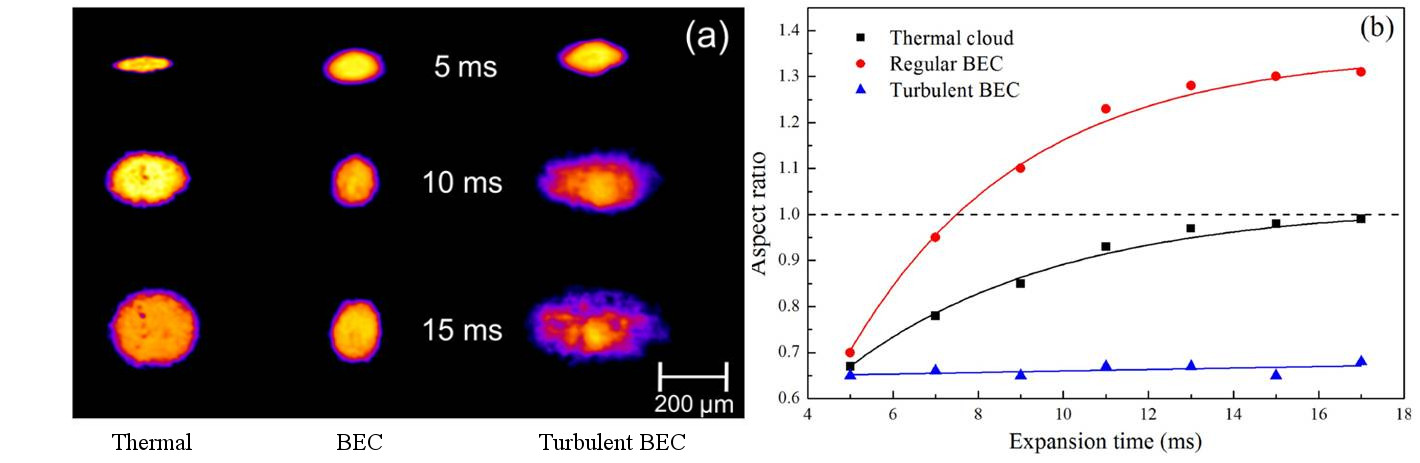}
\caption{TOF images: (a) Absorption images of a thermal cloud, a regular BEC and a turbulent BEC for three different expansion times. (b) Aspect ratio as a function of the expansion time for the different clouds. Source: thesis of Seman \cite{SemanThesis}, papers of Henn \emph{et al.} \cite{Henn2009} and Caracanhas \emph{et al.} \cite{Caracanhas2012}. }
\label{fig:AspRatio}
\end{center}
\end{figure}

One way to quantify the cloud expansion is by measuring its aspect ratio (ratio of the extension of the gas along the initially tight direction over the extension along the initially loose direction) in time. As the time of flight increases, the aspect ratio for a thermal cloud converges to one. For a BEC, depending on the initial symmetry of the confinement of the cloud, an aspect ratio initially below one, exceeds unity after a certain TOF.  In sharp distinction, we observed that for a turbulent BEC the aspect ratio remains constant during the whole expansion \cite{Henn2009}. The above are presented in Fig.~\ref{fig:AspRatio}, where (a) shows the free expansion of the three clouds (thermal cloud, regular BEC and a turbulent condensate) and (b) shows the aspect ratio evolution of each system.
The evolution of the aspect ratio of a turbulent condensate does not display the inversion observed in a TF condensate. In fact, for the cigar shape trap used in \cite{Henn2009}, the turbulent condensate expands keeping the same relative dimensions once released from its confinement. This property of the self-similar expansion of the turbulent cloud may be a possible signature of the emergence of quantum turbulence in a trapped BEC and its origin and deeper consequences are currently under investigation. A particular theoretical hydrodynamic model to treat the regular and turbulent superfluid free expansion obtained in Ref.~\cite{Caracanhas2013} is briefly reviewed in what follows.

To characterize the anomalous expansion of the turbulent sample, a generalized Lagrangian approach \cite{Pitaevskii2003bose} was applied, and the associated Euler-Lagrange equations were derived to describe the dynamics of the condensate cloud (i.e. collective modes and free expansion).
Considering the Feynman's description of the turbulent regime as a tangle vortex configuration \cite{Feynman1955}, we produced a hydrodynamic model for the turbulent cloud. The contribution of a vortex tangle to the kinetic energy was added to the system Lagrangian, through a phase in the variational ansatz, as will be shown.

Assuming a GP ansatz $\psi(\mathbf{r},t)=|\psi(\mathbf{r},t)| e^{iS(\mathbf{r},t)}$ for a system of $N$ particles of mass $m$, the energy functional of our system is given by
\begin{equation}
\varepsilon[\psi(\mathbf{r},t)] = \frac{\hbar^{2}|\nabla\psi(\mathbf{r},t)|^{2}}{2m} + V_{\text{trap}}|\psi(\mathbf{r},t)|^{2}+\frac{1}{2}g|\psi(\mathbf{r},t)|^{4}
\end{equation}
where $g = 4 \pi \hbar^{2}a_{s}/m$, and $V_{\text{trap}}(\mathbf{r})$ is the trap potential. The Lagrangian  of the system is the functional
\begin{equation}\label{Lagrangian}
\mathcal{L} =\int d^{3}r \left[i \frac{\hbar}{2}\left(\psi^{*}(\mathbf{r},t)\frac{\partial\psi(\mathbf{r},t)}{\partial t}-\psi(\mathbf{r},t)\frac{\partial\psi^{*}(\mathbf{r},t)}{\partial t}\right)-\varepsilon[\psi(\mathbf{r},t)]\right]
\end{equation}
for some function $\psi(\mathbf{r},t)$ that can depend on various variational parameters (here, the time-dependent TF radii and the expansion parameters).

Let us now choose the Thomas-Fermi (TF) trial function
\begin{eqnarray}\label{eq:Ansatz}
\psi_{TF}(\mathbf{r},t)&=&e^{iS(\mathbf{r},t)}\,\prod_{j=x,y,z}e^{i\beta_{j}m\,x_{j}^{2}/2\hbar}\,\sqrt{n_{0}}
\left(1-\sum_{j=x,y,z}\frac{j^2}{R_{j}^{2}}\right)^{1/2} \Theta \left(1-\sum_{j=x,y,z}\frac{j^2}{R_{j}^{2}}\right),
\end{eqnarray} where $\Theta(\dots)$ is the Heaviside step function and $n_{0} =15 N/(8 \pi R_{x}R_{y}R_{z})$, for the normalization to be satisfied. The phase $\phi_{j} =\beta_{j}\,m\,x_{j}^{2}/2\hbar$ of this trial function gives rise to an irrotational linear velocity field $u_{j} =\beta_{j}\,x_{j}$, with $j=x,y,z$. With this trial function, Eq.~\eqref{Lagrangian} yields
\begin{eqnarray}\label{L0}
\mathcal{L}=-\frac{Nm}{14}\left( \sum_{j=x,y,z} R_{j}^{2} \dot{\beta}_{j}
 +  \sum_{j=x,y,z}\beta_{j}^{2}R_{j}^{2} \right)
-\frac{Nm}{2}\langle\upsilon^2\rangle-\frac{15}{7}\frac{\hbar^{2}\,a_s\,N^{2}}{m}\frac{1}{R_{x}R_{y}R_{z}},
\end{eqnarray}
where we rewrote the additional velocity field as $|\frac{\hbar}{m}\nabla S(\mathbf{r},t)| = \upsilon$, that will be calculated in the following, considering the local velocity associated with the individual vortices. The brackets $\langle...\rangle$ represent spatial integration.
We calculated the term $\langle\upsilon^{2}\rangle$ for the energy of the individual vortices by integrating
the local vortex line energy density \cite{Pethick2008,Pitaevskii2003bose}. Associated with any given single vortex is an energy per unit length:
\begin{equation} \label{eq:density}
\varepsilon_{vort} = \frac{\pi n \hbar^2}{m}\ln\left(\frac{\ell}{\xi}\right),
\end{equation}
where $\ell$ is the ef{}fective intervortex separation, $\xi$ is the healing length (and, hence, the approximate vortex core size) and $n$ is the local number density. Assuming a mean vortex line length $L$ per unit volume, the total energy from the random vortex configuration becomes $E_{vort}^t=L N \varepsilon_{vort}/n$.

From Eq.~\eqref{L0} we immediately derive the Euler-Lagrange equations with respect to each of the parameters of Eq.~\eqref{eq:Ansatz}.
We chose the scaling $L \propto (R_{x}^{2}+R_{y}^{2}+R_{z}^{2})^{-1}$, so that the resulting dynamical equations be in accordance to the equations obtained previously in Ref.~\cite{Caracanhas2012}. Finally, the Lagrangian equations of motion are \cite{Caracanhas2013}:
\begin{eqnarray}
\ddot{R}_{j}=\;
14\pi\;\frac{\hbar^2}{M^{2}}\;\frac{L_{0}(R^{2}_{0x}+R^{2}_{0y}+R^{2}_{0z})R_{j}}{(R_{x}^{2} +R_{y}^{2}+ R_{z}^{2})^{2}}\ln \left( \frac{\ell}{\xi }\right)+\;15\;\frac{\hbar^2\;a_s}{M^{2}}\;\frac{N}{R_{j}R_{x}R_{y}R_{z}}\; , ~~~j=x,y,z,
\label{eq:Dyn}
\end{eqnarray} 
where $R_0$ and $L_{0}$ corresponds to the initial dimensions and the vortex line density of the cloud before the expansion.

\begin{figure}[htb]
\centering
\includegraphics[width=0.80\textwidth]{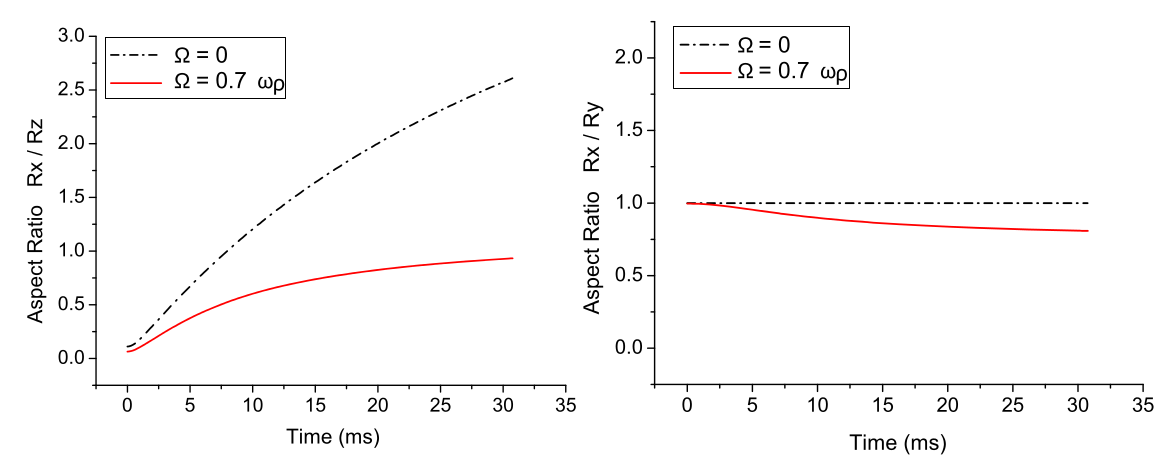}
\caption{Aspect ratio evolution during expansion of the gas: comparison between the free expansion of a regular (dashed line) and turbulent (filled line) BEC, with the corresponding cloud vorticity $\Omega$ given in terms of the highest trap frequency $\omega_r$. We see that for relatively large $\Omega$, hence large vortex number, the aspect ratio of the gas remains almost constant. Source: paper of Caracanhas \emph{et al.} \cite{Caracanhas2012}}
\label{fig:AspectRatio2}
\end{figure}

In Fig.~\ref{fig:AspectRatio2} we present the prediction of the present hydrodynamic model for the evolution of the cloud size (the ratios of the axial to radial cloud extensions) during the expansion of the gas, for two different values of $\Omega$. The latter is an adjustable parameter of our model [see Eq.~\eqref{Eq:OmVort}], related with the total number of vortices inside the cloud, and it will be discussed further in the next section. The calculations for the aspect ratio agreed with the experimental curves, showing that the physics behind the anomalous expansion of the turbulent system is captured by the generalized hydrodynamic model.

A close analysis of Eq.~\eqref{eq:Dyn} reveals the sources of the anomalous expansion. These are the extra kinetic energy $E_{vort}^t$ of the tangled vortices (the first term of the right-hand side of Eq.~\eqref{eq:Dyn}), which comes from the velocity field of the phase $S(\mathbf{r},t)$ and the repulsive interparticle interaction (the second term of the right-hand side of this same equation).
For long expansion times, that is, large values of the cloud dimensions, the vortex term dominates, since it scales with three inverse powers of $R_j$, whereas the interaction term has four inverse powers of these mean radii. When the vortex term dominates over the interaction (asymptotic limit), the equations acquire a simple form with $\ddot{R}_{j}/R_{j}$ independent of the index $j$, as can be verified from Eq.~\eqref{eq:Dyn}. That can explain the apparent scale invariance (self-similarity) of the turbulent cloud aspect-radio, within the time interval of the experimental observations. 

The hydrodynamic model also allows for a comparison between the interaction and kinetic energy of the cloud during the expansion, with the energy functional calculated from the solution of the dynamic equations \cite{Thompson2014}. For a vortex-free condensate, it is well known that the (interparticle) interaction energy determines the aspect ratio of the cloud \textit{in situ} and in free expansion \cite{Pitaevskii2003bose}. In a turbulent system, however, there is a competition between the repulsive interaction and the extra kinetic energy provided by the vortices.

In this same context, it is also possible to demonstrate that the asymmetry of the trap has an important contribution. As assumed before, the kinetic energy has now a crucial role in the expansion: it balances the effects of the interaction energy that, for a cigar shaped trap, affect mainly the tightly confined dimension of the trapped atomic cloud. The inhibition of the aspect ratio inversion in the turbulent cloud is related to a higher asymptotic velocity field along the less confined direction (axial direction of the cigar shape trap). That can be explained based on the preferential alignment of the vortex lines along the short axis, since the latter has lower probability for the bending the vortex lines. Then, the centrifugal contribution from the vortex velocity field is expected to completely change the expansion of the axial dimension.

\subsection{Frozen Modes}
Another important aspect worth attention is how the turbulent regime affects the dynamics of the collective modes of the parabolically trapped condensates. Again, this system can be modelled through the generalized Lagrangian formalism for the turbulent condensate of Sec.~\ref{SubSec:self-similar}. Equations~\eqref{eq:Dyn} can be linearized and from there the normal frequencies of the low energy collective modes extracted. Analytical solutions for the normal-mode frequencies can be obtained, assuming the cigar shape geometry of the harmonic trap. The main steps of this calculation are summarized as follows.

With the Feynman's relation \cite{Feynman1955,Fetter2009} for a uniform vortex distribution, we can define the vorticity as a function of the vortex number
\begin{equation}
\Omega = \frac{h}{2m}\frac{N_{vort}}{Area},
\label{Eq:OmVort}
\end{equation}
where we assume $N_{vort}=8\pi\,N\,a_{s}/(e\,\bar{R}_{0})$, with $\bar{R}_{0}=\sqrt{R_{0x}^{2} +R_{0y}^{2}+ R_{0z}^{2}}$ and $e$ being the Euler's number. We scale the time and space dimensions in Eq.~\eqref{eq:Dyn}, respectively with $\omega_r$ and $l_{r}=\sqrt{\hbar/M\omega_r}$, to obtain the dimensionless system of differential equations ($i,j,k$ taking the values $x,y,z$):
\begin{eqnarray}
\tilde{\ddot{R}}_{i}+\lambda^{2}\tilde{R}_{i} =
\frac{P_{r}}{\tilde{R}_{i}^{2}\,\tilde{R}_{j}\,\tilde{R}_{k}}\;+\;\frac{P_{v}\,(\tilde{R}_{0j}^{2}+\tilde{R}_{0k}^{2}+\tilde{R}_{0i}^{2})}{(\tilde{R}_{0j}\,\tilde{R}_{0k}\,\tilde{R}_{0i})}\;\; \frac{\tilde{R}_{i}}{(\tilde{R}_{j}^{2}+\tilde{R}_{k}^{2}+\tilde{R}_{i}^{2})^{2}},
\label{eq:Dyn2}
\end{eqnarray}
where $P_{r} =15 \, N\,a_s/l_{r}$, $P_{v} = \left(48\pi/e\right)\,\left(N\,a_s/l_{r}\right)$ and the trap aspect ratio $\lambda = \omega_{z}/\omega_r$. We can linearize the system of Eq.~\eqref{eq:Dyn2} by substituting $\bar{R}_i = \bar{R}_{0i} + \delta_i(t)$ and keeping first-order terms only. Here $\bar{R}_{0i}$ corresponds to the stationary solution in the trap, calculated with $\ddot{R}_i = 0$ in Eq.~\eqref{eq:Dyn2} and $\delta_i$ is the deviation around this equilibrium value.

Assuming an extremely elongated trap ($\lambda\ll1$), we can substitute $(\tilde{R}_{0x}^2+\tilde{R}_{0y}^{2}+\tilde{R}_{0z}^{2})\sim\tilde{R}_{0z}^{2}$ to simplify the stationary equations
\begin{eqnarray}
\label{eq:Din}\nonumber
\tilde{R}_{0r} = \frac{P_{r}}{\tilde{R}_{0r}^3\,\tilde{R}_{0z}}\;+\;\frac{P_{v}}{\tilde{R}_{0r}\,\tilde{R}_{0z}^{3}}
\\
\lambda^{2}\tilde{R}_{0z} =
\frac{P_{r}}{\tilde{R}_{0z}^{2}\,\tilde{R}_{0r}^{2}}\;+\;\frac{P_{v}}{\tilde{R}_{0^{2}}^{2} .\tilde{R}_{0z}^{2}},
\end{eqnarray}
This gives the approximate solution
\begin{eqnarray}
\label{eq:Sol1}
\tilde{R}_{0r}=\left[\frac{P_{r}\,\lambda^{2/3}}{(P_{r}+P_{v})^{1/3}}\,\frac{1}{1-\frac{P_{v}\,\lambda^{2}}{(P_{r}+P_{v})}}\right]^{3/10}\sim\frac{P_{r}^{3/10}\,\lambda^{1/5}}{(P_{r}+P_{v})^{1/10}}
\nonumber \\ \tilde{R}_{0z} =\frac{(P_{r}+P_{v})^{1/3}}{\lambda^{2/3}}
\left[\frac{P_{r}\,\lambda^{2/3}}{(P_{r}+P_{v})^{1/3}}\,\frac{1}{1-\frac{P_{v}\,\lambda^{2}}{(P_{r}+P_{v})}}\right]^{5}
\sim\frac{(P_{r}+P_{v})^{2/5}}{P_{r}^{1/5}\,\lambda^{4/5}},
\end{eqnarray} where cylindrical coordinates are used.

Conserving now the linear terms in $\delta_i$, we can construct the coefficient matrix ($\delta_{r}$, $\delta_{z}$) of the resulting system. We can then find the characteristic polynomial of this matrix, whose roots will give the frequencies of the collective modes
\begin{equation}
\tilde{\omega}_b = \sqrt{2\,-\frac{P_{v}\,\lambda^{2}}{(P_{v}+P_{r})}}\sim\sqrt{2},
\label{eq:wb}
\end{equation}
and
\begin{equation}
\tilde{\omega}_q = \frac{1}{\sqrt{2}}\frac{1}{(P_{v}+P_{r})}\left[A_{1}-\sqrt{A_{2}}\right]^{1/2},
\label{eq:wq}
\end{equation}
with $A_{1} = (3\,\lambda^2+4)\,P_{v}^2+(6\,\lambda^2+8+8\,\lambda^4)\,P_{r}\,P_{v}+(3\,\lambda^2+4)\,P_{r}^2$ and $\; A_{2} =(16-40\,\lambda^2+25\,\lambda^4)\,P_{v}^4+(-136\,\lambda^2+48\,\lambda^6+208\,\lambda^4+64)\,P_{r}\,P_{v}^3
+(350\,\lambda^4+96-168\,\lambda^2+64\,\lambda^8)\,P_{r}^2\,P_{v}^2 +(-48\,\lambda^6-88\,\lambda^2+64+176\,\lambda^4)\,P_{r}^3\,P_{v}+(-16\,\lambda^2+9\,\lambda^4+16)\,P_{r}^4$.
\normalsize
Those solutions are associated with the breathing ($\omega_b$) and the quadrupole ($\omega_q$) modes respectively \cite{Pethick2008,Pitaevskii2003bose}. While the breathing mode corresponds to an in-phase oscillation of the radial and axial directions of the condensate, the quadrupole mode is associated with an out-of-phase oscillation of these widths. For $P_{v}=0$ we recover the expected values for a regular BEC \cite{Stringari2003}
\begin{equation}
\tilde{\omega}_b = \sqrt{2}
\end{equation}
\noindent and
\begin{eqnarray}
\tilde{\omega}_q = \frac{1}{\sqrt{2}}\left[3\,\lambda^2+4\pm\sqrt{-16\,\lambda^2+9\,\lambda^4+16}\right]^{1/2}.
\end{eqnarray}
Finally in the limit $\lambda\ll1$, we can simplify Eq.~\eqref{eq:wq} as
\begin{equation}
\tilde{\omega}_q \rightarrow  \frac{1}{\sqrt{2}}\,\left[5+ \frac{3\,P_v}{(P_{v}+P_{r})}\right]^{1/2}\lambda,
\end{equation}
or
\begin{equation}
\omega_q = \tilde{\omega}_q \,\omega_r = \frac{1}{\sqrt{2}}\,\left[5+ \frac{3\,P_v}{(P_{v}+P_{r})}\right]^{1/2}\omega_z.
\end{equation}
Using the values for $P_{v}$ and $P_{r}$ that correspond to the experimental conditions, we find $\,\omega_q\, $ $\,20\%\,$ higher than the value expected for a normal BEC in this same trap conditions. This indicates that it is energetically more costly to excite these modes in the turbulent cloud.

\subsection{Momentum distribution of a turbulent trapped BEC \label{SubSec:MomDist}}
Not only the collective modes are affected by turbulence, but also the momentum distribution are expected to be considerably altered. This topic has been explored in two recent paper \cite{Bahrami2015,Fritsch2015}, that continued studies commenced in Ref.~\cite{Thompson2014}. Some of the new achievements will be briefly described in this Section.

In the above experimental works, the free expansion of the atomic cloud was used to evaluate the momentum distribution \textit{in situ}. It was assumed that after release of the gas from the trap, the atoms expand ballistically. With this, it is possible to relate the radial position $r$ after some time $\tau_f$ (TOF) with the \textit{in situ} momentum, $k$, using the relation \cite{Thompson2014} 
\begin{equation}
r = \frac{\hbar~\tau_f}{m}k,
\end{equation}
(see also Sec.~\ref{SubSec:WhatExp}). Performing an angular integration of the number of particles in a concentric spherical shell with radius $r$ around the center of mass, the radial density distribution $n(r)$ is obtained. Converting $r$ into $k$ with the above expression yields the momentum distribution $n(k)$. However, as the momentum distribution is extracted from the absorption image of the expanded cloud (i.e. a 2D projected image of the atomic cloud), the result is a projected momentum distribution of the cloud, $n'(k')$.

\begin{figure}[htb]
\centering
\includegraphics[width=0.5\textwidth]{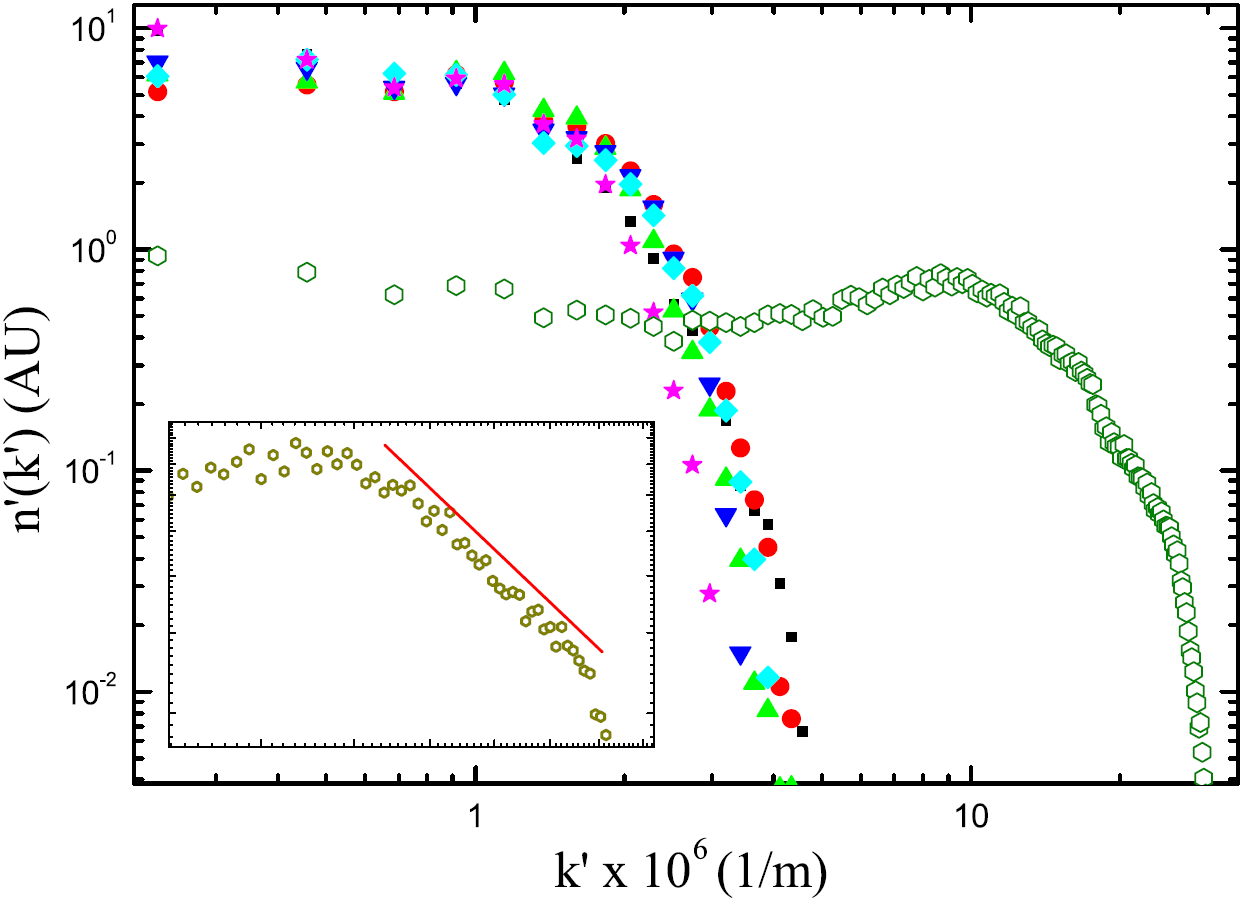}
\caption{Two-dimensional projected momentum density, $n'(k')$, on a log-log plot. The Thomas-Fermi condensates and condensates with a low number of vortices are shown in closed symbols. The $\blacksquare$, $\bullet$, $\blacktriangle$, $\blacktriangledown$, $\blacklozenge$ and $\bigstar$ symbols represent condensates with 0, 1, 2, 3, 4, and 5 vortices, respectively, and the open symbol is the averaged data from all the different realizations of turbulent states. The inset is a zoom of the linear regime seen in the turbulent data shown with a line of a slope of $-2$ to guide the eye. Source: paper of Thompson \emph{et al.} \cite{Thompson2014}.}
\label{fig:nk}
\end{figure}

Using this procedure, the authors of Ref.~\cite{Thompson2014} could evaluate the momentum distribution of the BEC in the three dif{}ferent regimes: equilibrium (no excitation), regular vortices and turbulent regime. The results  are shown in Fig.~\ref{fig:nk}. The behavior of the momenta for the BEC at equilibrium and the cloud with vortices are similar: on a logarithmic scale, it is a nearly constant distribution for small $k'$ values, followed by a smooth and steep drop-off at larger momenta. The turbulent cloud momentum distribution, though, has a very different behavior: a nearly constant distribution for momenta lower than a critical value $k'=1\times 10^7$~m$^{-1}$ followed by an abrupt change from a constant to a linearly (on a log-log plot) decreasing density. In the linear region, the density sharply decreases with a slope of $-2$ as the inset of Fig.~\ref{fig:nk} shows, with a behavior resembling a Kolmogorov power law. It is important to note, however, that the particular experimental configuration -- the method used to inject energy -- strongly affects the measured momentum distributions, not allowing a direct connection of the extracted power law in this trapped system to the Kolmogorov law. 
This is due to the finite size of the sample: the range of experimentally available length scales is much smaller compared to those of other classical or quantum fluids (see also discussion in Secs.~\ref{SubSec:QT} and \ref{SubSec:Differences}). Another peculiarity is the uncontrolled turbulent injection mechanism, that prohibits the settlement to a steady turbulent state and the establishment of an appropriate inertial range in the spectra.

In order to further the studies of the turbulent state, the authors of Refs.~\cite{Bahrami2015,Fritsch2015} realized a series new experiments, using the same techniques as before. Changing the excitation coils position, they could optimize the new apparatus to excite the dipole and quadrupole collective modes, and showed how the amplitude $A_{\text{exc}}$ and the time of the applied excitation $t_{\text{exc}}$ affected the mode frequencies and amplitudes \cite{Fritsch2015}. The novelty in this series of results is that the hold time $t_{\text{hold}}$ was varied and the reproducibility of the cloud dynamics thus enhanced.

\begin{figure}
\centering
\includegraphics[width=0.5\textwidth]{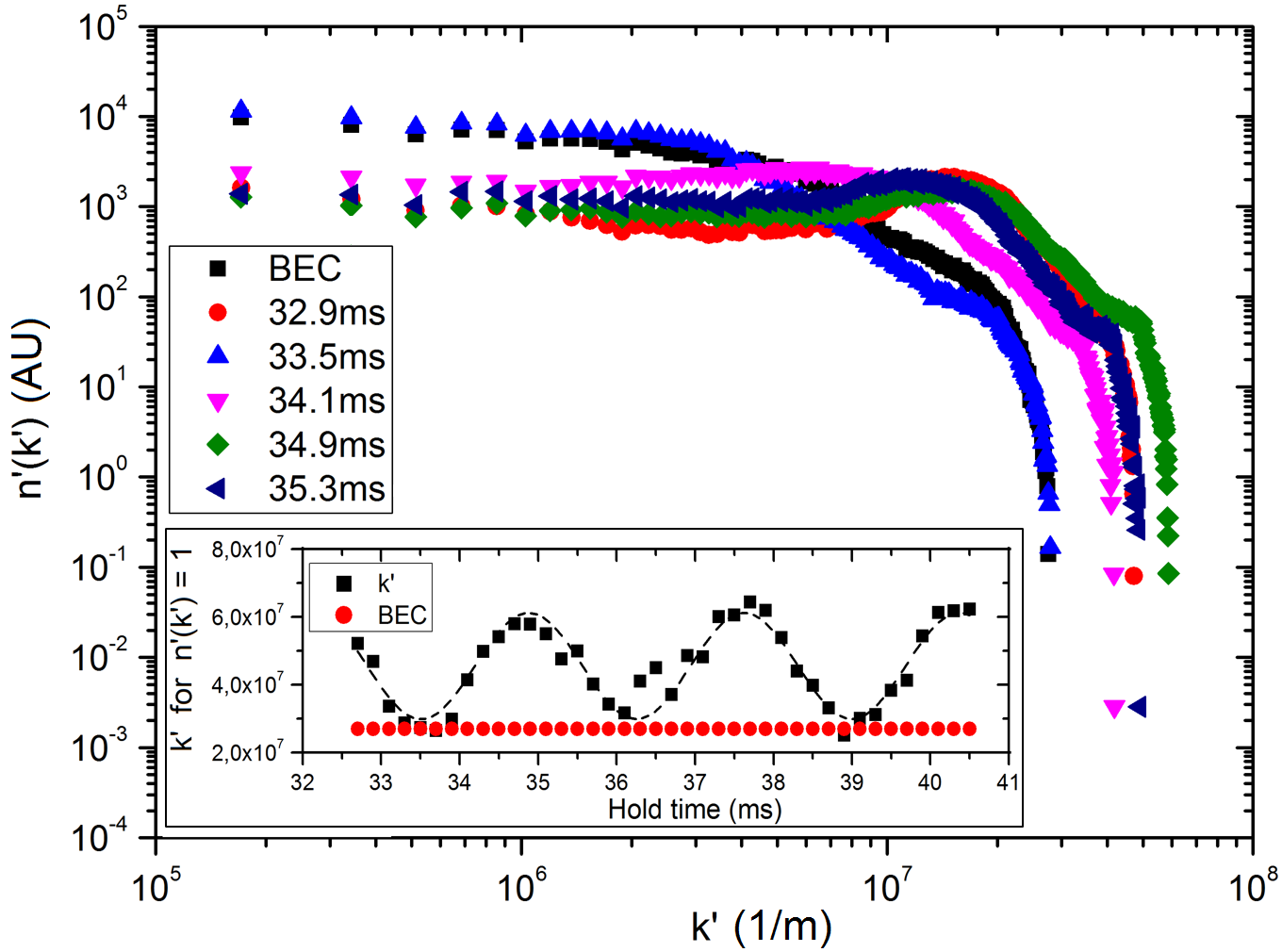}
\caption{2D momentum distribution extracted for different \emph{in situ} evolution times $t_{\text{hold}}$ for excitation amplitudes 470 mG/cm. The inset is the oscillation of the maximum momenta of the cloud $k'_{n'(k')=1}$ as a function of the evolution time showing a sinusoidal behavior with characteristic frequency matching the quadruple mode frequency of the BEC. Source: paper of Bahrami \emph{et al.} \cite{Bahrami2015}.}
\label{fig:nkXthold}
\end{figure}

Increasing the energy input to the system (excitation time and amplitude of the perturbation), the presence of irregularities in the density profile as well as the anomalous cloud expansion were confirmed (conservation of the initial aspect ratio). Both are related to the presence of vortices and turbulence in the sample. Finally, momentum distributions of the excited cloud were extracted following the same procedure as before \cite{Thompson2014}. In contrast to the previous data, where the hold time $t_{\text{hold}}$ was fixed, in Ref.~\cite{Fritsch2015} the momentum distribution was studied during different hold times, as seen in Fig.~\ref{fig:nkXthold}. 
It can be seen that there are two extreme curves in this graphic: one is a `stretched' momentum distribution (the green curve at $t_\text{hold} = 34.9$~ms) and exhibits a power-law behavior with higher slope; the second, is a squeezed distribution (the blue curve at $t_\text{hold} = 33.5$~ms), which is close -- but not equal -- to the equilibrium BEC momentum distribution, revealing a power-law behavior with a lower slope. As the hold time is changed, the behavior of the momentum distribution oscillates between these two extremes. In order to characterize this oscillation, the authors of Ref.~\cite{Bahrami2015} analyzed the quantity $k'_{n'(k')=1}$, i.e. the value $k'$ of the momentum where the momentum distribution crosses the level $n'(k')=1$, versus the hold time (see inset of Fig.~\ref{fig:nkXthold}). They found that this oscillation has the same frequency as it was found for the quadrupolar mode \cite{Fritsch2015}. Furthermore, the behavior of the amplitude and the frequency of this oscillation, when the excitation amplitude $A_{\text{exc}}$ is increased, is the same as the one obtained for the quadrupolar mode \cite{Fritsch2015}. These results showed that the momentum distribution does not reveal only the turbulent regime but is also coupled to the cloud collective modes. This is another important effect in trapped gases that has to be considered when determining the power laws from the momentum distribution of a turbulent BEC.

\section{New directions and challenges \label{Sec:Discussion}}
As in the preceding we presented the principal features of quantum turbulence in trapped atomic gases, it is timely to summarize, give concluding remarks and also suggest new lines of research. 

\subsection{What we have learnt \label{SubSec:WhatHaveWeLearnt}}
\vspace{-4mm}
~\\
\emph{Quantum turbulence is real!}~~Dilute atomic gases forming a superfluid Bose-Einstein condensate at ultralow temperatures inside a trap can indeed show turbulent behavior. This appears as an excited out-of-equilibrium state with an increasing (in time) number of interacting, bent and non-parallel vortices. Turbulence gives its signature in the anomalous expansion of the gas, once the confining potential has been turned off. Moreover, evidence of power laws in the distribution of momenta has been measured. However, these are indirect time-of-flight measurements and more precise future studies are needed in order to shed more light.

Due to their unique features and degree of controllability BECs can serve as laboratories for testing theories of quantum physics. In this context, turbulence has a new platform where it can be controllably created and engineered offering insight that would otherwise be not within approach. Indeed, a turbulent BEC inherits characteristics from classical and superfluid turbulence. By changing, for instance, the strength of the interparticle interaction one effectively changes the size of the system and the size of the vortex filaments, resulting thus in the variation of the available spatial scales. Quantities as healing length, vortex core size and intervortex separation can in principle be modulated in the laboratory, by changing the applied external magnetic fields. Hence, the extension over which the various (ultra)quantum or (quasi)classical manifestations of turbulence appear can be varied at will. 

Spontaneous or driven nucleation of a large number of vortices -- a central entity in any turbulent system -- is possible with present-day techniques and this is precisely why ultracold gases have become a unique platform for manifestation of QT. As we discussed, vortices can be nucleated under various circumstances and following different methods. In the laboratory of S\~ao Carlos vortices result from the interaction of the thermal cloud with the superfluid, when the system is subject to external oscillatory perturbations. Vortex nucleation, proliferation, reconnection and oscillations are all mechanisms that assist turbulence and hence merit further scrutinizing.

\paragraph{Some concrete criterion for distinction}
As discussed in detail in Sec.~\ref{Sec:Experimental}, a turbulent gaseous BEC will expand in a self-similar fashion once the trapping potential is switched off. This is a unique and novel characteristic that pertains to gases only and distinguishes a BEC from other quantum or classical turbulent systems. Note that, the anomalous expansion (see Sec.~\ref{SubSec:self-similar}) as contrasted to the regular expansion of a gaseous BEC is \emph{purely} a quantum phenomenon with no classical analogue.

\paragraph{The observed QT can be of mixed type} Trapped BECs can -- in principle -- show hydrodynamic (owing to chaotic motion of vortex tangles), quasiclassical or ultraquantum  turbulence (due to the restricted available length scales and method of vortex nucleation). Since nonlinear wave excitations are involved, it is probable that wave turbulence plays a role too, even though experimental evidence and also theoretical clarification is lacking.

\subsection{Open questions \label{SubSec:OpenQuestions}}

\paragraph{Beyond-mean-field aspects of QT decay}
A deterministic study of $T=0$ gaseous BECs is most of the times performed by applying the Gross-Pitaevskii equation. According to it, a system is always coherent and condensed in one single mode. However, as we have seen, energy transfer in turbulent gases involves and requires a plethora of acoustic excitations, even at $T=0$. This can, under circumstances, mean that highly excited gases are not coherent any more and one cannot neglect quantum effects as fragmentation and nontrivial correlations. Studying turbulence in a true many-body context, i.e. beyond mean-field, even if it is computationally challenging, can help in elucidating the behavior of turbulent systems. For instance, it has been shown that for undercritical rotation of a 2D trapped gas vortices are seen in the (first-order) correlation function before they are formed in the density and fragmentation evolves rapidly \cite{Weiner2014,Tsatsos2015}. 

Of particular theoretical interest would be the derivation of the hydrodynamic equations of motion [Eqs.~\eqref{Eq:continuityQuant} and ~\eqref{Eq:QuantumEuler}] for the full many-body quantum state $\Psi=\Psi(\mathbf r_1,\mathbf r_2,\dots,\mathbf r_N)$. One could even look for the corresponding hydrodynamic form of the MCTDHB equations of motion, which formalize the relationship of the full many-body state $\Psi$ to a finite set of orbitals. In such a way, one could study the contributions of beyond-mean-field physics to the onset of turbulence. 

Moreover, there are situations where fragmentation is known to appear in the ground state of trapped gases, as for instance in spinor BECs \cite{Mueller2006,Ho2000}. There, a mean-field treatment will fail and more elaborate methods are needed in order to study highly excited and turbulent states. 

\paragraph{More length scales available} In Sec.~\ref{Sec:Scales} we saw that the length scales available in trapped gases are limited both by the size of the system $L$ and the healing length $\xi$. The quantity $\log (L/\xi)$ gives us an estimation of how many decades (or length scales) are available in the trapped system; typically 1 or 2. The healing length $\xi$ however can be controlled by changing the externally applied magnetic fields. The reasoning is that, at equilibrium, the healing length is a function of the interaction strength $g$: the stronger the particles interact the tighter they sit together and so the less space it takes for a perturbation to be `healed'. By utilizing Feshbach resonances one can change the interaction strength and hence $\xi$. Furthermore, the Thomas-Fermi radius of the equilibrated system, i.e. the extension $L$ of the gas, increases as the interparticle interaction grows larger. Hence, two ends achieved with a single effort: increase $g$ to increase the numerator of $L/\xi$ and decrease its denominator at the same time. So, for trapped systems with enormously strong interacting particles one expects a very big number of scales to be able to participate in the various energy-transfer processes. Such a system would consist a valuable playground for exploring manifestations of quasiclassical turbulence and self-similar energy cascade, along with ultraquantum turbulence. Naturally, other parameters as well (trap frequencies and particle number) can be utilized to this end. Since a very large interaction strength will deplete the ground state causing fragmentation, this situation relates to the preceding paragraph. 

\paragraph{Visualization and direct measurements} Maybe the most important and urgent advances needed in the area of trapped ultracold gases are these related to the visualization of turbulent and perturbed gas. Technological progress along with pioneering ideas can assist in a direct measurement of the momentum distribution \emph{in situ} and the specific spectral dependence of the energy on the momenta. For instance, more elaborate experimental techniques could achieve direct observations, either \emph{in situ} or after expansion, of the velocity distribution of systems that are believed to be turbulent. Thus, the actual occupation of different wavelengths and momentum scales could be extracted.

Another advance would be the examination of the time-of-flight images of the BECs and their relation to the \emph{in situ} state. To date, it has been assumed that the vortex relative positions and configuration of a BEC involving vortices and other excitations will be `frozen' once the confining trap of the gas has been released. Hence, the time-of-flight measurements are considered to be reliable sources of information of the \emph{in situ} turbulence and its level of development. However, a more strict quantification of this process, presumably by comparing \emph{phase contrast} images taken inside the trap with ones taken after expansion, is needed. Related is the work of Ref.~\cite{Gerbier2008} on an optical lattice, where deviations of the momentum distribution from the time-of-flight images are reported.

In principle any imaging method that can distinguish and measure long-scale flow could help in determining the quasiclassical or not nature of the observed turbulence. In the absence of tracing particles in the case of trapped gases, any innovative technique for visualizing turbulence at its various stages would consist a significant advance in our understanding of QT.

\subsection{Related topics and new directions \label{SubSec:Related}}

\paragraph{Equivalence of the turbulent cloud to the propagation of a speckle light pattern}
Associated with the demonstration of the emergence of a turbulent regime in a sample of trapped superfluid $^{87}\rm{Rb}$ atoms and the study of its characteristics \cite{Henn2009a,Henn2010} is the observed self-similar expansion. Contrary to a regular BEC, as discussed in Sec.\ref{SubSec:self-similar}, when a turbulent cloud of  atomic superfluid is let to expand, the aspect ratio inversion is completely suppressed \cite{Henn2009}. Using the analogy between an atomic mater wave and a light field, we can associate the turbulent cloud in expansion with a speckle pattern of a light field that propagates. While an ordinary matter wave `diffracts' during expansion causing the inversion of the aspect ratio, the turbulent cloud has a random distribution of both density and phase, decreasing thus considerably its coherence length: the propagation is equivalent to a speckle pattern. Investigation of the density-density correlations shows signatures equivalent to the intensity-intensity correlations in the light field. Those aspects demonstrate that a turbulent drop of superfluid can be considered as a speckle field of matter wave.

\paragraph{Bosonic mixtures} 
Bosonic gases consisting of two (or more) different species -- for instance $\rm K$ and $\rm Na$ particles -- can have qualitatively different properties than the single-species gases studied so far. In Refs. \cite{Takeuchi2010,Ishino2011} counterflow turbulence in binary bosonic mixtures was investigated. Such a mixture can be considered as the analogue of quantum turbulence in $^4$He at relatively high temperatures, in which superfluid and normal fluid are turbulent at the same time. The importance of developing the study of quantum turbulence in this direction is also that a mixture of two different condensates can be used to simulate classical systems such as magnetohydrodynamic (MHD) ones  \cite{Brandenburg2005} which are well-known for their difficulty and lack of universality \cite{Lee2010,Dallas2013}. In MHD turbulence, a magnetic, electrically conducting, turbulent fluid is coupled to a turbulent magnetic field. There, self-organization processes and other exotic phenomena, absent in hydrodynamic turbulence, manifest \cite{Dallas2015}. Even more interestingly, a mixture of three species or more will go \emph{de facto} beyond any classical analogue.

Another interesting situation arising in mixtures, is the use of one particle species as means to visualize the second. In specific, one could utilize the first species as a matrix, that can take fingerprints of turbulence created in the other species. In other words, the circumstances and experimental settings should be assessed, where turbulence (or some of its properties) can be controllably transferred between the two species. This approach could open the new ways in visualizing quantum turbulence.

Moreover, one can ask whether the miscibility of the composite gas changes in the turbulent phase: will turbulence homogenize parts of the, otherwise separated, system?

\paragraph{Spin turbulence} Another interesting possibility that has been suggested is this of turbulence in a quantum mixture of different spin states. That is, a Bose-Einstein condensate consisting of particles that have a spin degree of freedom (see Ref.~\cite{Stamper-Kurn2013} for a review). The principle and qualitative difference between a spinor BEC and a mixture of different bosonic species is that the former permits population transfer from one spin-state to the other (spin-exchange collisions) whereas in the latter the particle number in each state is fixed. A hydrodynamic description of such spinor BECs has been delivered by Kudo, Yukawa and collaborators \cite{Kudo2010,Yukawa2012}. Soon after, it was shown that turbulence can exist in the spectrum of the (ferromagnetic) spin-dependent interaction energy, that -- in contrast to the traditional Kolmogorov spectrum -- obeys a $-7/3$ power law \cite{Fujimoto2012,Fujimoto2012b}. Turbulence in that case is triggered by driving the system far from the ground state (to a highly excited state) and then watching its dynamics. In that situation the spin density vectors are spatially disordered and, moreover, frozen in time resembling thus spin glass patterns \cite{Tsubota2013}. The mechanisms that transfer spin turbulent energy are not the same as in the hydrodynamical turbulence, since vortices or topological excitations were not created. It is probable that this is a manifestation of wave turbulence (see Sec.~\ref{SubSec:WT}), where cascades of spin waves play the main role in transferring energy. The work of Ref.~\cite{Tsubota2014b} summarizes the known, very recent results towards this direction.

In an independent work, the authors of Ref.~\cite{Villasenor2014} showed how a gas of two spin states driven by an oscillating magnetic field shows Kolmogorov power spectra. However, this is still hydrodynamic turbulence and not of the type of spin turbulence proposed in the references above.
 
It is worth mentioning the work of Cornell and collaborators of 2004 \cite{Schweikhard2004}, where a vortex lattice was created in a gas of $\sim 10^6$ $^{87}$Rb particles, occupying either of two internal spin states $|1\rangle\equiv|F=1,m_F=-1\rangle$ and $|2\rangle\equiv|F=2,m_F=1\rangle$. The initial configuration was that of a hexagonal vortex lattice with a relative population of $20-80\%$ in each spin state. This prepared state of coinciding vortex cores is not energetically favorable and hence the gas evolves into a state where vortices are shifted so that the density of the state $|1\rangle$ fills in the cores of the state $|2\rangle$. Interestingly, the vortex separation process passes through intermediate states, where the vortices fade out temporarily and reappear while the particles and vortices relocate within the cloud. These intermediate state are reported to be turbulent. 

In the above cases of turbulence in spinor systems the spin turbulence is actually not coupled to the density (mass) turbulence. A coupling and controlled transfer of energy (density-to-spin) can be a challenging future topic for both experimental and theoretical research.

Last, very recent numerical calculations suggest that a non-Abelian type of turbulence is possible in a spin-2 BEC \cite{Mawson2015}. The authors of this work present a scenario where vortex-antivortex lattices of fractional charge can emerge. 

\paragraph{Fermionic superfluids} 
Superfluidity can exist in ultracold atomic systems of fermions as well. It is manifested either as formation of molecules and eventual condensation of these or through a Bardeen-Cooper-Schriefer (BCS) mechanism, i.e. loosely bound pairs of fermions; the same ones that explains the emergence of superconductivity in metals. The former case has been achieved experimentally in ultracold lithium-6 \cite{Zwierlein2003,Zwierlein2004} whereas the latter in potassium gases \cite{Regal2004}. Interestingly, it has been possible, by tuning the external magnetic fields appropriately, to cross a Feshbach resonance and watch the crossover from the BEC to the BCS state \cite{Regal2004}. One then expects that, if a fermionic gas becomes superfluid, vortex states should also be feasible \cite{Bulgac2003}. Indeed, a landmark proof of the superfluidity of fermionic gases came with the work of Ref.~\cite{Zwierlein2005} where an array of vortices was successfully created on an ultracold gas of $^6$Li particles by a stirring process, similar to this followed in ordinary BECs. More interestingly, the arrays (or lattices) of vortices were seen to persist and exist in either possible manifestations of superfluidity: the BEC and the BCS states. Naturally, one raises the question whether fermionic ultracold gases can exhibit quantum turbulence as well. Would the emergence of tangled vortices and QT destroy superfluidity? Can one observe a BEC-BCS transition of a turbulent fermionic gas?

Partial answers came from recent theoretical studies. That a superfluid unitary Fermi gas below the critical temperature can sustain quantum turbulence was noted in Ref.~\cite{Wlazlowski2015b}, while in Ref.~\cite{Bulgac2014} vortices were studied in the BEC-BCS crossover. There the authors, using a mean-field local density approximation showed how vortex rings can form and propagate and compared their finding to the experiment of Ref.~\cite{Yefsah2013}. In the work of Ref.~\cite{Wlazlowski2015} vortex propagation and reconnections in unitary Fermi gases were studied. Moreover, a scenario where turbulence is generated has been proposed, where two vortex lattices -- initially separated in space and with a $\pi$ phase difference imprinted -- are let to mix and reconnect. 

Finally, the very recent appearance of the multiconfigurational time-dependent Hartree for Fermions (MCTDHF) \cite{Fasshauer2015} paves the way for accurate explorations of many-body dynamics of fermionic ultracold gases.

\vspace{2mm}
To conclude, it is worth mentioning that in the context of plasma physics \cite{Antar2003}, as well as nonlinear optics \cite{Dyachenko1991}, turbulence has been studied too.

\subsection{Discussion \label{SubSec:Discuss}}
Returning to classical fluids, turbulence still poses many questions in that domain. Physicists and applied mathematicians have been trying with a great deal of efforts to come up with precise formulations of the problem. Despite many difficulties encountered in classical fluids, there are many predictions and interpretations one can make based on a large variety of physical arguments. 

In quantum fluids, we characterize turbulence by the tangle configurations of vortex filaments. Due to the lack of a better definition most parts of the theory and experiments are looking for analogies to classical fluids. That, clearly, is a safe good track to follow because, independently of anything, analogies to classical fluids are always helpful. Nevertheless, we should not restrict ourselves to analogies only. It seems necessary to go deep into the quantum nature of fluids in order to investigate specific quantum structures like correlations and quantification of nonequilibrium states and more, in order to be capable of quantifying turbulence in quantum fluids in an unambiguous fashion. 

Much of what scientist are doing nowadays in quantum systems is equivalent to the twenty century fluid dynamics, focused on tracking vorticity distribution. While in classical fluids this might make sense, certainly in their quantum counterparts it loses meaning. As a long-standing tradition, any formal definition of classical turbulence has been avoided, seemingly as if there is a fear that a strict definition of turbulence would constitute the observed phenomena illusionary. In quantum fluids the same line of thought needs not be followed and a better or adequate definition might be encountered, without missing the concepts of QT. To this end, a lot of theoretical and -- even more -- experimental effort is needed.

In conclusion, turbulence in gaseous BECs is a relatively new field of research with promising future that gains increasing attention. The irregular flow regimes that characterize turbulence appear in many important processes in nature. Vorticity, central in superfluid systems, may present the most elementary ingredient of turbulence, and be a vehicle to the extensive study of turbulence in quantum gases that go beyond ordinary atomic BECs.  

\section*{Acknowledgements \label{Sec:ackn}}
\addcontentsline{toc}{section}{\nameref{Sec:ackn}}
We acknowledge financial support from FAPESP and fruitful discussions with Vassilios Dallas and Axel Lode. We thank Daniel Kleppner for useful comments. We also thank Emanuel Henn, Kilvia Magalh\~aes, Gustavo Telles, Daniel Magalh\~aes and Giacomo Roati for their long-standing contributions in the experimental investigations of quantum turbulence. CFB acknowledges access to the N8 HPC computing facilities. 
\vspace{40mm}

All authors declare no conflicting financial or other interests.
\newpage

\bibliographystyle{elsarticle-num}

\end{document}